\documentclass[12pt]{article}
\pdfoutput=1
\usepackage{jheppub}
\usepackage{amsmath}
\usepackage{MnSymbol}
\usepackage{ulem}
\def\be{\begin{equation}}
\def\ee{\end{equation}}
\def\ba{\begin{aligned}}
\def\ea{\end{aligned}}
\def\ben{\begin{eqnarray}\displaystyle}
\def\een{\end{eqnarray}}

\def\nn{\nonumber}
\def\rd{{\rm d}}

\def\bs{\boldsymbol }
\def\i{{\rm i}}

\preprint{
{\small{\textsf{DMUS--MP--15/09}}}}

\title{Factorisation and holomorphic blocks in 4d}

\author[1]{Fabrizio Nieri, Sara Pasquetti}

\affiliation[1]{Department of Mathematics, University of Surrey, Guildford, Surrey, GU2 7XH, UK}
\emailAdd{fb.nieri@gmail.com, sara.pasquetti@gmail.com}

\abstract{
We study $\mathcal{N}=1$ theories on Hermitian manifolds of the form $M^4=S^1\times M^3$
with  $M^3$ a $U(1)$ fibration over $S^2$, and their 3d  $\mathcal{N}=2$  reductions.
These manifolds admit an Heegaard-like  decomposition in solid tori $D^2\times T^2$ and $D^2\times S^1$.
We prove  that when the 4d and 3d anomalies are cancelled, the matrix  integrands  in the Coulomb branch partition functions can be factorised in terms of 1-loop factors  on $D^2\times T^2$ and $D^2\times S^1$ respectively. 
By evaluating the  Coulomb branch matrix integrals  we show that the 4d and 3d  partition functions can be expressed  as sums of products
of  4d and 3d holomorphic blocks.
}    

\begin{document}
\bibliographystyle{utphys}

\maketitle

\section{Introduction}

In recent years thanks to the development of a new method to formulate SUSY gauge theories on curved spaces initiated by \cite{Festuccia:2011ws} and to the  application of  Witten's localisation technique to  the path integral of theories defined on compact  spaces, a plethora of new exact results for SUSY gauge theories in various dimensions  have been obtained.

The focus of this note is on 4d  theories defined on Hermitian manifolds of the form $M^4=S^1\times M^3$
where $M^3$ is a possibly non-trivial  $U(1)$ fibration over the 2-sphere, and their 3d  reductions.
 These 4-manifolds can preserve 2 supercharges with opposite R-charge
and a holomorphic Killing  vector generating the  torus action on $M^4$ \cite{Dumitrescu:2012ha}, \cite{Dumitrescu:2012at}, \cite{Klare:2012gn}.
General results  \cite{Closset:2013vra}, \cite{Closset:2012ru} state that partition functions on these spaces 
do not depend on the  Hermitian  metric but are  holomorphic functions of the complex structure parameters and of the background gauge fields through the corresponding vector bundles.  Similar results hold for the 3d  $\mathcal{N}=2$ reductions  of these theories.

For these spaces it has also been observed that the partition function can be expressed in terms of simpler building blocks.
It turns out that for 3-manifolds $M^3_g$, which can be realised by gluing two solid tori  $D^2\times S^1$ with an  element
$g\in SL(2,\mathbb{Z})$, and likewise for 4-manifolds $M^4_g$ constructed from the fusion of two
  solid tori $D^2\times T^2$ with appropriate  elements in $SL(3,\mathbb{Z})$, 
the geometric block  decomposition is  very non-trivially  realised  also at the level of the partition functions.

This phenomenon was first observed for 3d $\mathcal{N}=2$ theories on $M^3_S=S^3$ and $M^3_{id}=
S^2_{id}\times S^1$ which were shown in  \cite{sfa} and  \cite{hb} (see also \cite{Taki:2013opa}, \cite{Hwang:2012jh}, \cite{Hwang:2015wna}) to admit a block decomposition
\be
\label{one1}
Z[S^3]=\sum_c \Big\| \mathcal{B}^{\rm 3d }_c\Big\|^2_{S}~,\quad Z[S^2_{id}\times S^1]=\sum_c \Big\| \mathcal{B}^{\rm 3d }_c\Big\|^2_{id}~,
\ee
where the 3d holomorphic blocks $\mathcal{B}^{\rm 3d }_c$ are solid tori $D^2\times S^1$ partition functions.
The two blocks are  glued by the appropriate $SL(2,\mathbb{Z})$ element $S$ or $id$ acting on 
the modular parameter of the boundary torus and on the mass parameters.
The sum is over the supersymmetric Higgs vacua of the theory which remarkably are the only states contributing to the sums in (\ref{one1}),
even though  these partition functions, although  metric independent, are not properly topological objects.
In fact, in the case of   $M^3_S=S^3$, the factorisation was proved  to follow from a  stretching invariance argument  \cite{Alday:2013lba}.  Indeed in \cite{Alday:2013lba} it is shown that it is possible to deform the $S^3$ geometry into two cigars $D^2\times S^1$ connected by a long tube, which effectively projects the theory into the SUSY ground states,  without changing the value of the partition function.

In \cite{hb} it  was developed an integral formalism to compute the holomorphic blocks which build on the fact that they are solutions to a set of difference equations. The 3d blocks are obtained by integrating a meromorphic one-form $\Upsilon^{\rm 3d}$,
consisting of the mixed Chern-Simons,  vector and chiral multiplet contributions on $D^2\times S^1$, on
an appropriate basis  of  middle-dimensional cycles in $(\mathbb{C}^{*})^{|G|}$
\be
\label{two2}
\mathcal{B}^{\rm 3d }_c=\oint_{\Gamma_c}\Upsilon^{\rm 3d}~.
\ee
Later on, in   \cite{Yoshida:2014ssa},  block integrals were derived  from localisation on $D^2\times S^1$.
Curiously the integrand $\Upsilon^{\rm 3d}$ turns out to be the ``square" root of the 
integrand appearing in the Coulomb branch partition function on the compact space,
 so that by combining (\ref{one1}) and (\ref{two2}) one finds
\be
\label{three3}
 Z[M_g]= \sumint \Big\|\Upsilon^{\rm 3d}\Big\|^2_{g}= \sum_{c}\Big\|\mathcal{B}^{\rm 3d}_c\Big\|^2_g
  =\sum_{c}\Big\|\oint_{\Gamma_c}\Upsilon^{\rm 3d}\Big\|^2_g~,
\ee
where the gluing rule can be $g=S,id$.
The first term of the equality is a smart  rewriting of the partition function
on the Coulomb branch, where the localising locus may contain a continuous and 
a discrete part. As observed in  \cite{hb} this suggestive chain of equalities hints that factorisation commutes with integration.

The factorisation  of partition functions  has been observed also on lens spaces $L_r$ \cite{Imamura:2013qxa},  on $S^2_A\times S^1$ with R-flux (3d twisted index)  \cite{Benini:2015noa}, in 4d  $\mathcal{N}=1$ theories on $S^3\times S^1$ (4d index) \cite{Yoshida:2014qwa}, \cite{Peelaers:2014ima} and  in 2d $\mathcal{N}=(2,2)$ theories on $S^2$, \cite{Benini:2012ui}, \cite{Doroud:2012xw}, \cite{Closset:2015rna}.
In fact for all these cases the block factorisation can be incorporated in the general analysis  of 2d, 3d and 4d $tt^*$ geometries
\cite{Cecotti:1991me}, \cite{Cecotti:2013mba}.
An alternative perspective on the factorisation is the localisation scheme known as 
the Higgs branch localisation considered in \cite{Benini:2012ui}, \cite{Doroud:2012xw}, \cite{Benini:2013yva}, \cite{Fujitsuka:2013fga}.
 
Results on block factorisation of partition functions have been obtained also for 5d  $\mathcal{N}=1$ theories on $S^5$ 
    \cite{Lockhart:2012vp}, \cite{Nieri:2013vba},
  $S^4\times S^1$    \cite{Kim:2012gu}, \cite{Terashima:2012ra}, \cite{Iqbal:2012xm}, on $Y^{p,q}$  \cite{Qiu:2013pta}, \cite{Qiu:2013aga}, general toric Sasaki-Einstein manifolds \cite{Qiu:2014oqa} and for 6d and 7d theories on  $S^6$, $S^7$ \cite{Minahan:2015jta}.

The goal of this note is to elucidate the block decomposition of partition functions for theories defined on $L_r$, $L_r\times S^1$, $S^2_A\times S^1$  and   $S^2\times T^2$.
The Coulomb branch partition functions on these spaces have been computed in 
\cite{Benini:2011nc}, \cite{Imamura:2012rq}, \cite{Benini:2015noa}  and   \cite{Closset:2013sxa}, \cite{Nishioka:2014zpa}, \cite{Honda:2015yha}. 

Our main result in 3d is the extension of the remarkable identity in (\ref{three3})
to  the lens space $M^3_r=L_r$ and to the twisted index  $M^3_A=S^2_A\times S^1$, which are respectively obtained through the  $r$-gluing
implementing the appropriate  $SL(2,\mathbb{Z})$ transformation on the boundary of one solid torus to obtain the  lens space geometry, and 
through  the $A$-gluing which realises the topological $A$-twist on $S^2$.

We then move to 4d, where for $M^4_S=S^3\times S^1$,  $M^4_r=L_r\times S^1$ and  $M^4_A=S^2\times T^2$   we are able to prove an identical relation
 \be
 \label{four4}
 Z[M^4_g]= \sumint \Big\|\Upsilon^{\rm 4d}\Big\|^2_{g}= \sum_{c}\Big\|\mathcal{B}^{\rm 4d}_c\Big\|^2_g
  =\sum_{c}\Big\|\oint_{\Gamma_c}\Upsilon^{\rm 4d}\Big\|^2_g~.
\ee
In the case of the index $S^3\times S^1$ and lens index $L_r\times S^1$, the  factorised form  of the integrand
emerges after we perform a modular transformation on  the complex structure parameters by means of the remarkable property of the elliptic Gamma function discovered in \cite{felderv}.
This transformation generates a term which can be identified with the 4d anomaly polynomial and represents an obstruction to factorisation. However, for anomaly free theories this factor is one and we can express  the integrand as $\|\Upsilon^{\rm 4d}\|^2_{r}$. It is then fairly easy to check that the $S^2\times T^2$ integrand can also be expressed in terms of the same meromorphic function $\|\Upsilon^{\rm 4d}\|^2_{A}$.
The second step in (\ref{four4}) is the actual  evaluation of the Coulomb branch sum and  integral on a suitable integration contour yielding
the factorisation into
4d holomorphic blocks $\mathcal{B}^{\rm 4d}_c$ which we compute in some explicit cases.
The last step in (\ref{four4})   introduces the 4d block integrals.
In general determining the integration  contours $\Gamma_c$ is
harder than the 3d case,  here we give a prescription in few examples based on physical considerations such as periodicity/invariance under large gauge transformations.

The paper is organised as follows. We begin section \ref{3dlens} with the  study of $\mathcal{N}=2$ theories on the lens space where, thanks to a new identity for the generalised double Sine function, we can  prove the integrand factorisation. 
 We then show the  block factorisation for  two interacting cases.
We take a small detour to discuss the $T[SU(2)]$ theory. In this case, thanks to the transformation
properties of the  holomorphic blocks, we are able to prove that partition functions 
on generic 3-manifolds admitting a block decomposition are 
invariant  under mirror symmetry. 
In section \ref{3dtwistedindex} we discuss the 3d twisted index.
In section \ref{4dlens} we introduce the lens index partition function and show that the integrand can be 
expressed in a factorised form after cancelling the anomalies.
We then show two examples of block factorisation.
We  check the analogue factorisation of  $S^2\times T^2$ partition functions  in  section \ref{other}.
Finally in section \ref{4dhb} we introduce the 4d block integrals.
The paper is supplemented by several appendices where we discuss many technical details and computations.

\section{3d $\mathcal{N}=2$ partition functions on $S^3/\mathbb{Z}_r$ }\label{3dlens}
We consider  the free orbifold $S^3/\mathbb{Z}_r$ of the squashed 3-sphere  $S^3=\{(x,y)\in \mathbb{C}^2|~ b^2|x|^2+b^{-2}|y|^2=1\}$, with the identification
\be
(x,y) \sim (e^{\frac{2\pi  \i}{r}} x,  e^{-\frac{2\pi  \i}{r}} y)~.
\ee
The resulting smooth 3-manifold is the squashed lens space  $L_r$.

The partition function  of $\mathcal{N}=2$ theories on $L_r$  has been first obtained in \cite{Benini:2011nc} and revised  in  \cite{Imamura:2012rq}. 
The localising locus is labelled by the continuous  variables $ \bs Z$  in the Cartan of the gauge group $G$
and discrete holonomies   $\bs \ell$ in the maximal torus.
The integer variables $0\leq \ell_1\leq \ldots \leq \ell_{|G|}$, $\ell_n \in [0,r-1]$, parameterise  the topological sectors.
The holonomy  is non-trivial since 
the fundamental group of the background manifold is $\pi_1(L_r)=\mathbb{Z}_r$ and
breaks the gauge group  to \footnote{Throughout this paper we  restrict to $U(N)$ or $SU(N)$ gauge  groups, so we don't have to worry about global issues \cite{Razamat:2013opa}.}
\be
G\to \prod_{k=0}^{r-1} G_k\,,
\ee
where the subgroup $ G_k$ has rank given by the number of $\ell_n=k$. We also turn on continuous $\bs \Xi$ 
and discrete  $\bs H$  variables for the non-dynamical symmetries.

The partition function reads
\ben
Z[{L_r}]&=&\sum_{\bs \ell}\int\frac{\rd \bs Z}{2\pi\i \prod_k |\mathcal{W}_k|}~Z_{\rm cl}\times Z_{\rm 1-loop}^V \times Z^{\rm matter}_{\rm 1-loop}~,
\een
where $|\mathcal{W}_k|$ is the order of the Weyl group of $G_k$. The classical terms is given by the mixed    Chern-Simons action (CS). 
For example, a pure $U(N)$ CS term contributes as\footnote{In \cite{Imamura:2013qxa} it has been suggested to add the sign factor $e^{\i\pi\kappa\sum_n\ell_n^2}$  in eq. (\ref{CSaction}).}\be\label{CSaction}
e^{-\frac{\i\pi}{r} \kappa \sum_n
 Z_n^2}  ~ e^{\frac{\i\pi}{r} \kappa\sum_{n}\ell^2_n} ~.
 \ee
For  $U(1)$ factors we can also turn on an FI term $\xi$
\be
e^{-\frac{2\pi\i}{r} \sum_n
 Z_n \xi}  ~ e^{\frac{2\pi\i}{r} \sum_{n}\ell_n \theta}~,
\ee
where we have considered a background holonomy $\theta$ also for the topological $U(1)$. The 1-loop contribution of matter multiplets is given by
\be
Z^{\rm matter}_{\rm 1-loop}=
\prod_{i}\prod_{\rho_i}\prod_{\phi_i}
\hat s_{b,-\rho_i(\bs \ell)-\phi_i(\bs H)} 
\left(\i\frac{Q}{2} (1-\Delta_i)-\rho_i(\bs Z)-\phi_i(\bs \Xi)\right)~,
\ee
where $i$ runs over the chiral multiplets,  $\rho_i,\phi_i$, are respectively the weights of the representation of the   gauge  and  flavour groups and $\Delta_i$ the Weyl weight. For convenience we will absorb the Weyl weight  into the mass parameter, and we will be denoting the squashing parameter by $b=\omega_2=\omega_1^{-1}$, with $Q=\omega_1+\omega_2$. The 1-loop contribution of the vector multiplet is given by
\be\label{3dvector}
Z_{\rm 1-loop}^V=\prod_{\alpha}\frac{1}{\hat s_{b,\ell_\alpha} 
\left(\i\frac{Q}{2}+ Z_\alpha\right)}=\prod_{\alpha>0}4\sinh\frac{\pi}{r}\left(\frac{ Z_\alpha}{\omega_1} + i \ell_\alpha \right)
\sinh\frac{\pi}{r}\left(\frac{ Z_\alpha}{\omega_2} - i \ell_\alpha \right)~,
\ee
where the product is over the positive roots $\alpha$ of $G$ and we set $Z_\alpha=\alpha(\bs Z)$, $\ell_\alpha=\alpha(\bs \ell)$.

The function  $\hat s_{b,H}$ is the projection of the (shifted) double Sine function improved by a sign factor $\sigma$, and it is defined as the $\zeta$-regularised product
\be
\hat s_{b,-H}(X)=\sigma(H)\prod_{\substack{n_1,n_2\geq 0\\ n_2-n_1=H\mod r}}\frac{n_1\omega_1+n_2\omega_2+Q/2-\i X}{n_2\omega_1+n_1\omega_2+Q/2+\i X}~,
\ee
where the sign factor is given by
\be\label{sign}
\sigma(H)=e^{\frac{\i\pi}{2r}([H](r-[H])-(r-1)H^2)}~.
\ee
In appendix \ref{appSpecial} we have derived a new expression for $\hat s_{b,H}$  in terms of ordinary double Sine functions 
\be
\hat s_{b,-H}(X)=\sigma(H)S_2(\omega_1(r-[H])+X|Q,r\omega_1)S_2(\omega_2[H]+X|Q,r\omega_2)\,.
\ee
This expression  allows us to  easily evaluate the asymptotic, locate zeros and poles,  take the residues and express it in  a  factorised form
\be
\label{dd}
\hat s_{b,-H}(X)= e^{-\frac{\i\pi}{2r}(r-1)H^2}e^{\frac{i\pi}{2}\Phi_2(Q/2-\i X)}\Big\|(e^{\frac{2\pi}{r\omega_1}(\i Q/2+X)} e^{-\frac{2\pi \i }{r}H} ;e^{2\pi \i\frac{Q}{r\omega_1}})_\infty\Big\|^2_{\substack{\omega_1\leftrightarrow\omega_2\\
 H\leftrightarrow r-H}}~,
\ee
where $\Phi_2$ is a combination of quadratic Bernoulli polynomials defined in (\ref{phi2}). Notice that inside the $q$-Pochhammer symbols we can take  $[H]\sim H$ because of the periodicity. Moreover,  the sign factor erases the residual dependence on $[H]$  so that the function $\hat s_{b,-H}(X)$ depends only on  $H$.

\subsection{Factorisation}

We will now show that by using our expression (\ref{dd}) 
 the partition function of   theories with integer effective CS couplings (parity anomaly free) can be expressed
 in terms of a suitable set   of holomorphic variables and factorised in 3d holomorphic blocks.

We begin with the simplest parity anomaly free theory,  the free chiral with $-1/2$ CS unit
\be\label{Zdelta1}
Z_\Delta(X,H)= e^{\frac{\i\pi}{2r}(r-1)H^2}e^{-\frac{i\pi}{2}\Phi_2(Q+\i X)} 
\hat s_{b,-H}(\i Q/2-X) \,.
\ee
The subscript $\Delta$ is due the fact that, in the  context of the 3d-3d correspondence
relating 3d $\mathcal{N}=2$ theories to analytically continued CS on hyperbolic 3-manifolds, this
theory is associated to  the ideal tetrahedron \cite{Dimofte:2014zga}.
In this context the fundamental Abelian mirror duality relating the anomaly free chiral to
the $U(1)$ theory with 1 chiral and $1/2$ CS unit
is interpreted as a change of polarisation. At the level of  lens space partition functions this duality reads
\be
\sum_{\ell=0}^{r-1}\int_\mathbb{R}\frac{\rd Z}{2\pi \i}\,e^{-\frac{\i\pi}{r}(Z^2+2Z(X-\i Q/2))}e^{-(r-1)\frac{\i\pi}{r}(\ell^2+2H\ell)}\,Z_\Delta(Z,\ell)=Z_\Delta(X,H)~.
\ee
We prove this equality in appendix \ref{far}.\footnote{This identity has also been derived from the pentagon identity on the lens space in \cite{Dimofte:2014zga}.}

The  half CS unit in  (\ref{Zdelta1}) has the effect to  cancel the quadratic factor in (\ref{dd}) so that the anomaly free result can be written in a block factorised form\footnote{The block factorised form (\ref{Zdelta2})  for the tetrahedron theory on the lens space
was derived via projection in \cite{Imamura:2013qxa} and appeared as the fundmanetal building block for the state integral model for analytically continued CS at level $r$
\cite{Dimofte:2014zga}.
} 
\ben\label{Zdelta2}
Z_\Delta(X,H)&=&\frac{(qx^{-1};q)_\infty}{(\tilde x^{-1};\tilde q^{-1})_\infty}=
\Big\|\mathcal{B}_\Delta^{\rm 3d}(x;q) \Big\|^2_r~,
%=\nn\\
%&=& (q x^{-1} ;q)_\infty (\tilde q \tilde x^{-1} ;\tilde q)_\infty =\Big\|(q x^{-1} ;q)_\infty \Big\|^2_r=
%\Big\|\mathcal{B}_\Delta^{\rm 3d}(x;q) \Big\|^2_r~,
\een
in terms of holomorphic  variables
\be
\label{df}
\begin{array}{lll}x=e^{\frac{2\pi }{r\omega_1}X} e^{\frac{2\pi \i }{r}H}=
 e^{2\pi\i\chi} e^{\frac{2\pi \i}{r}H},&\quad & \tilde x=e^{\frac{2\pi }{r\omega_2}X} 
 e^{-\frac{2\pi \i }{r}H}=e^{2\pi\i\frac{\chi}{r\tau-1}} e^{-\frac{2\pi \i }{r}H}~,\\
q=e^{2\pi\i\frac{ Q}{r\omega_1}}=e^{2\pi \i \tau},&\quad & \tilde q=e^{2\pi\i\frac{ Q}{r\omega_2}}=e^{2\pi\i\frac{\tau}{r\tau-1}}~.\end{array}
\ee
The  3d holomorphic block
\be\label{Deltaindex}
\mathcal{B}_\Delta^{\rm 3d}(x;q) =(q x^{-1} ;q)_\infty ~,
\ee
is the  partition function  on $D^2\times_\tau S^1$ of the tetrahedron theory  defined in \cite{hb}.
 Notice that when $|q|<1$ we have $|\tilde q|>1$ and 
\begin{align} \label{zqinf}
(x;q)_\infty &= \sum_{n=0}^\infty \frac{(-1)^n q^{\frac{n(n-1)}{2}}x^n}{(q;q)_n} =
 \begin{cases} \prod_{r=0}^\infty (1-q^rx) &{\rm if}\quad |q|<1 \\[.1cm]
 \prod_{r=0}^\infty (1-q^{-r-1}x)^{-1} &{\rm if}\quad |q|>1~. \end{cases}
\end{align}
Basically blocks in $x,q$, and $\tilde x, \tilde q$,  share the same series expansion but
they converge to different functions.
This is actually  a key feature of holomorphic blocks which has been extensively discussed in
\cite{hb} and will play an crucial role in the  example we discuss in section \ref{tsu2}.

The two blocks are glued through the $r$-pairing  acting as
\be
\tau\to\tilde \tau= -\hat r(\tau)=\frac{\tau}{r\tau-1}~,\quad \hat r=\left(\begin{array}{cc}1&0\\[-2pt]-r&1\end{array}\right)~,  
\ee
where $\tau$ is to be identified with the modular parameter of the boundary $T^2$, while the flavour fugacity and holonomy transform as 
\be
\chi\to \tilde \chi= \frac{\chi}{r\tau-1}~,\quad H\to\tilde H= r-H~.
\ee
This gluing rule as expected coincides with  the $\hat r\in SL(2,\mathbb{Z})$ element (composed with the inversion) realising   the $L_r$ geometry from a pair of solid tori.

CS terms at integer level and FI terms can be expressed in terms of periodic variables as
 $r$-squares of Theta functions defined in (\ref{Theta}) by means of  (\ref{thetamod})\footnote{
For the improved  CS term proposed in \cite{Imamura:2013qxa} we simply have
$e^{-\frac{\i\pi}{r}  Z^2} e^{-\frac{\i\pi}{r}  (r-1)\ell^2} = \Big\| \Theta(q^{\frac{1}{2}}s ;q) \Big\|^{-2}_r$.}
 \begin{align}
\label{fif}
e^{-\frac{\i\pi}{r} 
 Z^2} e^{\frac{\i\pi}{r}\ell^2} \propto 
\Big\| \Theta(-q^{\frac{1}{2}}s ;q) \Big\|^{-2}_r~,
\qquad
e^{-\frac{2\pi\i}{r}  
Z \xi}  e^{\frac{2\pi\i}{r} \ell \theta} &\propto
% \prod_n\Big\| \frac{\Theta(e^{-\frac{2\pi}{r\omega_1}( Z+\xi)}e^{-\frac{2\pi\i}{r}(\ell+\theta)};e^{2\pi\i\frac{Q}{r\omega_1}})}{\Theta(e^{-\frac{2\pi}{r\omega_1}Z}e^{-\frac{2\pi\i}{r}\ell};e^{2\pi\i\frac{Q}{r\omega_1}})\Theta(e^{-\frac{2\pi}{r\omega_1}\xi}e^{-\frac{2\pi\i}{r}\theta};e^{2\pi\i\frac{Q}{r\omega_1}})} \Big\|^{-2}_{\substack{\omega_1\leftrightarrow \omega_2\\ \ell_n\leftrightarrow r-\ell_n}}=\nn\\
\Big\| \frac{\Theta(s^{-1}u;q)}{\Theta(s^{-1};q)\Theta(u;q)} \Big\|^{-2}_r~,
 \end{align}
with $s=e^{\frac{2\pi}{r\omega_1}Z}e^{\frac{2\pi\i}{r}\ell}$
 and $u=e^{-\frac{2\pi}{r\omega_1}\xi}e^{-\frac{2\pi\i}{r}\theta}$. 
Similarly, the vector multiplet can be factorised as
\be
\label{3dvhb}
Z_{\rm 1-loop}^V=\prod_{\alpha>0}4\sinh\frac{\pi}{r}\left(\frac{ Z_\alpha}{\omega_1} + i \ell_\alpha \right)
\sinh\frac{\pi}{r}\left(\frac{ Z_\alpha}{\omega_2} - i \ell_\alpha \right)\propto
\Big\|\prod_{\alpha>0} \left(s_\alpha^{\frac{1}{2}}-s_\alpha^{-\frac{1}{2}}\right) \Big\|^{2}_r~.
\ee
The $\propto$ means that we are dropping background contact terms depending on $\omega_{1,2}$ and $r$ only. From now on we will assume equalities up to these constants.

Obviously the factorised expressions  are not unique. As pointed out in \cite{hb} the  ambiguity amounts to the  freedom to multiply the blocks by ``$q$-phases" (elliptic ratios of Theta functions with unit $S,id,r$-squares).
For example another possibility is to factorise the vector multiplet contribution as in \cite{hb}\footnote{
The vector multiplet factorised form in \cite{hb} differs from ours  by a sign factor $(-1)^\ell$. Notice that  $\| \Theta(-q^{\frac{1}{2}}s_\alpha;q)\|^{2}_r =(-1)^{\ell_\alpha}
\| \Theta(q^{\frac{1}{2}}s_\alpha;q) \|^{2}_r$.   }
\be
\label{secop}
Z_{\rm 1-loop}^V=\Big\|\prod_{\alpha>0} \frac{  \Theta(q^{\frac{1}{2}}s_\alpha;q) }{(q s_\alpha ;q)_\infty(q s_\alpha^{-1};q)_\infty} \Big\|^{2}_r\,.\ee

These observations imply that on parity anomaly free theories, where the total effective CS couplings are integers,
we can replace each 1-loop  vector multiplet with (\ref{3dvhb}), each chiral contribution with $\|\mathcal{B}_\Delta^{\rm 3d}(x;q)\|^2_r$ and
then factorise the remaining integer CS units using (\ref{fif}).  This procedure allows us to rewrite the
partition function as
 \ben
\label{infam}
Z[{L_r}]=e^{-\i\pi\mathcal{P}}  \sum_{\bs\ell}\int\frac{\rd \bs Z}{2\pi\i \prod_k |\mathcal{W}_k|}~\Big\| \Upsilon^{\rm 3d} \Big\|^2_r~,
\een
with exactly the same  integrand $\Upsilon^{\rm 3d}$  appearing in the analogous factorisation  observed in \cite{hb} for $S^3$ and $S^2_{id}\times S^1$.
The three cases differ only for the integration measure which can include also a summation over a discrete set and for the gluing rule. The prefactor $e^{-\i\pi\mathcal{P}}$ is the contribution of background mixed CS terms which can have half-integer coupling preventing their factorisation. 

The integrand $\Upsilon^{\rm 3d}$ appears also in the definition 3d blocks via block integrals proposed in \cite{hb}
\be
\label{bo}
\mathcal{B}^{\rm 3d}_c=\oint_{\Gamma_c} \frac{\rd\bf{s}}{2\pi\i {\bf s}} \Upsilon^{\rm 3d}~,
\ee
where $\Gamma_c$ is  an appropriate basis  of  middle-dimensional cycles in $(\mathbb{C}^{*})^{|G|}$.
Recently block integrals were rederived via localisation on $D^2\times S^1$ by  \cite{Yoshida:2014ssa}.
In their analysis the $\mathcal{B}^{\rm 3d}_\Delta(x;q)$ block corresponds to imposing Dirichlet (D) boundary conditions
\be
\mathcal{B}^{\rm 3d}_\Delta(x;q)=(qx^{-1};q)_\infty=\mathcal{B}^{\rm 3d}_{\rm D}(x;q)~,
\ee
whereas by imposing  Neumann (N) boundary conditions leads to
\be\label{B3dN}
\mathcal{B}^{\rm 3d}_{\rm N}(x;q)=  \frac{1}{(x;q)_\infty}~,
\ee
the two choices being related by
\be 
\mathcal{B}^{\rm 3d}_{\rm D}(x;q)=\Theta(x;q)\mathcal{B}^{\rm 3d}_{\rm N}(x;q)~\,.
\ee
In our language on the l.h.s. we have a chiral of charge $+1$, R charge $0$ with added $-1/2$ CS units.
On the r.h.s. we have a chiral  of charge $-1$, R charge $2$ with added $+1/2$ CS units.
From the  perspective of \cite{Yoshida:2014ssa}, 
the Theta functions represent the elliptic genus  of  
a Fermi multiplet on the  boundary torus.

We are then able to extend to the lens space the remarkable Riemann bilinear-like relation discovered 
 for $S^3$ and $S^2_{id}\times S^1$ \cite{hb}:
\be
 \sum_{\bs \ell}\int\frac{\rd \bs Z}{2\pi\i \prod_k |\mathcal{W}_k|}~\Big\| \Upsilon^{\rm 3d} \Big\|^2_r=e^{-\i\pi\mathcal{P}}\sum_c \Big\| \mathcal{B}^{\rm 3d}_c \Big\|^2_r= e^{-\i\pi\mathcal{P}}
\sum_{c} \Big\| \int_{\Gamma_c} \frac{\rd\bf{s}}{2\pi\i {\bf s}}   \Upsilon^{\rm 3d} \Big\|^2_r\,.
\ee
The intermediate step,  the block factorisation of the partition function, 
is checked for two specific examples in the next subsections, for earlier results   see \cite{Imamura:2013qxa}.
Notice that, while  the parity  anomaly cancellation condition is a sufficient condition to factorise the integrand
in the first step, in the second step it is only a necessary condition. The actual evaluation of the integral might require
additional  conditions to ensure convergence. However as we already mentioned, there are 
other ways to prove factorisation besides explicit integral evaluation. For example,
Higgs branch localisation, stretching/projection arguments or the existence of a commuting set of difference operators in $x,q$ and $\tilde x,\tilde q$
 acting on the partition functions.

\subsection{SQED}
We now consider the $U(1)$ theory  with $N_f$ charge $+1$ and $N_f$ charge $-1$ chirals (SQED), for which we turn on masses $X_a,\bar X_b$, and background holonomies $H_a,\bar H_b$. We also turn on the FI $\xi$ and the associated holonomy $\theta$. The $L_r$ partition function reads
 \begin{align}
 Z_{\rm SQED}&=\sum_{\ell=0}^{r-1}\int_\mathbb{R}\frac{\rd Z}{2\pi\i}\;e^{-\frac{2\pi\i}{r}Z\xi}e^{\frac{2\pi\i}{r}\ell\theta}\prod_{a,b=1}^{N_f}\hat s_{b,-\ell-H_a}(-Z-X_a+\i Q/2) \hat s_{b,\ell+\bar H_b}(Z+\bar X_b+\i Q/2)=\nn\\
 &=
 \sum_{\ell=0}^{r-1}\int_\mathbb{R}\frac{\rd Z}{2\pi\i}\;e^{\frac{2\pi \i}{r}Z\xi}e^{\frac{2\pi\i}{r}\ell\theta}\;\prod_{a,b=1}^{N_f}\frac{\hat s_{b,-\ell-H_a}(Z-X_a+\i Q/2)}{\hat s_{b,-\ell-\bar H_b}(Z-\bar X_b-\i Q/2)}~,
 \end{align}
where in the last step we simply sent $Z\to -Z$ and used the reflection property (\ref{sbreflection}). In order to evaluate the integral we can close the contour in the upper-half plane (assuming $\xi >0$) and take the sum of the residues at the poles of the numerator
 \be\label{SQEDpoles}
 \begin{array}{l}
 Z=Z_{(1)}=X_c+\i\omega_1[\ell+H_c]+\i jQ+\i k r\omega_1~,\\[5pt]
Z=Z_{(2)}=X_c+\i\omega_2(r-[\ell+H_c])+\i jQ+\i kr\omega_2~,\end{array}\quad
c=1,\ldots, N_f~,\quad j,k \in \mathbb{Z}_{\geq 0}~.
 \ee
The details of the computation and notations are given in appendix \ref{qedcomp}, the result is
\begin{multline}
\label{touse}
 Z_{\rm SQED}=e^{-\i\pi\mathcal{P}}\sum_{c=1}^{N_f}e^{\frac{2\pi \i}{r}(X_c\xi_{\rm eff}-H_c\theta_{\rm eff})}\times\\
 \times\Big\|\prod_{a,b=1}^{N_f}\frac{( q e^{\frac{2\pi }{r\omega_1}X_{ca}}
 e^{\frac{2\pi\i }{r}H_{ca}} ;q)_\infty}
 {(e^{\frac{2\pi }{r\omega_1}X_{c\bar b} } 
  e^{\frac{2\pi\i }{r}H_{c \bar b}} ;q)_\infty}{}_{N_f}\Phi_{N_f-1}\left(\begin{array}{l}e^{\frac{2\pi }{r\omega_1}X_{c \bar b}}   e^{\frac{2\pi\i }{r}H_{c\bar b}}
\\ q e^{\frac{2\pi }{r\omega_1}X_{ca}}  e^{\frac{2\pi\i }{r}H_{ca}} \end{array};u\right)\Big\|^2_r~,
\end{multline}
where we introduced the notation
\be 
X_{ca}=X_c-X_a~,\quad X_{c\bar b}=X_c-\bar X_b~,\quad H_{ca}=H_c-H_a~,\quad H_{c\bar b}=H_c-\bar H_b~,
\ee
and set
\be
u=e^{-\frac{2\pi}{r\omega_1}\xi_{\rm eff}}e^{-\frac{2\pi\i}{r}\theta_{\rm eff}}~,\quad 
\tilde u=e^{-\frac{2\pi}{r\omega_2}\xi_{\rm eff}}e^{\frac{2\pi\i}{r}\theta_{\rm eff}}~.
\ee

We can finally express everything in terms of the ``holomorphic" variables
\be
x_a=e^{\frac{2\pi }{r\omega_1}X_a } e^{\frac{2\pi \i }{r}H_a}\,,
\quad \bar x_b=e^{\frac{2\pi }{r\omega_1}\bar X_b}  e^{\frac{2\pi\i }{r}\bar H_b}~,
\ee
factorising  the classical part as
\be
e^{\frac{2\pi\i}{r}(X_c\xi_{\rm eff}-H_c\theta_{\rm eff})}= \Big\|\frac{\Theta(x_c^{-1}u ;q)}{\Theta(u;q)\Theta(x_c^{-1};q)}\Big\|^2_r~,
\ee
where we used (\ref{fif}). Therefore, we finally obtain
\be
Z_{\rm SQED}=e^{-\i\pi\mathcal{P}}\sum_{c=1}^{N_f} \Big\| \mathcal{B}^{\rm 3d}_{c} \Big\|^2_r~,
\ee
where
\be
\mathcal{B}^{\rm 3d}_{c}=  \frac{\Theta(x_c^{-1}u;q)}{\Theta(u;q)\Theta(x_c^{-1};q)}
\prod_{a,b=1}^{N_f} \frac{( q x_{c}x_a^{-1}  ;q)_\infty}
 {(x_c \bar x_{b}^{-1} ;q)_\infty}
 {}_{N_f}\Phi_{N_f-1}\left(\begin{array}{l} x_c \bar x_{b}^{-1}\\ q x_{c}x_a^{-1} \end{array};u\right)
\ee
are the same SQED holomorphic blocks derived for $S^3$ and  $S^2_{id}\times S^1$.

\subsection{$T[SU(2)]$}\label{tsu2}

As an application of the result obtained in the previous section we consider the 
 mass deformed $T[SU(2)]$ theory.
This is a   $U(1)$ theory with 2  charge $+1$ and 2 charge $-1$ chirals and a neutral chiral.
We turn on vector and axial  masses $\frac{m}{2}, \frac{\mu}{2}$, the FI parameter $\xi$ and their respective  holonomies $\frac{H_V}{2}, \frac{H_A}{2},\theta \in \mathbb{Z}_r$. 

The $T[SU(2)]$ theory is part of a family of theories $T[G]$ introduced in \cite{Gaiotto:2008ak} as
 boundary field theories coupled to the  bulk 4d $\mathcal{N}=4$ SYM with gauge group $G$ for which they provide $S$-dual of Dirichlet boundary conditions.
$T[G]$ are 3d $\mathcal{N}=4$ theories with $G\times G^L$ global symmetry  rotating the Coulomb and Higgs branches. 3d mirror symmetry acts by exchanging Higgs and Coulomb branches hence swapping $T[G]$ to $T[G^L]$.

In \cite{Hosomichi:2010vh} it was shown that the $S^3$ partition function of the  mass deformed $T[SU(2)]$ theory 
(the axial mass $m$ coincides with the mass of the 4d adjoint breaking the 4d SYM to  $\mathcal{N}=2^*$)  coincides with the $S$-duality kernel in Liouville theory acting on the torus conformal blocks.
It was also explicitly  proved that the $S^3$ partition function is  invariant under the action of mirror symmetry. Actually, as we are about to see,  the self mirror property can proved on generic 3-manifolds that can be decomposed in solid tori. This result follows from the highly  non-trivial tranformations of holomorphic blocks across mirror frames.

The lens space partition function of $T[SU(2)]$ reads
\be
Z^I= Z(m,\xi,\mu;H_V,\theta,H_A)=\frac{1}{ \hat s_{b,H_A}(\mu)}\sum_{\ell=0}^{r-1}~\int_\mathbb{R}\frac{\rd Z}{2\pi\i} ~e^{\frac{2\pi \i}{r} (Z\xi+\ell \theta)}~ 
\frac{\hat s_{b,-\ell\pm \frac{H_V}{2}+ \frac{H_A}{2 }}(Z\pm \frac{m}{2}+\frac{\mu}{2}+\i Q/4)}
{\hat s_{b,-\ell\pm \frac{H_V}{2}- \frac{H_A}{2 }}(Z\pm \frac{m}{2}-\frac{\mu}{2}-\i Q/4)}~,
\ee
where we used the notation $f_{\pm h}(\pm x)=f_h(x)f_{-h}(-x)$. Introducing
\be
z=e^{\frac{2\pi}{r\omega_1}\mu}e^{\frac{2\pi \i}{r}H_A},\quad 
x=e^{\frac{2\pi}{r\omega_1}m}e^{\frac{2\pi \i}{r}H_V},\quad
y=e^{\frac{2\pi}{r\omega_1}\xi}e^{\frac{2\pi \i}{r}\theta}~,
\ee
and using the result (\ref{touse}), we can write\footnote{We introduced the index $I$ to distinguish the theory from its mirror as it will be clear later.} 
\be
Z(m,\xi,\mu;H_V,\theta,H_A)=
e^{-\i\pi\mathcal{P}}\left(\Big\|\mathcal{B}^{{\rm 3d},I}_{1}\Big\|^2_r+\Big\|\mathcal{B}^{{\rm 3d},I}_{2}\Big\|^2_r\right)\,,
\ee
with
\be
\begin{split}
\mathcal{B}^{{\rm 3d},I}_{1}&=\frac{(q x^{-1};q)_\infty}{(q^{\frac{1}{2}}x^{-1}z^{-1};q)_\infty}{}_2\Phi_1\left(\begin{array}{cc}q^{\frac{1}{2}}z^{-1}&q^{\frac{1}{2}}x^{-1}z^{-1}\\
q&qx^{-1}\end{array};q^{\frac{1}{2}}zy^{-1}\right)~,\\
\mathcal{B}^{{\rm 3d},I}_{2}&=
\frac{\Theta(y;q) \Theta(q^{\frac{1}{2}}xz^{-1};q)}{\Theta(yx^{-1};q) \Theta(q^{\frac{1}{2}}z^{-1};q)}
\frac{(qx;q)_\infty}{(q^{\frac{1}{2}}xz^{-1};q)_\infty}{}_2\Phi_1\left(\begin{array}{cc}q^{\frac{1}{2}}z^{-1}&q^{\frac{1}{2}}xz^{-1}\\q&q x\end{array};q^{\frac{1}{2}}zy^{-1}\right)~,
\end{split}
\ee
and 
\be
e^{-\i\pi\mathcal{P}}=
e^{-\frac{\i\pi}{2r}\left((r-1)H_A^2 +\mu^2+2(  m+\mu-\i Q/2)(\xi-\mu-\i Q/2)-(H_V+H_A)(\theta+(r-1)H_A) \right)}~,
\ee
is the contribution of background CS terms.

Mirror symmetry acts by exchanging Higgs and Coulomb branches, correspondently  the vector mass and the FI parameter
are swapped while the axial mass is inverted, and similarly for the associated holonomies
\be
\xi\to m\,,\quad \mu\to -\mu\,,\quad \theta\to H_V\,,\quad H_A\to -H_A~,
\ee
so that the partition function in the mirror frame reads
\be
Z^{II}= Z(\xi,m,-\mu;-\theta,-H_V,-H_A)=
e^{-\i\pi\mathcal{P}}\left(\Big\|\mathcal{B}^{{\rm 3d},II}_{1}\Big\|^2_r+\Big\|\mathcal{B}^{{\rm 3d},II}_{2}\Big\|^2_r\right)~,
\ee
where we used that $\mathcal{P}$ is invariant under the mirror map and 
obtained the blocks in phase $II$ from the ones in phase $I$ by applying the mirror map $x\to y$,  $y\to x$, $z\to z^{-1}$
\be\begin{split}
\mathcal{B}^{{\rm 3d},II}_{1}&=\frac{(qy^{-1};q)_\infty}{(q^{\frac{1}{2}}y^{-1}z;q)_\infty}{}_2\Phi_1\left(\begin{array}{cc}q^{\frac{1}{2}}z&q^{\frac{1}{2}}y^{-1}z\\q&qy^{-1}\end{array};q^{\frac{1}{2}}z^{-1}x^{-1}\right)~,\\
\mathcal{B}^{{\rm 3d},II}_{2}&=
\frac{\Theta(x;q) \Theta(q^{\frac{1}{2}}yz;q)}{\Theta(xy^{-1};q) \Theta(q^{\frac{1}{2}}z;q)}
\frac{(qy;q)_\infty}{(q^{\frac{1}{2}}yz;q)}_\infty{}_2\Phi_1\left(\begin{array}{cc}q^{\frac{1}{2}}z&q^{\frac{1}{2}}y z\\q& qy\end{array};q^{\frac{1}{2}}z^{-1}x^{-1}\right)~.\end{split}
\ee

At this point proving that the partition function  is invariant under  mirror symmetry
amounts to prove the following equality
%\be
%Z(m,\xi,\mu;H_V,\theta,H_A)=Z(\xi,m,-\mu;-\theta,-H_V,-H_A)~,
%\ee
%or at the level of blocks 
\be
\Big\|\mathcal{B}^{{\rm 3d},I}_{1}\Big\|^2_r+\Big\|\mathcal{B}^{{\rm 3d},I}_{2}\Big\|^2_r=\Big\|\mathcal{B}^{{\rm 3d},II}_{1}\Big\|^2_r+\Big\|\mathcal{B}^{{\rm 3d},II}_{2}\Big\|^2_r~.
\label{topro}
\ee
As we already mentioned the two sets of blocks inside an $r$-square (with $|\tilde q|>1$ if $|q|<1$) share the same series expansion but converge to different functions which crucially  have different transformation properties. 
Indeed by using identities (\ref{id1}), (\ref{id2}), (\ref{id4}), (\ref{id3}) we can show  that
\be
|q|<1:~\left\{
  \begin{array}{l  l}
   \mathcal{B}^{{\rm 3d},II}_{1} = \mathcal{B}^{{\rm 3d},I}_{1} & \\
      \mathcal{B}^{{\rm 3d},II}_{2} = \mathcal{B}^{{\rm 3d},I}_{1}-\mathcal{B}^{{\rm 3d},I}_{2}
  \end{array}\right.~, \quad
   |q|>1:~ \left\{
 \begin{array}{l  l}
    \mathcal{B}^{{\rm 3d},II}_{1} = \mathcal{B}^{{\rm 3d},I}_{1}+\mathcal{B}^{{\rm 3d},I}_{2} & \\
     \mathcal{B}^{{\rm 3d},II}_{2} = -\mathcal{B}^{{\rm 3d},I}_{2}
  \end{array}~,
  \right.
  \ee
which ensures (\ref{topro}). The transformations of the blocks across mirror frames  has the characteristic 
structure of a  jump across a Stokes wall. The interplay between mirror symmetry and
Stokes phenomenon for 3d blocks  and its relation  to analytically continued CS theory
 has been extensively discussed in \cite{hb}.

Notice that our proof relies only  on the blocks transformation properties and  makes no reference to the specific gluing rule, hence it can be extended to all the cases in which the partition function can be block factorised.

\subsection{SQCD}
We now continue our examples with the $SU(2)$ theory with $N_f$ fundamentals and $N_f$ antifundamentals chirals (SQCD). The partition function reads 
\be
\label{su2}
Z_{\rm SQCD}=\sum_{\ell=0}^{r-1}\int_{\mathbb{R}}\frac{\rd Z}{2\pi\i}\;4\sinh\frac{2\pi}{r\omega_1}(Z- \i\omega_1\ell)\sinh\frac{2\pi}{r\omega_2}(Z+\i\omega_2\ell)
\times \prod_{a',b'=1}^{2N_f}\frac{\hat s_{b,-\ell-H_{a'}}(Z-X_{a'}+\i Q/2)}{\hat s_{b,-\ell-\bar H_{b'}}(Z-\bar X_{b'}-\i Q/2)}~,
\ee
where we defined
\be
X_{a'}=(X_a,-\bar X_b)=-\bar X_{b'};\quad H_{a'}=(H_a,-\bar H_b)=-\bar H_{b'}~.
\ee
In this form the matter sector reads formally the same as the previous abelian theory with the replacements $a\to a'$, $b\to b'$.
In fact also the vector multiplet contribution is equivalent to   a pair of  charge $\pm 2$  chirals. Therefore, there is a canonical Abelian theory $\widehat{Z}_{\rm SQCD}[\xi,\theta]$ associated to the $SU(2)$ theory, 
for which we also turn on an FI coupling $e^{\frac{2\pi\i}{r}Z\xi}e^{\frac{2\pi\i}{r}\ell\theta}$.
Since the  vector multiplet  does not bring any pole, the residue computation proceeds exactly as  in the SQED 
case  and the $SU(2)$ partition function can be obtained from  the limit 
\be\label{su2limit}
Z_{\rm SQCD}=\lim_{\xi,\theta\to 0}\widehat{Z}_{\rm SQCD}[\xi,\theta]~,
\ee
where
\begin{multline}
\widehat{Z}_{\rm SQCD}[\xi,\theta]=e^{-\i\pi\mathcal{P}}\sum_{c'=1}^{2N_f}e^{\frac{2\pi \i}{r}(X_{c'}\xi_{\rm eff}-H_{c'}\theta_{\rm eff})}
\Big\|\prod_{a',b'=1}^{2N_f}\frac{( q e^{\frac{2\pi }{r\omega_1}X_{c'a'}}
 e^{\frac{2\pi\i }{r}H_{c'a'}} ;q)_\infty}
 {(e^{\frac{2\pi }{r\omega_1}X_{c'\bar b'}} 
  e^{\frac{2\pi\i }{r}H_{c'\bar b'}} ;q)_\infty}\times \\
  \times\sum_{n\geq 0} 4\sinh\frac{2\pi}{r\omega_1}(- X_{c'}-\i\omega_1 H_{c'}-\i n Q) \frac{(e^{\frac{2\pi }{r\omega_1}X_{c'\bar b'}}   e^{\frac{2\pi\i }{r}H_{c'\bar b'}};q)_n}{
(q e^{\frac{2\pi }{r\omega_1}X_{c'a'}}  e^{\frac{2\pi\i }{r}H_{c'a'}};q)_n} u^n\Big\|^2_r~,
\end{multline}
with
\be
\xi_{\rm eff}=\xi+\sum_{a'}X_{a'}-\i N_f Q,\quad \theta_{\rm eff}=\theta-(r-1)\sum_{a'}H_{a'}~.
%u&=&e^{\frac{2\pi\xi_{\rm eff}}{\omega_1}}e^{\frac{2\pi\i}{r}\theta_{\rm eff}}~.
\ee

\section{3d twisted index}\label{3dtwistedindex}

We now consider  $\mathcal{N}=2$ theories with  R-symmetry on $S^2_A\times S^1$ with a topological $A$-twist on $S^2$.
This background has been recently reconsidered  in  \cite{Benini:2015noa}.
The topological twist  is performed by turning on a background for the R-symmetry
proportional to the spin connection with a quantised magnetic flux, as a consequence R-charges are integers.
Magnetic fluxes are also turned on for all the flavour symmetries.
 
The path integral on this space localises on  BPS configurations 
labelled by continuous variables $\bs Z$ in the Cartan  and discrete variables $\bs \ell$ in the maximal torus of the gauge algebra.
The integer variables $\bs \ell$ parameterise the magnetic flux while   ${\bs z=e^{2\pi\i \bs Z}}$ is 
the holomorphic combination of the $S^1$ holonomy and of the real scalar.
We also turn on analogous continuous and discrete variables for the non-dynamical symmetries.
The partition function reads
\ben
\label{tip}
Z[S^2_A\times S^1]&=&\sum_{\bs \ell}\int\frac{\rd \bs z}{2\pi\i    \bs z |\mathcal{W}|}~Z_{\rm cl}\times Z_{\rm 1-loop}^V \times Z^{\rm matter}_{\rm 1-loop}~.
\een
 The contributions to the classical part come from (mixed) CS terms. In particular, a pure CS and FI read 
\be
{\bs z}^{\kappa \bs{\ell}} ~, \quad   \bs z^\theta \xi^{\bs \ell}~,
\ee 
where $\xi,\theta$, are the holonomy and flux associated to the topological $U(1)$ symmetry.
The contribution of a chiral multiplet with R-charge $R$ is given by
\be
Z_{\chi}^{(B)}[S^2_A\times S^1]=\frac{{z^\frac{B}{2}}}{ (  q^{\frac{1-B}{2}} z;q )_B}~,
\ee
where the shifted R-charge  $B=\ell-R+1$  is quantised.
Finally the vector multiplet contribution is given by
\be
Z_V[S^2_A\times S^1]=\prod_{\alpha>0} q^{-\frac{|\ell_\alpha|}{2}}(1-q^{\frac{|\ell_\alpha|}{2}} z^{\pm}_\alpha)~,\ee
where we used the usual shorthand notation $f(x)f(x^{-1})=f(x^\pm)$. We refer the reader to  \cite{Benini:2015noa} for a detailed analysis of the integration contour in (\ref{tip}).

Geometrically, the twisted index background is realised by  gluing two solid tori twisted in the same direction so
to realise the A-twist on $S^2$.
We then expect that also in this case the partition function can be expressed in
terms of the universal blocks $\mathcal{B}^{{\rm 3d}}_c$.

We begin studying the  free chiral with R-charge 0 and $-1/2$ CS unit (the tetrahedron theory).
It is easy to see that by defining  the $A$-gluing acting as
\be
\tau\to -\tau~, \quad Z\to Z ~,\quad {\rm or}\quad q\to q^{-1} \quad z\to z~,
\ee
we obtain the twisted index of the tetrahedron theory by $A$-fusing two  3d blocks
\be
\label{thea}
\Big\| \mathcal{B}^{\rm 3d}_\Delta(x ;q)\Big\|^2_{A}=( q^{\frac{2+\ell}{2}} z;q )_\infty
( q^{-\frac{2+\ell}{2}}  z;q^{-1} )_\infty=\frac{1}{ (  q^{-\frac{\ell}{2}} z;q )_{\ell+1}}=
\frac{1}{ (  q^{\frac{1-B}{2}} z;q )_B}=Z_{\Delta}[S^2_A\times S^1]
~,
\ee
where the  the holomorphic variable $x$ is identified with the combination  $x=z^{-1}q^{-\ell/2}$.
As expected  
\be 
\label{rela}
Z_{\chi}^{(B)}[S^2_A\times S^1]=Z_\Delta[S^2_A\times S^1]
z^{\frac{B}{2}}~,
\ee
with the factor $z^{B/2}$  contributing the $+1/2$ CS unit. 

CS terms at integer level and FI terms can also be expressed as $A$-squares
of the same blocks appearing in  (\ref{fif}) 
\be\label{twistedCS}
\Big\|\Theta(-q^\frac{1}{2} x ;q)\Big\|^{-2}_{A}=z^{\ell}~,\quad \Big\|\frac{\Theta(x^{-1}u;q)}{\Theta(x^{-1} ;q)\Theta(u;q)}\Big\|^{-2}_{A}=z^\theta \xi^\ell~,
\ee
where $u=q^{\theta/2}\xi$. Finally  also the vector multiplet can be factorised as in (\ref{3dvhb}) 
\be
\label{vea}
\Big\|\prod_{\alpha>0} \left(s_\alpha^{\frac{1}{2}}-s_\alpha^{-\frac{1}{2}}\right) \Big\|^{2}_A=\prod_{\alpha>0} q^{-\frac{|\ell_\alpha|}{2}}(1-q^{\frac{|\ell_\alpha|}{2}} z^{\pm}_\alpha)=Z_V[S_A^2\times S^1]~,\ee
with $s_\alpha=q_\tau^{-\ell_\alpha/2}z^{-1}_\alpha  $
 or alternatively\footnote{
 Up to a  factor $(-1)^{\ell_\alpha}$, see discussion  in \cite{Benini:2015noa}.}
\be
\Big\|\prod_{\alpha>0} \frac{   \Theta(-  q^{\frac{1+\ell_\alpha}{2}}z_\alpha    ;q)     }{(q^{\frac{2+\ell_\alpha}{2}}z_\alpha;q)_\infty
(q^{\frac{2-\ell_\alpha}{2}}z^{-1}_\alpha;q)_\infty}
\Big\|^2_A=
\prod_{\alpha>0}q^{-\frac{|\ell_\alpha|}{2}}(1-q^{\frac{|\ell_\alpha|}{2}}z^\pm_\alpha)~.
\ee

From eqs. (\ref{thea}), (\ref{rela}), (\ref{twistedCS}) and (\ref{vea}) it follows  straightforwardly 
that  for parity anomaly free theories the integrand is  factorised
\be
Z[S^2\times S^1]=\sum_{\bs \ell }\oint\frac{\rd \bs z}{2\pi\i {\bs z} |\mathcal{W}|}\Big\| \Upsilon^{\rm 3d}\Big\|^2_A~.
\ee
Clearly one expects the result of the contour integral to take factorised form too. Indeed
in   \cite{Benini:2015noa} it has been observed that this is the case.
For example it is  an easy exercise to show that  the  SQED partition function can be written in terms of the  3d holomorphic blocks
\be
Z_{\rm SQED}=e^{-\i\pi\mathcal{P}}\sum_c   \Big\| \mathcal{B}^{\rm 3d}_c\Big\|^2_{A}~.
\ee
We will not show the details of the computation because we will perform an almost identical computation for the $S^2\times T^2$ case in section \ref{other}.

In the end  we can extend also to the twisted index case the identity
\be\sum_{\bs \ell}\oint\frac{\rd \bs z}{2\pi\i {\bs z} |\mathcal{W}|}\Big\| \Upsilon^{\rm 3d}\Big\|^2_A=
\sum_{c} \Big\| \int_{\Gamma_c} \frac{\rd \bs s}{2\pi\i \bs s}   \Upsilon^{\rm 3d} \Big\|^2_A~,
\ee
suggesting that the factorisation commutes with integration.

\section{4d $\mathcal{N}=1$ lens index}\label{4dlens}
In this section we consider   $\mathcal{N}=1$ theories formulated on $L_r\times S^1$.  The lens index of a
  chiral multiplet of R-charge $R$ and unit charge under a $U(1)$ symmetry is \cite{Razamat:2013opa}
\be
\hat{\mathcal{I}}_\chi^{(R)}(w,H)=\sigma(H)\mathcal{I}_{0,\chi}^{(R)}(w,H)\mathcal{I}_\chi^{(R)}(w,H)~,
\ee
with
\be
\mathcal{I}_\chi^{(R)}(w,H)=\Gamma((pq)^{\frac{R}{2}}wp^{[H]};pq,p^r)\Gamma((pq)^{\frac{R}{2}}wq^{r-[H]};pq,p^r)~,
\ee
where $w$ is the $U(1)$ fugacity and $H$ the  holonomy along the non-contractible circle of $L_r$.
$\mathcal{I}_0(w,H)$ is the  zero-point energy
\be 
\mathcal{I}^{(R)}_{0,\chi}(w,H)=\left((pq)^{\frac{R}{2}}w\right)^{-\frac{1}{2r}[H](r-[H])}\left(pq\right)^{\frac{1}{4r}[H](r-[H])}(pq^{-1})^{\frac{1}{12r}[H](r-[H])(r-2[H])}~,
\ee
and, as suggested in   \cite{Imamura:2013qxa}, we included the sign $\sigma(H)$ defined in (\ref{sign}). 
%Notice the lens index enjoys the important reflection property
%\be
%\hat{\mathcal{I}}^{(R)}_\chi(w,H)\hat{\mathcal{I}}^{(-R)}_\chi(w^{-1},-H)=1~.
%\ee

For a chiral multiplet in a given representation of a gauge group $G$ and global flavour group, the lens index reads
%
%, Cartan generators $T_n$, a representation $\mathcal{R}_G$ and a weight $\rho$, we have
%\be
%\mathcal{I}_\chi^{(R)}(u,h)\to\mathcal{I}_\chi^{(R)}(\rho(\bs z),\rho(\bs \ell));\quad \bs z= \prod_n z_n^{T_n};\quad \bs \ell=\sum_n \ell_nT_n~,
%\ee
%
%where $z_n$ and $\ell_n$ represent gauge fugacities and holonomies. Let us also consider a global symmetry group $F$ with Cartan generators $F_a$, a representation 
%$\mathcal{R}_F$ and a weight $\phi$. 
\be
\prod_{\rho,\phi}\hat{ \mathcal{I}}_\chi^{(R)}(\rho(\bs z) \phi(\bs\zeta),\rho(\bs \ell)+\phi(\bs H))~,
 \ee
where  $\bs z,\bs \zeta$, are respectively the  gauge and global fugacities associated to the Cartan, $\rho,\phi$, the weights of the gauge and flavour representations,  while $\bs \ell , \bs H$,  are respectively the gauge and background holonomies in the maximal torus, which can be represented by vectors with components in $\mathbb{Z}_r$. The gauge theory lens index is then obtained by summing over the dynamical holonomies $0\leq \ell_1\leq \ldots\leq \ell_{|G|}\leq r-1$, $\ell_n\in[0,r-1]$ and integrating the matter contribution with  integration measure given by the vector multiplet of the unbroken gauge group
\begin{multline}\label{lensindex}
I=\sum_{\bs \ell}\oint_{T^{|G|}}\frac{\rd \bs z}{2\pi\i \bs z\prod_k |\mathcal{W}_k|}~\prod_{\alpha}\hat{\mathcal{I}}_V(\alpha(\bs z),\alpha(\bs \ell))\times\\
\times\prod_{i}\hat{\mathcal{I}}_\chi^{(R_i)}(\rho_i(\bs z)\phi_i(\bs\zeta),\rho_i(\bs \ell)+\phi_i(\bs H))~,
\end{multline}
where  $\alpha$ denote the gauge roots, and we defined
\be
\hat{\mathcal{I}}_V(w,H)=\sigma(H)\mathcal{I}_{0,V}(w,H)\mathcal{I}_V(w,H)~,
\ee
with
\be
\mathcal{I}_V(w,H)=\frac{1}{\Gamma(w^{-1}p^{r-[H]};pq,p^r)\Gamma(w^{-1}q^{[H]};pq,q^r)}~,
\ee
and zero-point energy
\be
\mathcal{I}_{0,V}(w,H)=
w^{\frac{1}{2r}[H](r-[H])}\left(pq\right)^{-\frac{1}{4r}[H](r-[H])}(pq^{-1})^{-\frac{1}{12r}[H](r-[H])(r-2[H])}~.
\ee
If the gauge group has an abelian factor we can introduce an FI term which contributes to the partition function as
\be
 {\bs z}^{\frac{\xi^{\rm 4d}}{r}} e^{\frac{2\pi i}{r} {\bs \ell} \theta}\,,
\ee
where we turned on also a background holonomy $\theta$ for the topological $U(1)$ symmetry.
As argued in   \cite{Imamura:2011uw} the 4d FI parameter $\frac{\xi^{\rm 4d}}{r}$ needs to be quantised.
This allows the index, which is independent on continuous couplings, to actually depend on the FI parameter.

In the following we will show that by performing a modular transformation and cancelling the anomalies it is possible to
express the  lens index integrand in a very neat factorised form.

\subsection{Chiral multiplet}\label{4dchirev}
Let us consider the index of a single chiral
and introduce the following parametrisation
\be
w=e^{\frac{2\pi \i}{\omega_3}M}\,,\quad p=e^{2\pi\i \frac{\omega_1}{\omega_3}}\,,\quad q=e^{2\pi\i \frac{\omega_2}{\omega_3}}\,,\quad pq=e^{2\pi\i\frac{Q}{\omega_3}}\,,
\ee
where $Q=\omega_1+\omega_2$, and $\omega_3=\frac{2\pi}{\beta}$ measures the (inverse) $S^1$ radius $\beta$. For convergence, we also assume ${\rm Im} \left(\frac{\omega_{1,2}}{\omega_3}\right)>0$. Also, since it is going to appear quite often, we define the combination
\be\label{XU}
X=\frac{QR}{2}+M~.
\ee
By using the modular transformation 
 (\ref{gammamod2}) and the reflection properties of the elliptic Gamma function (appendix \ref{appSpecial})
 %Now we consider the following modular transformation 
%\ben
%\label{longmod}
%&&\mathcal{I}_\chi^{(R)}\left(u,h\right)=\Gamma(e^{\frac{2\pi\i}{\omega_3}(X+\omega_1[h])};e^{2\pi\i\frac{r\omega_1}{\omega_3}},e^{2\pi\i\frac{Q}{\omega_3}})\Gamma(e^{\frac{2\pi\i}{\omega_3}(X+\omega_2(r-[h]))};e^{2\pi\i\frac{r\omega_2}{\omega_3}},e^{2\pi\i\frac{Q}{\omega_3}})=\nn\\
%&&\quad=e^{-\frac{\i\pi}{3}\left(B_{33}(X+\omega_3+\omega_1[h]|Q,\omega_3,r\omega_1)+B_{33}(X+\omega_3+\omega_2(r-[h])|Q,\omega_3,r\omega_2)\right)}\times\nn\\
%&&\qquad\times\Gamma(e^{\frac{2\pi\i}{r\omega_1}(X+\omega_1[h])};e^{2\pi\i\frac{Q}{r\omega_1}},e^{-2\pi\i\frac{\omega_3}{r\omega_1}})\Gamma(e^{\frac{2\pi\i}{r\omega_2}(X+\omega_2(r-[h]))};e^{2\pi\i\frac{Q}{r\omega_2}},e^{-2\pi\i\frac{\omega_3}{r\omega_2}})~,
%\een
%which can be derived from (\ref{gammamod2}) and the reflection properties of the elliptic Gamma function (appendix \ref{appSpecial}). It is not difficult to verify that
%% since
%%\be
%%\frac{e^{-\frac{\i\pi}{3}\left(B_{33}(Z+\omega_3+\omega_1[h]|Q,\omega_3,r\omega_1)+B_{33}(Z+\omega_3+\omega_2(r-[h])|Q,\omega_3,r\omega_2)\right)}}{e^{-\frac{\i\pi}{3}\left(B_{33}(Z+\omega_3|Q,\omega_3,r\omega_1)+B_{33}(Z+\omega_3+r\omega_2)|Q,\omega_3,r\omega_2)\right)}}=e^{\frac{\i\pi}{2r}[h](r-[h])}\mathcal{I}_{0,\chi}^{(R)}(u,h)^{-1}~,
%%\ee
we can rewrite 
\ben\label{4dchiral}
\hat{\mathcal{I}}^{(R)}_\chi(w,H)=e^{-\i\pi\left(\frac{1}{3}\Phi_3(X)+\frac{1}{2}\Phi_2(X)\right)}\times
\hat Z^{\rm 4d}_\chi(X,H)\,,
\een
where
\be
\label{coc}
\hat Z^{\rm 4d}_\chi(X,H)=\frac{e^{-\frac{i\pi}{2r}H^2(r-1)}e^{\frac{\i\pi}{2}\Phi_2(Q-X)}}{\mathcal{G}(Q-X,-H)}~.
\ee
The cubic polynomial $\Phi_3(X)$ is defined in (\ref{phi3}). As we will see in  section  \ref{sec_anomaly}, these polynomials contribute to the 4d gauge and global anomalies. In the above expression we introduced the function\footnote{For $r=1$, $\mathcal{G}$ coincides with the so-called modified elliptic Gamma function, see for example \cite{Spiridonov:2012ww}.}
\be
\label{modga}
\mathcal{G}(X,H)=\Gamma(e^{\frac{2\pi\i}{r\omega_1}(X+\omega_1[H])};e^{2\pi\i\frac{Q}{r\omega_1}},e^{-2\pi\i\frac{\omega_3}{r\omega_1}})\Gamma(e^{\frac{2\pi\i}{r\omega_2}(X+\omega_2(r-[H]))};e^{2\pi\i\frac{Q}{r\omega_2}},e^{-2\pi\i\frac{\omega_3}{r\omega_2}})~,
\ee
satisfying 
\ben\label{refl}
\mathcal{G}(X,H)\mathcal{G}(Q-X,-H)=e^{-\frac{\i\pi}{r}H(r-H)}e^{\i\pi \Phi_2(X)}~,
\een
and which can be factorised as
\begin{align}
\label{gfac}
\mathcal{G}(X,H)&=\Gamma(x ;q_\tau,q_\sigma)\Gamma(\tilde x ;\tilde q_\tau,\tilde q_\sigma )=  \Big\| \Gamma(x ;q_\tau,q_\sigma)\Big\|^2_{r}~,
%\\
%\mathcal{G}(Q-X,-h)&=\Gamma(q_\tau x^{-1} ;q_\tau,q_\sigma)\Gamma(\tilde q_\tau\tilde x^{-1} ;\tilde q_\tau,\tilde q_\sigma )=  \Big\| \Gamma(q_\tau x^{-1} ;q_\tau,q_\sigma)\Big\|^2_{r}~,
\end{align}
where the 4d $r$-pairing acts according to
%\footnote{We remind that $X=\frac{QR}{2}+U$, and we rescale $\chi=\frac{X}{r\omega_1}$.}
\be\label{conrule}
\begin{array}{lll}
q_\tau=e^{2\pi\i\frac{Q}{r\omega_1}}=e^{2\pi\i\tau}~,&\quad & \tilde q_\tau=e^{2\pi\i\frac{ Q}{r\omega_2}}=e^{2\pi\i\tilde\tau}~,\\[5pt]
q_\sigma=e^{-2\pi\i\frac{ \omega_3}{r\omega_1}}=e^{2\pi\i\sigma}~,&\quad &\tilde q_\sigma=e^{-2\pi\i\frac{ \omega_3}{r\omega_2}}=e^{2\pi\i\tilde\sigma}~,\\[5pt]
x=e^{\frac{2\pi \i }{r\omega_1}X} e^{\frac{2\pi \i }{r}H}=e^{2\pi\i \chi}e^{\frac{2\pi \i }{r}H}~,&\quad &
\tilde x=e^{\frac{2\pi \i}{r\omega_2}X}e^{-\frac{2\pi \i }{r}H}=e^{2\pi\i\tilde{\chi}}e^{\frac{2\pi \i }{r}\tilde H}~,
\end{array}
\ee
with
\be\label{conrulemod}
\tilde \tau=\frac{\tau}{r\tau-1}~,\quad \tilde\sigma=\frac{\tilde \sigma}{r\tau-1}~,\quad \tilde{\chi}=\frac{\chi}{r\tau-1}~,\quad \tilde H=r-H~.
\ee
%It is manifest from the above expression that the $\mathcal{G}$ function, in terms of which we can express
%the``anomaly free" lens index of a chiral multiplet, is factorised into two blocks fused by the $SL(3,\mathbb{Z})$ element implementing the gluing of two solid tori $D^2\times T^2$ into the  space $L_r$ \FN{Clarify}

Notice that in the  3d limit $\omega_3\to+\infty_\mathbb{R}$  (or $q_\sigma \to 0$), we have
\ben\label{Zdeltalimit}
\frac{1}{\Gamma(q_\tau x^{-1} ;q_\tau,q_\sigma)}& \stackrel{q_\sigma\to 0}{\longrightarrow}&(q_\tau x^{-1} ;q_\tau)_\infty=\mathcal{B}^{\rm 3d}_{\rm D}(x;q_\tau)~,
%\\\mathcal{G}(Q-X,-h)^{-1}&\stackrel{q_\sigma\to 0}{\longrightarrow}&\Big\|(q_\tau x^{-1} ;q_\tau)_\infty\Big\|^2_r=Z_\Delta(\i X,h)~
\een
and 
\be
\hat Z^{\rm 4d}_\chi(X,H)\stackrel{\omega_3\to+\infty}{\longrightarrow}\hat s_{b,-H}(\i Q/2-\i X)~,
\ee
with the  quadratic polynomial $\Phi_2(Q-X)$ in (\ref{coc}) contributing the correct half CS unit in 3d.
The function $\hat Z^{\rm 4d}_\chi(X,H)$ satisfies 
%we can also write
%\be\label{4dchiral2}
%\hat Z^{\rm 4d}_\chi(X,h)=
%e^{\frac{i\pi}{2r}h^2(r-1)}e^{-\frac{\i\pi}{2}\Phi_2(X)}\mathcal{G}(X,h)~,
%\ee
%with
 \be
\hat Z^{\rm 4d}_\chi\left(X,H\right)
\hat Z^{\rm 4d}_\chi\left(Q-X,-H\right)
=1~,
\ee
compatible with a superpotential term $\mathcal{W}\propto \Psi_1 \Psi_2$ for two chiral superfields  $\Psi_{1,2}$, which disappear from the IR physics. In the case $r=1$, $\hat Z^{\rm 4d}_\chi$ can be shown to reduce to  the result for  a  chiral multiplet found  in \cite{Closset:2013sxa,Assel:2014paa}.\footnote{In order to compare with the result of \cite{Assel:2014paa}, we need $\zeta_3(0,x|\omega_1,\omega_2,\omega_3)=-\frac{1}{6}B_{33}(x|\omega_1,\omega_2,\omega_3)$ and some property of the Bernoulli polynomials and elliptic Gamma function summarised in appendix \ref{appSpecial}.}

We see that there are two natural ways to  rewrite the lens index for a chiral 
\ben
\label{4dchiral1}
\hat{\mathcal{I}}^{(R)}_\chi(w,H)&=&e^{-\i\pi\left(\frac{1}{3}\Phi_3(X)+\frac{1}{2}\Phi_2(X)\right)}\times
e^{\frac{i\pi}{2r}H^2(r-1)}e^{-\frac{\i\pi}{2}\Phi_2(X)}
 \Big\| \mathcal{B}^{\rm 4d}_{\rm N}(x ;q_\tau,q_\sigma)\Big\|^2_{r}~,
\een
or
\ben
\label{4dchiral2}
\hat{\mathcal{I}}^{(R)}_\chi(w,H)&=&e^{-\i\pi\left(\frac{1}{3}\Phi_3(X)+\frac{1}{2}\Phi_2(X)\right)}\times
e^{-\frac{i\pi}{2r}H^2(r-1)}e^{\frac{\i\pi}{2}\Phi_2(X)}
 \Big\| \mathcal{B}^{\rm 4d}_{\rm D}(x ;q_\tau,q_\sigma)\Big\|^2_{r}~,
\een
where, in analogy with the 3d case, we defined the 4d holomorphic blocks for the anomaly free chiral
\be\label{4dDN}
\mathcal{B}^{\rm 4d}_{\rm D}(x;q_\tau,q_\sigma)= \frac{1}{\Gamma(q_\tau x^{-1} ;q_\tau,q_\sigma)}~, \quad
\mathcal{B}^{\rm 4d}_{\rm N}(x;q_\tau,q_\sigma)= \Gamma(x ;q_\tau,q_\sigma)~,
\ee
with
\be 
\mathcal{B}^{\rm 4d}_{\rm D}(x;q_\tau,q_\sigma)=\Theta(x;q_\tau)\mathcal{B}^{\rm 4d}_{\rm N}(x;q_\tau,q_\sigma)~.
\ee
We interpret the 4d blocks as partition functions on $D^2\times_\tau T^2_\sigma$, where $\epsilon=\tau/R_1$ is the cigar equivariant parameter and $\sigma$ is the torus modular parameter. From (\ref{4dchiral1}) and (\ref{4dchiral2})  we 
see that the polynomials $\Phi_3, \Phi_2$, which we will identify with anomaly contributions, are obstructions to factorization,
while the  anomaly free chiral indexes
\be
\label{dfac}
Z_{\rm D}[L_r\times S^1]= \Big\| \mathcal{B}^{\rm 4d}_{\rm D}(x ;q_\tau,q_\sigma)\Big\|^2_{r}~, \quad Z_{\rm N}[L_r\times S^1]= \Big\| \mathcal{B}^{\rm 4d}_{\rm N}(x ;q_\tau,q_\sigma)\Big\|^2_{r}~,
\ee
have a neat geometric realisation as 4d blocks glued through the 4d $r$-pairing (\ref{conrulemod}), which implements the gluing of two solid tori $D^2\times_\tau T^2_\sigma$ to form the $L_r\times S^1$ geometry. 

Similarly to the 3d case, 4d holomorphic blocks are annihilated by a set of difference equations which can be interpreted as Ward identities for surface operators  wrapping the torus $T^2_\sigma$ and acting at the tip of the cigar. 

For example for  $\mathcal{B}^{\rm 4d}_{\rm D}$ we find\footnote{For the free chiral case, there is an apparent symmetry between $q_\sigma$ and  $q_\tau$, for example   we  also have $\left(T_{q_\sigma,x}-\Theta(x^{-1};q_\tau)\right)\frac{1}{\Gamma(q_\sigma x^{-1};q_\tau,q_\sigma)}=0$. However there is a profound difference between  $q_\sigma$ and  $q_\tau$. This clearly visible  if we realise these 4d theories as defects in  6d theories engineered on elliptic Calabi-Yau's. In that setup $q_\sigma$ corresponds to  a  K\"ahler parameter while $q_\tau$ is related to the topological  string coupling.
}

\be
\left(T_{q_\tau,x}-\Theta(x^{-1};q_\sigma )\right)\mathcal{B}^{\rm 4d}_{\rm D}(x;q_\tau,q_\sigma)=
\frac{1}{\Gamma(x^{-1};q_\tau,q_\sigma)}-\frac{\Theta(x^{-1};q_\sigma )}{\Gamma(q_\tau x^{-1};q_\tau,q_\sigma)}=0~,
\ee
where $T_{q,x} f(x)=f(q x)$ is the $q$-shift operator acting on $x$. 
The lens index is annihilated also by another equation for the tilde variables
\be
\left(T_{q_\tau,x}-\Theta(x^{-1};q_\sigma )\right)Z_{\rm D}[L_r\times S^1]=
\left(T_{\tilde q_\tau,\tilde x}-\Theta(\tilde x^{-1};\tilde q_\sigma )\right)Z_{\rm D}[L_r\times S^1]=0~,
\ee
and similarly for $\mathcal{B}^{\rm 4d}_{\rm N}$, $Z_{\rm N}[L_r\times S^1]$. 

The existence of two commuting  sets of difference operators annihilating the lens index  indicates
 that it might be expressed in a block  factorised form.
Indeed we will shortly  see that anomaly free interacting theories can also be factorised in 4d holomorphic blocks.
We also  expect that our 4d holomorphic blocks will be the building blocks to construct partition functions on more general geometries through suitable pairings. For example, in section  \ref{other} we will discuss the $S^2\times T^2$ case.

We close this section by  observing  that our definition of  the  blocks $\mathcal{B}^{\rm 4d}_{\rm D}$  and  $\mathcal{B}^{\rm 4d}_{\rm N} $ via factorisation or as solutions to difference equations suffers from an obvious ambiguity.
It is clear that we have the freedom to multiply our blocks by $q_\tau$-phases
$c(x;q_\tau)$ satisfying 
\be 
c(q_\tau x;q_\tau)=c(x;q_\tau)\,, \qquad
\Big\| c(x;q_\tau)\Big\|^2_r=1~.
\ee 
The first condition ensures that the  $c(x;q_\tau)$ is a $q_\tau$-constant passing through the difference operator while the second condition ensures that these ambiguities disappear once two  blocks are glued. 4d  blocks for more complicated theories will be also defined up to $q_\tau$-phases, which can be expressed as elliptic ratios of theta functions.

\subsection{Vector multiplet}\label{4dvecrev}
Repeating the steps we have done for the chiral multiplet, we can also bring the vector multiplet contribution to the following form
\be\label{4dvector}
\prod_\alpha\hat{\mathcal{I}}_V(\alpha(\bs z),\alpha(\bs \ell))=e^{\i\pi\sum_\alpha\left(\frac{1}{3}\Phi_3(\alpha(\bs Z))+ \frac{1}{2} \Phi_2(\alpha(\bs Z))\right)}\times \hat Z^{\rm 4d}_V(\bs Z,\bs \ell)~,
\ee
with
\be\label{Zv4d}
\hat Z^{\rm 4d}_V(\bs Z,\bs \ell)=\prod_\alpha\frac{e^{-\frac{\i\pi}{2r}(r-1)\alpha(\bs \ell)^2}e^{\frac{\i\pi}{2} \Phi_2(\alpha(\bs Z))}}{\mathcal{G}(\alpha(\bs Z),\alpha(\bs \ell))}
%=\hat Z^{\rm 4d}_V(Q-\alpha(\bs Z),-\alpha(\bs \ell))^{-1}
~,
\ee
%where in the second equality we used the reflection property (\ref{refl}) and $\sigma^2=1$. 
where $\bs z=e^{\frac{2\pi\i}{\omega_3}\bs Z}~.$
Also in this case the prefactor of (\ref{4dvector}) is an exponential of a cubic  polynomial  contributing to the anomaly,
which we will  discuss in subsection (\ref{sec_anomaly}).
%Also
%\be\label{4dvector2}
%\hat Z^{\rm 4d}_V(\alpha(\bs Z),\alpha(\bs \ell))=\hat Z^{\rm 4d}_\chi(Q-\alpha(\bs Z),-\alpha(\bs\ell))~,
%\ee
%which is equivalent to a chiral with weight $-\alpha$ and $R=2$. 
In the 3d limit $\omega_3\to+\infty_\mathbb{R}$ we have
\be
\hat Z^{\rm 4d}_V(\bs Z,\bs \ell)\stackrel{\omega_3\to+\infty}{\longrightarrow} \prod_\alpha \frac{1}{\hat s_{b,\alpha(\bs \ell)}(\i Q/2+\i \alpha(\bs Z))}~,
\ee
matching the 3d vector contribution (\ref{3dvector}) with the  identifications $(\alpha(\bs Z),\alpha(\bs \ell))=(\i Z_\alpha,\ell_\alpha)$.
 It  the case $r=1$, $\hat Z^{\rm 4d}_V$ reduces to the contribution of the vector multiplet in \cite{Assel:2014paa}.
By using the  factorised form of the $\mathcal{G}$ function we can  express $ \hat Z^{\rm 4d}_V$ as
\be
 \hat Z^{\rm 4d}_V(\bs Z,\bs \ell)= \prod_{\alpha>0}\Big\|s^{\frac{1}{2}}_\alpha \frac{\Gamma(q_\tau s^{-1}_\alpha;q_\tau,q_\sigma)}{\Gamma(s^{-1}_\alpha;q_\tau,q_\sigma)}\Big\|^2_r=\prod_{\alpha>0}\Big\|s_\alpha^{\frac{1}{2}}\Theta(s_\alpha^{-1};q_\sigma)\Big\|^2_r~,
\ee

where we used (\ref{thetashift}), (\ref{thetamod}), (\ref{gammashiftpiu}), and defined the holomorphic variables 
\be
s_\alpha=e^{\frac{2\pi\i}{r\omega_1}\alpha(\bs Z)}e^{\frac{2\pi\i}{r}\alpha(\bs \ell)}~.
\ee
%=e^{2\pi\i\chi_\alpha}e^{\frac{2\pi\i}{r}\ell_\alpha}~.
%\quad \tilde s_\alpha=e^{\frac{2\pi\i}{r\omega_2}\alpha(\bs Z)}e^{-\frac{2\pi\i}{r}\alpha(\bs \ell)}=e^{2\pi\i\tilde\chi_\alpha}e^{\frac{2\pi\i}{r}\tilde\ell_\alpha}~,
%\ee
%where
%\be
%\tilde \chi_\alpha=\frac{\chi_\alpha}{r\tau-1}~,\quad \tilde \ell_\alpha=r-\ell_\alpha~.
%\ee
In this form we immediately see that in the 3d limit $q_\sigma\to 0$, $\hat Z^{\rm 4d}_V$ matches the 3d vector contribution (\ref{3dvhb})
%\be
%\hat Z^{\rm 4d}_V(\bs Z,\bs \ell)\stackrel{q_\sigma\to 0}{\longrightarrow}\prod_{\alpha>0} (2\i)^2\sin \pi (\chi_\alpha+\ell_\alpha/r)\sin \pi(\tilde\chi_\alpha+\tilde\ell_\alpha/r)~,
%\ee
(notice that $\Theta(x;0)=1-x$).
 We then define
\be\label{Bvec}
\mathcal{B}^{\rm 4d}_{\rm vec}(\{s_\alpha\};q_\tau,q_\sigma)=
%\prod_{\alpha>0} s_\alpha^{\frac{1}{2}}\frac{\Gamma(q_\tau s^{-1}_\alpha;q_\tau,q_\sigma)}{\Gamma(s^{-1}_\alpha;q_\tau,q_\sigma)}=
\prod_{\alpha>0} 
s_\alpha^{\frac{1}{2}}\Theta(s_\alpha^{-1};q_\sigma)
~,
\ee
such that
\be
\hat Z^{\rm 4d}_V(\bs Z,\bs \ell)=\Big\| \mathcal{B}^{\rm 4d}_{\rm vec}(\{s_\alpha\};q_\tau,q_\sigma)\Big\|^2_r~.
\ee
Other choices of $\mathcal{B}^{\rm 4d}_{\rm vec}$ are clearly possible possible. 
For example we can also write
\be
\hat Z^{\rm 4d}_V(\bs Z,\bs \ell)=\prod_{\alpha>0}\Big\|  \Theta(q_\tau^{\frac{1}{2}}s_\alpha;q_\tau)  \Gamma(q_\tau s_\alpha^\pm;q_\tau,q_\sigma)\Big\|^2_r~,
\ee
with
\be
 \mathcal{B}^{\rm 4d}_{\rm vec}(\{s_\alpha\};q_\tau,q_\sigma)=\prod_{\alpha>0} \Theta(q_\tau^{\frac{1}{2}}s_\alpha;q_\tau)  \Gamma(q_\tau s^{\pm}_\alpha;q_\tau,q_\sigma)~,\ee
 which in the 3d limit $q_\sigma\to 0$ reduces to the 3d block (\ref{secop}).
% \be
%\prod_{\alpha>0} \Theta(q_\tau^{\frac{1}{2}}s_\alpha;q_\tau)  \Gamma(q_\tau s^{\pm}_\alpha;q_\tau,q_\sigma)
%\stackrel{q_\sigma\to 0}{\longrightarrow} \prod_{\alpha>0} \frac{  \Theta(q_\tau^{\frac{1}{2}}s_\alpha;q_\tau) }{(q_\tau s_\alpha^{\pm};q_\tau)_\infty}\ee

Finally, we observe that the FI terms can also be naturally factorised as in 3d (\ref{fif})
\be
\label{4dfi}
e^{\frac{2\pi\i}{\omega_3}Z\frac{\xi^{\rm 4d}}{r} }e^{\frac{2\pi\i}{r}\theta \ell}= \Big\|\frac{\Theta( s^{-1} u_{\rm 4d} ;q_\tau)}{\Theta( u_{\rm 4d};q_\tau)\Theta( s^{-1};q_\tau)}\Big\|^{-2}_r~,
\ee
with
\be 
s=e^{\frac{2\pi\i}{r\omega_1}Z}e^{\frac{2\pi\i}{r}\ell}~,\quad u_{\rm 4d}=e^{-\frac{2\pi \i }{r\omega_1}\frac{\omega_1\omega_2}{\omega_3}\xi^{\rm 4d}}e^{-\frac{2\pi\i}{r}\theta}\,.
\ee

\subsection{Anomalies and factorisation}\label{sec_anomaly}

We now return to  the polynomials $\Phi_3$, $\Phi_2$  appearing in the modular transformations (\ref{4dchiral}), (\ref{4dvector}).
We will see that their total  contributions reconstructs the 4d anomaly polynomial. 
This  interplay between modular transformations  and anomalies  was first observed in \cite{Spiridonov:2012ww} (see also \cite{Assel:2014paa}, \cite{Ardehali:2015hya}).

%In the previous subsection we found it is possible to write chiral and vector multiplet lens indexes such that some holomorphic and factorisation properties are manifest. Moreover, one can also identify contributions which behave nicely with respect to the usual 3d limit $L_r\times S^1\to L_r$. However, as a byproduct one generates (exponential of) cubic polynomials spoiling some of these features, and, even worst, gauge invariance. As we are going to see explicitly, such contributions can be mapped to anomaly polynomials, and hence they have to vanish in a well defined theory (at least the dynamical ones). Let us begin by reminding that, upon the modular transform (\ref{4dchiral}), each chiral contributes to the lens index (\ref{lensindex}) with the exponential factor $e^{-\i\pi\mathcal{P}_I}$, where we defined the anomaly polynomial (see (\ref{4dchiral}))

Collecting the contribution of the  chiral multiplets we find
\ben
\mathcal{P}_{i}(\bs Z,\bs \Xi)&=&\frac{1}{3}\Phi_3\left(\frac{QR_i}{2}+\rho_i(\bs Z)+\phi_i(\bs \Xi)\right)+\frac{1}{2} \Phi_2\left(\frac{QR_i}{2}+\rho_i(\bs Z)+\phi_i(\bs \Xi)\right)~,
\een
where we introduced the exponentiated flavour fugacities $\bs \zeta=e^{\frac{2\pi\i}{\omega_3}\bs\Xi}$. Similarly, the vector contributes with
a factor  $e^{-\i\pi\sum_\alpha\mathcal{P}_\alpha}$, where
\ben
\sum_\alpha\mathcal{P}_\alpha(\bs Z)&=&-\sum_\alpha\left( \frac{1}{3}\Phi_3(\alpha(\bs Z))+\frac{1}{2} \Phi_2(\alpha(\bs Z))\right)~.
\een
In total we find
\be
\mathcal{P}_{\rm tot}(\bs Z,\bs \Xi)=\sum_i\sum_{\rho_i,\phi_i}\mathcal{P}_{i}(\bs Z,\bs\Xi)+\sum_{\alpha}\mathcal{P}_\alpha(\bs Z)=\mathcal{P}_{\rm loc}(\bs Z,\bs \Xi)+\mathcal{P}_{\rm gl}(\bs \Xi)~,
\ee
where in the last step we further distinguished between local (gauge (G)) and global (flavour (F), R-symmetry (R) and gravity (g)) contributions.

\textbf{$\bullet$ Gauge and mixed gauge anomalies.}
Collecting the various powers of $\bs Z$ we get
\begin{align}\label{firstanomalyeqn}
\text{GGG}&:\sum_{i}\sum_{\rho_i,\phi_i}\frac{\rho_i(\bs Z)^3}{3r\omega_1\omega_2\omega_3}\\
\text{GGR}&:\sum_{i}\sum_{\rho_i,\phi_i}\frac{\rho_i(\bs Z)^2}{2r\omega_1\omega_2\omega_3}Q(R_i-1)+\sum_\alpha\frac{\alpha(\bs Z)^2}{2r\omega_1\omega_2\omega_3}Q\cdot 1\\
\text{GGF}&:\sum_{i}\sum_{\rho_i,\phi_i}\frac{\rho_i(\bs Z)^2}{r\omega_1\omega_2\omega_3}\phi_i(\bs\Xi)\\
\text{GRR}&:\sum_{i}\sum_{\rho_i,\phi_i}\frac{\rho_i(\bs Z)}{4r\omega_1\omega_2\omega_3}(Q(R_i-1))^2\\
\text{GRF}&:\sum_{i}\sum_{\rho_i,\phi_i}\frac{\rho_i(\bs Z)}{r\omega_1\omega_2\omega_3}Q(R_i-1)\phi_i(\bs \Xi)\\
\text{GFF}&:\sum_{i}\sum_{\rho_i,\phi_i}\frac{\rho_i(\bs Z)}{r\omega_1\omega_2\omega_3}\phi_i(\bs \Xi)^2\\
\text{Ggg}&:\sum_{i}\sum_{\rho_i,\phi_i}\frac{\rho_i(\bs Z)}{12r\omega_1\omega_2\omega_3}(2\omega_3^2-\omega_1^2-\omega_2^2+2\omega_1\omega_2(r^2-1))~.\label{lastanomalyeqn}
\end{align}
All these terms have to vanish on physical  theories theories, leading to
conditions on the $R$-charge and on the flavour fugacities.

\textbf{$\bullet$ Global anomalies.}
For the $\bs Z$ independent terms we have
\begin{align}
\text{FFF}&:\sum_{i}\sum_{\rho_i,\phi_i}\frac{\phi_i(\bs \Xi)^3}{3r\omega_1\omega_2\omega_3}\\
\text{RRR}&:\sum_{i}\sum_{\rho_i,\phi_i}\frac{(Q(R_i-1))^3}{24r\omega_1\omega_2\omega_3}+\sum_\alpha \frac{(Q\cdot 1)^3}{{24r\omega_1\omega_2\omega_3}}\\
\text{FFR }&:\sum_{i}\sum_{\rho_i,\phi_i}\frac{\phi_i(\bs \Xi)^2}{2r\omega_1\omega_2\omega_3}Q(R_I-1)\\
\text{FRR}&:\sum_{i}\sum_{\rho_i,\phi_i}\frac{\phi_i(\bs \Xi)}{4r\omega_1\omega_2\omega_3}(Q(R-1))^2\\
\text{Fgg}&:\sum_{i}\sum_{\rho_i,\phi_i}\frac{\phi_i(\bs \Xi)}{12r\omega_1\omega_2\omega_3}(2\omega_3^2-\omega_1^2-\omega_2^2+2\omega_1\omega_2(r^2-1))\\
\text{Rgg}&:\left(\sum_{i}\sum_{\rho_i,\phi_i}\frac{Q(R_i-1)}{24r\omega_1\omega_2\omega_3}+\sum_\alpha \frac{Q\cdot 1}{24r\omega_1\omega_2\omega_3}\right)(2\omega_3^2-\omega_1^2-\omega_2^2+2\omega_1\omega_2(r^2-1))~.
\end{align}

In \cite{DiPietro:2014bca} it was observed that partition functions on  $M^3\times S^1_\beta$ have a divergent limit when  the $S^1$ radius $\beta$ shrinks to zero. The leading term is
\be
\ln Z[M^3\times S^1_\beta]\stackrel{\beta\to 0}{\sim} -\frac{\pi^2}{\beta}{\rm Tr}(R)L_R[M^3]-\frac{1}{12\beta}{\rm Tr}(U(1))L_F[M^3]+\text{subleading}~,
\ee
where $L_{R,F}[M^3]$ are integrals of local quantities which can be computed for the given 3d (Seifert) manifold $M^3$ and supergravity background. In the $M^3=S^3_b$
case in particular 
\be\label{komar}
\ln Z[S^3_b\times S^1_\beta]\stackrel{\beta\to 0}{\sim} -\frac{\pi^2 r_3(b+b^{-1})}{6\beta}{\rm Tr}(R)-\i m \frac{\pi^2 r_3^2}{3\beta}{\rm Tr}(U(1))~,
\ee
where $m$ is a real mass for the $U(1)$ symmetry and $r_3$ the $S^3_b$ scale. By using the asymptotics of $\Phi_3$, $\Phi_2$,
%\be
%-\frac{\i\pi}{3}\Phi_3\left(\frac{QR}{2}+X\right)-\frac{\i\pi}{2}\Phi_2\left(\frac{QR}{2}+X\right)\stackrel{\omega_3\to+\infty}{\sim}-\frac{\i\pi\omega_3}{12r\omega_1\omega_2}(Q(R-1)+2X)+\textrm{finite}~,
%\ee
it is not difficult to verify that 
\be
\label{univa}
\ln e^{-\i \pi \mathcal{P}_{\rm gl}}\stackrel{\omega_3\to+\infty}{\sim}-\frac{\i\pi\omega_3}{12r\omega_1\omega_2}\left(\sum_{i}\sum_{\rho_i,\phi_i}Q(R_i-1)+\sum_\alpha Q\cdot 1\right)-\frac{\i\pi\omega_3}{6r\omega_1\omega_2}\sum_{i}\sum_{\rho_i,\phi_i}\phi_i(\bs \zeta)\,,%\nn\\
%&&=-\frac{\i\pi\omega_3}{12r\omega_1\omega_2}Q\sum_{\mathcal{R}}{\rm Tr}_{\mathcal{R}}(R)+\frac{\i\pi\omega_3}{6r\omega_1\omega_2}(-\Xi^a)\sum_{\mathcal{R}}{\rm Tr}_{\mathcal{R}}(F_a)\equiv \frac{\mathcal{A}}{\beta}
\ee
reproducing  the expected  universal divergent factor with the identifications $\beta=\frac{2\pi}{\omega_3}$, $\i\omega_1=-\frac{b}{r_3}$, $\i\omega_2=-\frac{b^{-1}}{r_3}$, the volume being rescaled by $1/r$.

Finally we consider the  extra  exponential quadratic terms appearing in the definition of  $\hat Z^{\rm 4d}_\chi$ in (\ref{coc}). 
We already observed that in the   3d  limit $\omega_3 \to +\infty_\mathbb{R}$,
these polynomials  contribute the expected half CS  units. 
 These polynomials are actually $\omega_3$ independent, and for convenience
  we refer to their total  contribution as  3d anomaly contribution.
Each chiral of weights $\rho_i$, $\phi_i,$ contributes with
\be
\label{mcst1}
\mathcal{P}^{\rm 3d}_i=\mp\left(\frac{1}{2}\Phi_2\left(\frac{QR_i}{2}+\rho_i(\bs Z)+\phi_i(\bs \Xi)\right)-\frac{r-1}{2r}\left(\rho_i(\bs \ell)+\phi_i(\bs H)\right)^2\right)~,
%=\frac{\omega_3}{2r\omega_1\omega_2\omega_3}\left(\phantom{\frac{.}{.}}\rho(\bs Z)^2+\rho(\bs Z)(Q(R-1)+2\phi(\bs\Xi))+\right.\\
%\left.+\phi(\bs\Xi)^2+Q(R-1)\phi(\bs\Xi)+\frac{(Q(R-1))^2}{4}+\frac{r^2-1}{6}\omega_1\omega_2\right)
\ee
where the sign $\mp$ depends on the choice (\ref{4dchiral1}) or (\ref{4dchiral2}) respectively. 
%while the vector multiplet  yields 
%\ben 
%\label{mcst2}
%\sum_\alpha\mathcal{P}^{3d}_\alpha=- \sum_\alpha \frac{1}{2}\Phi_2\left(\alpha(\bs Z)\right)~.
%%&=&-\frac{\omega_3}{2r\omega_1\omega_2\omega_3}\left(\phantom{\frac{.}{.}}\alpha(\bs Z)^2-\alpha(\bs Z)Q+\frac{Q^2}{4}+\frac{r^2-1}{6}\omega_1\omega_2\right)\,.
%\een
In total we find
\be\label{3dan}
\mathcal{P}^{\rm 3d}_{\rm tot}(\bs Z,\bs \Xi)=\sum_{i}\sum_{\rho_i,\phi_i}\mathcal{P}^{\rm 3d}_{i}(\bs Z,\bs\Xi)=\mathcal{P}^{\rm 3d}_{\rm loc}(\bs Z,\bs \Xi)+\mathcal{P}^{\rm 3d}_{\rm gl}(\bs \Xi)~.
%+\sum_{\alpha}\mathcal{P}^{\rm 3d}_\alpha(\bs Z) 
\ee
On physical 4d  theories, where the 4d gauge anomaly is cancelled,   the would be 3d parity anomaly is also automatically cancelled,  namely in the 3d limit $e^{-\i\pi\mathcal{P}^{\rm 3d}_{\rm loc}}$  would contribute integer CS units.
This implies that the factor  $e^{-\i\pi\mathcal{P}^{\rm 3d}_{\rm loc}}$   can always be factorised in Theta functions as in (\ref{fif}).

We arrive at the conclusion that,  on physical theories where there is no obstruction from anomalies, the lens index integrand can be
expressed in terms of the holomorphic variables and arranged in the factorised form
\be
\label{infa}
I= e^{-\i\pi(\mathcal{P}_{\rm gl}+\mathcal{P}^{\rm 3d}_{\rm gl})}
  \times 
\sum_{\bs \ell}\oint\frac{\rd \bs z}{2\pi\i \bs z\prod_k |\mathcal{W}_k|} \Big\|\Upsilon^{\rm 4d} \Big\|_r^2~,
\ee
up to prefactors due to the non-dynamical anomalies.
As we will see in some explicit case, for anomaly free theories we also have
\be
I= e^{-\i\pi(\mathcal{P}_{\rm gl}+\mathcal{P}^{\rm 3d}_{\rm gl})}
  \times 
\sum_{c}\ \Big\|\mathcal{B}^{\rm 4d}_c \Big\|_r^2~.
\ee
We are thus led to try to use the integrand $\Upsilon^{\rm 4d}$  to define 
4d blocks via  block integrals as in the 3d case. We will return to this in section \ref{4dhb}.

In  \cite{Assel:2014paa}  it was pointed out
that the anomaly cancellation conditions are necessary to express the partition function on Hopf surfaces $\mathcal{H}_{p,q}\simeq S^1\times S^3$ in  terms of periodic variables (under  $S^1$ shifts) consistent with the invariance under large gauge transformations. 

To understand the effect of large gauge transformations at the level of the blocks, it is useful to 
look first at
the semiclassical  limit $\tau=R_1\epsilon \to 0$, where we remove the $\Omega$-deformation on the disk by turning off the equivariant parameter ($\epsilon \to 0$).
In this limit  the  theory  is effectively described by  a twisted superpotential
obtained by summing over the KK masses $\frac{\i}{R_1}$ and  $\frac{\i \sigma}{R_1}$
due to the torus compactification of the 4d theory \cite{Nekrasov:2009uh}. 
The contribution of a chiral multiplet  to the twisted superpotential is given by
 \be
\label{dpw}
\widetilde{ \mathcal{W}}(a)=\sum_{n,m\in \mathbb{Z}}\left(a+\frac{\i}{R_1}(\sigma n+ m)\right) \left(\ln(a+\frac{\i}{R_1}(\sigma n+ m))-1\right)~.
\ee
This  sum needs to be regularised, in  appendix \ref{twistreg}
 we briefly review how one can do that, the result is 
\be\label{Weff}
\widetilde{\mathcal{W}}(a)=\frac{\pi}{R_1}\mathcal{P}_3(\i R_1 a) +\frac{1}{2\pi R_1}\sum_{k\neq 0}\frac{e^{-2\pi R_1 a k}}{k^2(1-q_\sigma^k)}~,
\ee
where \be
\mathcal{P}_3(X)=\frac{X^3}{3 \sigma  }-\frac{ X^2 (1+\sigma )}{2 \sigma  }+\frac{X (1+\sigma  (3+\sigma ))}{6 \sigma  }-\frac{ (1+6 \sigma  (1+\sigma ))}{72 \sigma }~.
\ee
We can immediately identify in (\ref{Weff})  the semiclassical limit of the anomaly free chiral
\be 
\ln\mathcal{B}_{\rm N}(e^{-2\pi R_1 a};\tau,\sigma)=
\sum_{k\neq 0}\frac{e^{-2\pi R_1 ka}}{k(1-q_\tau^k)(1-q_\sigma^k)}
\stackrel{\tau\to0}{\sim} -\frac{1}{2\pi \i\tau}\sum_{k\neq 0}\frac{e^{-2\pi k R_1 a}}{k^2(1-q_\sigma^k)}=
\frac{\i\widetilde{\mathcal{W}}_N(a)}{\epsilon}~,
\ee
while $\mathcal{P}_3$  contributes to the anomaly polynomial on $\mathbb{R}^2\times T^2_\sigma$.

As it will become important later on, we observe that while the twisted superpotential as defined in (\ref{dpw})
is invariant under large gauge transformations being manifestly 
doubly periodic on the torus $T^2_\sigma$, i.e. invariant under $a\to a+\frac{\i}{R_1}(\sigma n+ m)$, the regularisation produces polynomial terms which explicitly break the periodicity. 
%This mechanism is similar to what happens when integrating out KK fermions on $S^3\times S^1$, where gauge non-invariant CS terms are produced because of regularization of the sum over the KK tower \cite{DiPietro:2014bca}. \SP{check}
%
Therefore the semiclassical analysis shows that  anomalies represent  an obstruction to the periodicity/gauge invariance of the superpotential.\footnote{See  \cite{Cecotti:2013mba} for a thorough analysis of the periodicity in the context of the 4d $tt^*$ equations.}

We then see that the block integrands of anomaly free theories defined in (\ref{infa}), 
in the semiclassical limit
\be
\log\Upsilon^{\rm 4d} \stackrel{\epsilon\to 0}{\longrightarrow}    \frac{\i \widetilde{ \mathcal{W}}}{\epsilon}~,
\ee
are  doubly periodic  on the torus.
In section \ref{4dhb} we will return to this point and see that at the quantum level,
the  invariance under  large gauge transformation will 
be preserved only up to $q_\tau$-phases.

\subsection{SQED}\label{isqed}

We will now study two interacting theories to illustrate the general mechanism of factorisation.
Our first example will be the  $U(1)$  theory with $N_f$ chirals and  $N_f$ antichirals, with 
R-charge $R$ and an FI terms (SQED). In this case the lens index  reads
 \be
I_{\rm SQED}=\sum_{\ell=0}^{r-1}\oint \frac{\rd z}{2\pi \i z} z^{-\frac{\xi^{\rm 4d}}{r}} e^{\frac{2\pi\i}{r}\ell\theta}
\prod_{a,b=1}^{N_f} 
\hat{\mathcal{I}}^{(R)}_\chi(z^{-1} \zeta_a,\ell+H_a)\hat{\mathcal{I}}^{(R)}_\chi(z \bar \zeta_b^{-1},-\ell-\bar H_b  )~,
\ee
where we  parametrise the fugacities as 
\be
z=e^{\frac{2\pi i}{\omega_3}Z}\,, \quad \zeta_a=e^{\frac{2\pi i }{\omega_3}M_a}\,, \quad \bar \zeta_b=e^{\frac{2\pi i}{\omega_3} \bar M_b}~,
\ee
with associated holonomies $\ell, H_a, \bar H_b$.
It is also useful to introduce the combinations
\be
X_a=\frac{QR}{2}+M_a\,,\quad \bar X_b= -\frac{QR}{2}+\bar M_b~.
\ee
We evaluate the lens index by taking the sum of the residues inside the unit circle at the poles 
\be
\label{poles4qsqed}
Z_{(1)}=jQ +kr\omega_1+ X_c +\omega_1[\ell+H_c]~,\quad Z_{(2)}=jQ+kr\omega_2 + X_c +\omega_2(r-[\ell+H_c])~,
\ee
where $j,k\in\mathbb{Z}_{\geq 0}$. The detailed computation is performed in appendix \ref{SQEDindex}, here we report the key steps. We first perform the modular transformation using (\ref{4dchiral1})  for the fundamentals and  (\ref{4dchiral2}) for the antifundamentals, and we get
\begin{multline}
 \label{modtra}
\prod_{a,b}\hat{\mathcal{I}}^{(R)}_\chi(z^{-1} \zeta_a,\ell+H_a)\hat{\mathcal{I}}^{(R)}_\chi(z\bar \zeta_b^{-1},-\ell-\bar H_b)=\\
=e^{-\i\pi\mathcal{P}_{\rm gl}}e^{-\i\pi\mathcal{P}_{\rm loc}}\prod_{a,b}\frac{e^{\frac{\i\pi}{2r}(\ell+\bar H_b)^2(r-1)}e^{-\frac{\i\pi}{2}\Phi_2(Z-\bar X_b)}}{e^{\frac{\i\pi}{2r}(\ell+ H_a)^2(r-1)}e^{-\frac{\i\pi}{2}\Phi_2(Q+Z-X_a)}}\frac{\mathcal{G}(Z-\bar X_b,-\ell-\bar H_b)}{\mathcal{G}(Q+Z-X_a,-\ell-H_a)}=\\=
e^{-\i\pi(\mathcal{P}_{\rm gl}+\mathcal{P}_{\rm gl}^{\rm 3d})}e^{-\i\pi(\mathcal{P}_{\rm loc}+\mathcal{P}^{\rm 3d}_{\rm loc})}
\prod_{a,b}\frac{\mathcal{G}(Z-\bar X_b,-\ell-\bar H_b)}{\mathcal{G}(Q+Z-X_a,-\ell-H_a)}~.
\end{multline}
As we discussed, the modular transformation produces polynomials contributing to the  global and local anomalies. The dynamical part of the 4d anomaly ($\mathcal{P}_{\rm loc}$) must vanish on this physical theory.
In fact, as this theory is non-chiral, the  GGG anomaly vanishes automatically, while the cancellation of  the GGF anomaly  requires the balancing of the $U(1)$ flavour charges of fundamentals and antifundamentals 
\be\label{barionanomaly}
\sum_i\sum_{\phi_i}\phi_i(\bs \Xi)=\sum_a M_a-\sum_b \bar M_b=0~.
\ee
This is actually automatic since the flavour symmetry group  is  $SU(N_f)\times SU(\bar N_f) \times U(1) $ with fundamentals and 
antifundamentals oppositely charged under the baryonic symmetry. Then we also have 
\be\label{Hbal}
\sum_a H_a-\sum_b \bar H_b=0\mod r~.
\ee
In order to cancel the GGR anomaly the condition is\footnote{We denote ${\rm Tr}_\mathcal{R}(T_n T_m)={\rm T}_2(\mathcal{R})\delta_{mn}$. For $SU(N_c)$ the fundamental and adjoint generators are normalised according to ${\rm T}_2(f)=1/2$, ${\rm T}_2(ad)=N_c$.}
\be\label{Z2anomaly_bis}
N_f{\rm T}_2(f)(R-1)+\bar N_f {\rm T}_2(\bar f)(R-1)+{\rm T}_2(ad)\cdot 1=0~,
\ee
which fixes  $R=1$.
For the vanishing of the GFF anomaly
we must require
\be\label{modanomaly}
\sum_i\sum_{\phi_i}\phi_i(\bs \Xi)^2=\sum_a M_a^2-\sum_b \bar M_b^2=0~,
\ee
\be
\label{modanomalyH}
\sum_a H_a^2-\sum_b \bar H^2_b=0~.
%\quad \sum_a H_a M_a-\sum_b \bar H_b\bar M_b=0~.
\ee
The other anomalies also vanish without imposing any further constraint. What is left of the 4d anomaly is the global part ($\mathcal{P}_{\rm gl}$), which reduces just to the FFF term.

Since we used  (\ref{4dchiral1})  for the fundamentals and  (\ref{4dchiral2}) for the antifundamentals,
 the  $Z^2$ terms in $\mathcal{P}^{\rm 3d}_{\rm loc}$ are  automatically cancelled. We could have also used (\ref{4dchiral2}) (or (\ref{4dchiral1})) for both fundamentals and antifundamentals as well. This would have led to a different but of course equivalent form of the integrand.
Altogether the 3d anomaly contributions yield the global factor  $\mathcal{P}^{\rm 3d}_{\rm gl}$  and a renormalisation of $\xi^{\rm 4d}$, $\theta$, which are however trivial once we impose (\ref{barionanomaly}), (\ref{Hbal}), (\ref{modanomaly}) and (\ref{modanomalyH}).

Finally we find
\be
\label{4dusqed}
I_{\rm SQED}=
e^{-\i\pi \mathcal{P}_{\rm gl}}\sum_{\ell=0}^{r-1}\oint \frac{\rd z}{2\pi \i z}\Big\|  \Upsilon^{\rm 4d}_{\rm SQED} \Big\|^2_r~,
\ee
with
\be
\label{sqedblock}
\Upsilon^{\rm 4d}_{\rm SQED}=\frac{\Theta( s^{-1}u_{\rm 4d} ;q_\tau)}{\Theta( u_{\rm 4d};q_\tau)\Theta(s^{-1};q_\tau)}
\prod_{a,b=1}^{N_f}\frac{\Gamma( s\bar x_b^{-1};q_\tau,q_\sigma)}
{\Gamma(q_\tau s x_a^{-1};q_\tau,q_\sigma)}~,
\ee
where we introduced the holomorphic variables
\be
s=e^{\frac{2\pi\i }{r\omega_1}Z}e^{-\frac{2\pi\i}{r}\ell}~,\quad x_a=e^{\frac{2\pi\i }{r\omega_1}X_a } e^{\frac{2\pi \i }{r}H_a}\,,
\quad \bar x_b=e^{\frac{2\pi \i}{r\omega_1}\bar X_b}  e^{\frac{2\pi\i }{r}\bar H_b}~,\quad
u_{\rm 4d}=e^{-\frac{2\pi \i }{r\omega_1}\frac{\omega_1\omega_2}{\omega_3}\xi^{\rm 4d}}e^{-\frac{2\pi\i}{r}\theta}~,
\ee
and used (\ref{thetamod}) to write
\be
e^{-\frac{2\pi\i}{\omega_3}\frac{\xi^{\rm 4d}}{r} Z}e^{\frac{2\pi\i}{r}\theta \ell}=\Big\|\frac{\Theta( s^{-1} u_{\rm 4d} ;q_\tau)}{\Theta( u_{\rm 4d};q_\tau)\Theta( s^{-1};q_\tau)}\Big\|^2_r~,
\ee
as in 3d.  Notice the integrand $\Upsilon^{\rm 4d}_{\rm SQED}$ in (\ref{sqedblock}) could have been assembled by adding a 4d block $\mathcal{B}^{\rm 4d}_{\rm D}$ for each chiral and  a block $\mathcal{B}^{\rm 4d}_{\rm N }$ for each anti-chiral plus the FI contribution. 
In this case the polynomial $\mathcal{P}^{3d}_{\rm loc}$ defined in  (\ref{3dan})
 vanishes.

Finally by taking the sum of the residues at the poles (\ref{poles4qsqed}), we obtain
\be\label{ZSQED4dbody}
I_{\rm SQED}=e^{-\i\pi\mathcal{P}_{\rm gl}}
\sum_{c=1}^{N_f} \Big\| \mathcal{B}^{\rm 4d}_{c} \Big\|^2_r~,
\ee
with
\be
\label{4dhbqed}
\mathcal{B}^{\rm 4d}_{c}=
\frac{\Theta(x_c^{-1}u_{\rm 4d} ;q_\tau)}{\Theta(u_{\rm 4d};q_\tau)\Theta(x_c^{-1};q_\tau)}
\prod_{a,b=1}^{N_f}\frac{\Gamma(x_c \bar x_b^{-1};q_\tau,q_\sigma)}
{\Gamma(q_\tau  x_c x_a^{-1};q_\tau,q_\sigma) }
 {}_{N_f}E_{N_f-1}\left(\begin{array}{c}
x_c \bar x_b^{-1} \\
q_\tau  x_c x_a^{-1} 
\end{array};q_\tau,q_\sigma;u_{\rm 4d}\right)~,
\ee
where the elliptic series ${}_N E_{N-1}$ is  defined in (\ref{ellseries1}). For $r=1$ our result agrees with \cite{Peelaers:2014ima} (after a modular transformation). Notice that the cancellation of the  GGF anomaly is  related to the balancing condition (\ref{ellbal1}) of the elliptic series, while the GFF anomaly cancellation to its modular properties (\ref{ellmod1}). The sum over $c$ runs over the supersymmetric vacua given by the minima of the the twisted superpotential discussed in the previous section.

It is easy  to write down a difference operator for these blocks.
We find that the elliptic hypergeometric series (\ref{ellseries1}) is annihilated by the operator
\be\label{ellop}
\hat H (\vec{x},\vec{y};u,T_{q_\tau,u})
=\left(\prod_{i=1}^N\Theta(q_\tau^{-1}y_i T_{q_\tau,u};q_\sigma)-u\prod_{i=1}^N\Theta(x_i T_{q_\tau,u};q_\sigma)\right)~.
\ee
Since 
\be
\mathcal{B}_c^{\rm 4d} \propto t(u_{\rm 4d};x_c){}_{N_f}E_{N_f-1}\left(\begin{array}{c}x_c\bar x_b^{-1}\\q_\tau x_c x_a^{-1}\end{array};q_\tau,q_\sigma;u_{\rm 4d}\right)~,
\ee
where for convenience we denoted
 \be
 t(u_{\rm 4d};x_c)=
 \frac{\Theta(x_c^{-1}u_{\rm 4d} ;q_\tau)}{\Theta( u_{\rm 4d};q_\tau)\Theta( x_c^{-1};q_\tau)}\,,\ee
 satisfying 
 \be
T_{q_\tau,u}^n t(u_{\rm 4d};x_c)^{-1}=x_c^{-n}t(u_{\rm 4d};x_c)^{-1}~,
\ee
we see that the  blocks $\mathcal{B}^{\rm 4d}_{c}$ are solutions to the difference operator
\be
t(u_{\rm 4d};x_c)\hat H(x_c\bar x_b^{-1},q_\tau x_c x_a^{-1};u_{\rm 4d},T_{q_\tau,u})t(u_{\rm 4d};x_c)^{-1}=\hat H(\bar x_b^{-1},q_\tau x_a^{-1};u_{\rm 4d},T_{q_\tau,u})~,
\ee
for $c=1,\ldots, N_f$. As we have already noticed in the case of the free chiral, if we define the blocks $\mathcal{B}^{\rm 4d}_{c}$ 
as solutions to this difference operator with the additional requirement that  their $r$-square reproduces the 
partition function (\ref{ZSQED4dbody}), we still have the $q_\tau$-phases ambiguity.
For example we can multiply the blocks by 
the  elliptic ratio of theta functions
\be 
\label{sqedqco}
c(u_{\rm 4d};q_\tau)=\prod_{a,b=1}^{N_f}\frac{\Theta(u_{\rm 4d}\bar x_b^{-1};q_\tau)}{\Theta(u_{\rm 4d} q_\tau x_a^{-1};q_\tau)}~,
\ee
which satisfies $c( q_\tau u_{\rm 4d};q_\tau)=c(u_{\rm 4d};q_\tau)$ and 
has unit $r$-square when the  anomaly cancellation conditions (\ref{barionanomaly}), (\ref{Hbal}), (\ref{modanomaly}), (\ref{modanomalyH}) are imposed.
It is also easy to check that since $\Theta(q_\tau^{1/2} x;q_\tau) \stackrel{\epsilon\to 0}{\longrightarrow}    e^{-\i \pi \frac{ (R_1X)^2   }{R_1\epsilon}},$ eq. (\ref{sqedqco}) has a trivial semiclassical limit. Indeed in  general 
$q_\tau$-phases are not  visible in the  the   semiclassical asymptotics.

We conclude by checking the 3d limit of our results.
At the level of the 4d blocks this amounts  to take $q_\sigma \to 0$, yielding
\be
\mathcal{B}^{\rm 4d}_{c}(\vec x; u_{\rm 4d};q_\tau,q_\sigma) \to  \mathcal{B}^{\rm 3d}_{c} 
(\vec x; u_{\rm 3d},q) ~,
\ee
with the obvious identifications
\be\label{3d4dmap}
q_\tau=q~,\quad  (\i X_a,H_a)\Big|_{\rm 4d}=(X_a,H_a)\Big|_{\rm 3d}~,\quad (\i X_b,\bar H_b)\Big|_{\rm 4d}=(\bar X_b,\bar H_b)\Big|_{\rm 3d}~.
\ee
Notice that the 3d mass parameters are still restricted to satisfy the 4d anomaly cancellation conditions.
As  explained in  \cite{Aharony:2013dha}, the 
 reduction  of the   4d  index to 3d  generates  theories
with the same gauge and matter content of the original theory but with a  compact Coulomb branch
and  with   non-trivial superpotential terms enforcing the  restriction on the masses  \cite{Aharony:2013dha}.
Moreover the relation between 4d and 3d FI parameters
% ($\omega_1\omega_2=1$ on $S^3_ b$)
\be
\i \frac{\xi^{\rm 4d}}{\omega_3}\stackrel{\omega_3\to +\infty}{\longrightarrow}\xi^{\rm 3d}
\ee
is consistent with a continuous   3d FI.

%
%
%\SP{remove below?}
%
%The exponential
%$\mathcal{P}_{\rm gl}$ yields the expected divergent term for the SQED \SP{is it zero??for sqcd?}. In the end we have
%\be
%I_{\rm SQED}= 
%e^{-\i\pi\mathcal{P}_{\rm gl}}
%\sum_{c=1}^{N_f} \Big\| \mathcal{B}^{\rm 4d}_{c} \Big\|^2_r~\stackrel{\omega_3\to +\infty}{\longrightarrow}~
%e^{\frac{\mathcal{A}}{\beta}} e^{-\i\pi \mathcal{P}^{\rm 3d}_{\rm gl} }
%\sum_{c=1}^{N_f} \Big\| \mathcal{B}^{\rm 3d}_{c} \Big\|^2_r=e^{\frac{\mathcal{A}}{\beta}}
%Z_{\rm SQED}[L_r]~.
%\ee
%
%
\subsection{SQCD}

We now move to the $SU(2)$ theory with $N_f$ chirals and  $N_f$ antichirals.
The lens index reads:
\be
I_{\rm SQCD}=\sum_{\ell=0}^{r-1}\oint \frac{\rd z}{2\pi \i z}\hat{\mathcal{I}}_V(z^{\mp 2},\pm 2\ell)\prod_{a,b=1}^{N_f} 
\hat{\mathcal{I}}^{(R)}_\chi(z^{\mp } \zeta_a,\pm \ell+H_a)\hat{\mathcal{I}}^{(R)}_\chi(z^\pm \bar \zeta_b^{-1},\mp \ell-\bar H_b  )~.
\ee
We can  collect the flavour fugacities and background holonomies into
\be
\zeta_{a'}=(\zeta_a,\bar \zeta_b^{-1})=\bar \zeta_{b'}^{-1}~,\quad H_{a'}=(H_a,-\bar H_b)=-\bar H_{b'}~,\quad a',b'=1,\dots ,2N_f~.
\ee
We also define
\be
X_{a'}=\frac{QR}{2}+M_{a'}= -\bar X_{b'}=\frac{QR}{2}-\bar M_{b'}~,
\ee
where $M_{a'}=(M_a,-\bar M_b)=-\bar M_{b'}$. In this notation the matter sector reads exactly the same as the  SQED theory with the replacements $a\to a'$ and $b\to b'$, the only differences being the different $R$ charge and the ``reality" constraints $X_{a'}=-\bar X_{b'}$, $H_{a'}=-\bar H_{b'}$. The set of poles inside the unit circle we will sum over is also formally unchanged with respect to the abelian case (\ref{poles4qsqed}) because the vector does not bring any pole. 

The first step is to perform the modular transformation, which upon imposing the anomaly cancellation allows us to
factorise the integrand as
\be
I_{\rm SQCD}=
e^{-\i\pi(\mathcal{P}_{\rm gl}+\mathcal{P}^{\rm 3d}_{\rm gl})}\sum_{\ell=0}^{r-1}\oint \frac{\rd z}{2\pi \i z}\Big\|  \Upsilon^{\rm 4d}_{\rm SQCD} \Big\|^2_r~,
\ee
with 
\be
\label{sqcdint}
\Upsilon^{\rm 4d}_{\rm SQCD}=  s\frac{\Gamma( q_\tau  s^{2};q_\tau,q_\sigma)}{\Gamma( s^{2};q_\tau,q_\sigma)}
\prod_{a',b'}\frac{\Gamma( s\bar x_{b'}^{-1};q_\tau,q_\sigma)}
{\Gamma( q_\tau s x_{a'}^{-1};q_\tau,q_\sigma)}~,
\ee
where
\be
s=e^{\frac{2\pi\i}{r\omega_1}Z}e^{-\frac{2\pi\i}{r}\ell}~,\quad x_{a'}=e^{\frac{2\pi\i}{r\omega_1}X_{a'}}e^{\frac{2\pi\i}{r}H_{a'}}~.
\ee
The GGF cancellation parallels the abelian case. The GGR anomaly cancellation 
\be\label{Z2anomaly}
N_f{\rm T}_2(f)(R-1)+\bar N_f {\rm T}_2(\bar f)(R-1)+{\rm T}_2(ad)\cdot 1=0~,
\ee
in this case yields $R=\frac{N_f-N_c}{N_f}$ for $SU(N_c)$. All other anomalies vanish without imposing further conditions.

Also in this case we observe that   the integrand $\Upsilon^{\rm 4d}_{\rm SQCD}$ in (\ref{sqcdint}) can be obtained by adding a 4d block $\mathcal{B}^{\rm 4d}_{\rm D/N}$ for each chiral/anti-chiral plus the vector multiplet contribution. In this case however we need to take into account 
 the polynomial $\mathcal{P}^{\rm 3d}_{\rm loc}$, which, once the 4d anomaly cancellation conditions are imposed, contributes a factor $\| s^2 \|^2_r$ to the partition function.

We then take the sum of the residues at the poles.
The detailed computation is performed in appendix \ref{SQCDindex}, here 
we give the final result in the fully factorised form
\be\label{ZSQCD4dbody}
I_{\rm SQCD}=e^{-\i\pi\mathcal{P}_{\rm gl}}\sum_{c'=1}^{2N_f}\Big\| \mathcal{B}^{\rm 4d}_{c'}\Big\|^2_r~,
\ee
with
\be\label{sqcdblock}
\mathcal{B}^{\rm 4d}_{c'}=x_{c'}
\Theta(x_{c'}^2;q_\sigma) 
\prod_{a'}\frac{\Gamma(x_{c'} x_{a'};q_\tau,q_\sigma)}
{\Gamma(q_\tau x_{c'} x_{a'}^{-1};q_\tau,q_\sigma)}
{}_{2N_f+4}E_{2N_f+3}\left(x_{c'};x_{c'} x_{a'};q_\tau,q_\sigma;1\right)~,
\ee
where we introduced the very-well-poised elliptic hypergeometric series defined in (\ref{ellseries2}). For $r=1$ our result agrees with \cite{Peelaers:2014ima} (after a modular transformation).

\section{$\mathcal{N}=1$  theories on $S^2\times T^2$}\label{other}

We now turn to the manifold $S^2\times T^2$  which supports $\mathcal{N}=1$ 
supersymmetric theories with R-symmetry.  
To preserve supersymmetry the theories need to be topologically twisted on $S^2$
and the R-charges need to be quantised. This background has been studied in 
\cite{Closset:2013sxa},\cite{Nishioka:2014zpa} and more recently in   \cite{Benini:2015noa} and   \cite{Honda:2015yha}.

As in the twisted index case reviewed in section \ref{3dtwistedindex}, the localising locus is parameterised by
 continuous variables $\bs Z$ in the Cartan  and discrete variables $\bs \ell$ in the maximal torus of the gauge algebra.
The integer variables $\bs \ell$ parameterise the quantised magnetic flux while   ${\bs z=e^{2\pi\i \bs Z}}$ 
is a combination of the two holonomies on the torus.
We also turn  on analogous continuous and discrete variables for the non-dynamical symmetries.
The partition function reads
\ben
Z[S^2\times T^2]&=&\sum_{\bs\ell}\oint_{\rm J.K.}\frac{\rd \bs z}{2\pi\i \bs z |\mathcal{W}| }~Z_{\rm cl}\times Z_{\rm 1-loop}^V \times Z^{\rm matter}_{\rm 1-loop}~.
\een
The contributions to the classical part come only from possible FI  terms for $U(1)$ factors
\be
e^{-{\rm Vol}(T^2)\zeta {\bs\ell}}=\xi^{\bs \ell}\,.
\ee 

The contribution of a chiral multiplet with R-charge $R$,  $U(1)$ fugacity $z$ and  flux $H$
 is given by\footnote{ 
The relation between our Theta function $\Theta(x;q_\sigma)$ and the theta function $\vartheta_1(x;q_\sigma)$
appearing in \cite{Closset:2013sxa,Nishioka:2014zpa,Benini:2015noa,Honda:2015yha} is $\vartheta_1(x;q_\sigma)=\i \eta(q_\sigma) q_\sigma^\frac{1}{12}x^{-\frac{1}{2}}\Theta(x;q_\sigma)$, $\eta(q_\sigma)=q_\sigma^\frac{1}{24}(q_\sigma;q_\sigma)_\infty$.}
\be 
\label{s2t2chi}
Z_{\chi}^{(B)}[S^2\times T^2]=  q_\sigma^{-\frac{B}{12}}z^{\frac{B}{2}}
\prod_{k=-\frac{|B|-1}{2}}^{\frac{|B|-1}{2}}\frac{1}{\Theta(q_\tau^k z;q_\sigma)^{\textrm{sign}(B)}}
=
\frac{  q_\sigma^{-\frac{B}{12}}z^{\frac{B}{2}}}{\Theta( q_\tau^{\frac{1-B}{2}} z;q_\tau,q_\sigma )_{B}}
~,
\ee
where we used the definition of $\Theta$-factorials in (\ref{thetafac}) and defined $B=H-R+1$.
The vector multiplet contribution is given by\footnote{Up to a zero-point energy contribution $\eta(q_\sigma)^{2|G|}\prod_{\alpha}q_\sigma^{\frac{1}{12}}$ which can be absorbed  in the integration measure. 
In  \cite{Benini:2015noa} an  extra $(-1)^{\ell_\alpha}$ appears in the definition of the vector multiplet.}
\be
\label{s2t2v}
Z_V[S^2\times T^2]=\prod_{\alpha>0}  q_\tau^{-\frac{|\ell_\alpha|}{2}}  \Theta( q_\tau^{\frac{|\ell_\alpha|}{2}}z_\alpha^{\pm};q_\sigma)~.
\ee

In the above expressions $q_\sigma=e^{2\pi\i \sigma}$ is identified with   the torus complex modulus and $ q_\tau=e^{2\pi\i\tau}$ with  the angular momentum fugacity.
By using that  $\Theta(x;0)=1-x$, it is immediate to check that, in the   $q_\sigma\to 0$  limit, the 1-loop contributions (\ref{s2t2chi}) and (\ref{s2t2v}) tend to their counterpart on $S^2_A\times S^1$ (up to the zero-point energy factor).

Geometrically, the $S^2\times_\tau T^2_\sigma$ background is realised by  gluing two solid tori $D^2\times_\tau T^2_\sigma$ twisted in the same direction so that  to realise the A-twist on $S^2$. We then expect that also in this case partition functions can be expressed in
terms of the universal blocks $\mathcal{B}^{{\rm 4d}}_c$ fused with the $A$-gluing defined  by
\be
\tau\to -\tau~,  \quad \sigma\to \sigma~,\quad Z\to Z~, \quad {\rm or}\quad q_\tau\to q^{-1}_\tau~, \quad q_\sigma\to q_\sigma~, \quad z\to z~.
\ee

As clear from our discussion on anomalies, the free chiral alone is not expected to factorise, we need instead to look at an anomaly free object,
for example
\begin{multline}
 \label{ZDs2t2}
\Big\| \mathcal{B}^{\rm 4d}_{\rm D}(x ;q_\tau,q_\sigma)\Big\|^2_{A}=
\frac{1}{ \Gamma( q_\tau^{\frac{2+h}{2}} z;q_\tau,q_\sigma )
\Gamma( q_\tau^{-\frac{2+h}{2}} z;q^{-1}_\tau,q_\sigma )
   }=\\ 
   %\frac{1}{\Theta( q_\tau^{-\frac{h}{2}} z;q_\tau,q_\sigma )_{h+1}}
=\frac{1}{\Theta( q_\tau^{\frac{1-B}{2}} z;q_\tau,q_\sigma )_{B}}
=Z_{\rm D}[S^2\times T^2]
~,
\end{multline}
where we identified  the holomorphic variable $x$ with the combination $x=z^{-1}q_\tau^{-H/2}$.
As expected 
\be 
\label{atc2}
Z_{\chi}^{(B)}[S^2\times T^2]=Z_{\rm D}[S^2\times T^2]\times z^{\frac{B}{2}} q_\sigma^{-\frac{B}{12}}~,
\ee
showing that  we need to multiply the anomaly free chiral  by the factor $z^{B/2}$, which in the 3d twisted index limit 
we identified with a  half CS unit, and by  the zero-point energy. 

 FI terms can also be expressed as $A$-squares as in   (\ref{twistedCS}).
Similarly, the vector multiplet contribution can be re-obtained by fusing two  4d blocks $\mathcal{B}^{\rm 4d}_{\rm vec}$ (\ref{Bvec}) 
with $s_\alpha=z^{-1}_\alpha q_\tau^{-\ell_\alpha/2} $
\be
\label{atv}
\Big\|\prod_{\alpha>0}s_\alpha^{\frac{1}{2}}  \Theta(s_\alpha^{-1};q_\sigma)\Big\|^2_{A}=
 \prod_{\alpha>0}q_\tau^{-\frac{|\ell_\alpha|}{2}}
 \Theta(q_\tau^{\frac{|\ell_\alpha|}{2}}z_\alpha^{\pm} ;q_\sigma)
=Z_V[S^2\times T^2]~.
\ee

So we arrive at the conjectured relation
\be
Z[S^2\times T^2]=e^{-\i\pi\mathcal{P}}\sum_{\bs\ell}\oint\frac{\rd \bs z}{2\pi\i \bs z|\mathcal{W}|}\Big\|\Upsilon^{\rm 4d}\Big\|^2_A=e^{-\i\pi\mathcal{P}}
 \sum_c   \Big\| \mathcal{B}^{\rm 4d}_c\Big\|^2_{A}\,.\ee
The first equality states the factorisation of the integrand of the Coulomb branch partition function. This follows from the above discussion on chiral and vector multiplets. For anomaly free theories, the induced effective half CS units  either cancel between chirals and antichirals or add up to integer values  and can be factorised  as in   (\ref{twistedCS}).

The second non-trivial equality states the factorisation of the $S^2\times T^2$ partition   function in terms of the very same 4d
 blocks $\mathcal{B}^{\rm 4d}_c$ found in the $L_r\times S^1$ case.
 
Let us explicitly check this relation  in  the SQED case. The partition function is given by
\be
\begin{split}
Z_{\rm SQED}[S^2\times T^2]=\sum_{\ell\in\mathbb{Z}}\oint_{\rm J.K.}\frac{\rd z}{2\pi\i z}\xi^\ell Z_{\rm 1-loop}(z,\zeta,\bar \zeta,B,\bar B)~,\\
Z_{\rm 1-loop}(z,\zeta,\bar \zeta,B,\bar B)=\prod_{a,b=1}^{N_f}\frac{q_\sigma^{-\frac{B_a}{12}}(z\zeta_a)^{\frac{B_a}{2}}q_\sigma^{-\frac{\bar B_b}{12}}(z^{-1}\bar \zeta_b^{-1})^{\frac{\bar B_b}{2}}}{\Theta(q_\tau^\frac{1-B_a}{2} z\zeta_a;q_\sigma,q_\tau)_{B_a}\Theta(q_\tau^\frac{1-\bar B_b}{2} z^{-1}\bar \zeta_b^{-1};q_\sigma,q_\tau)_{\bar B_b}}~,
\end{split}
\ee
where
\be
B_a=1+h_a+\ell~,\quad \bar B_b=1+\bar{h}_b-\ell~.
\ee
In this case the anomaly cancellation conditions are
\be
\prod_{a,b}\zeta_a\bar \zeta_b^{-1}=1~,\quad \sum_{a,b}(h_a+\bar h_b)+2N_f=0~,\quad \prod_{a,b}\zeta_a^{h_a+1}\bar\zeta_b^{\bar h_b+1}=1~.
\ee
By using  the definition of $\Theta$-factorials in (\ref{thetafac}) 
it is easy to show that we can equivalently rewrite the partition function as
\be
Z_{\rm SQED}[S^2\times T^2]=e^{-\i\pi\mathcal{P}_{\rm SQED}}\sum_{\bs \ell}\oint\frac{\rd  z}{2\pi\i z}\Big\|\Upsilon^{\rm 4d}_{\rm SQED}\Big\|^2_A~,
\ee
with the SQED integrand defined in (\ref{sqedblock}) with the identifications
\be
\label{s2t2parap}
s=q_\tau^{\frac{\ell}{2}}z\, \quad  x_a=q_\tau^{-\frac{h_a}{2}}\zeta_a^{-1}~,\quad \bar x_b=q_\tau^{\frac{\bar{h}_b}{2}}\bar\zeta_b^{-1}~,\quad u_{\rm 4d}=(-1)^{N_f}\xi~,
\ee
and
\be
 e^{-\i\pi\mathcal{P}_{\rm SQED}}=(-1)^{\frac{1}{2}\sum_{a,b}(h_a-\bar h_b)}~.
\ee
The integration contour is determined by the Jeffrey-Kirwan residue prescription, which in this case simply amounts in taking the contribution from the simple poles associated to the fundamental matter (mod $q_\sigma^\mathbb{Z}$). Such factors have poles only for $B_c=\ell+h_c+1>0$, which are then at
%\be
%\begin{split}
%Z_{\rm SQED}[S^2\times T^2]=\sum_{\ell\in\mathbb{Z}}e^{2\pi\i\xi\ell}\sum_{z_*\in \mathcal{M}^*_{\rm sing}}{\rm JKRes}_{z=z_*}(\bs Q(z_*,\eta))Z_{\rm 1-loop}(z,\zeta,\bar \zeta,B,\bar B)~,\\
%Z_{\rm 1-loop}(z,\zeta,\bar \zeta,B,\bar B)=\prod_{a,b=1}^{N_f}\frac{q_\sigma^{-\frac{B_a}{12}}(z^{-1}\zeta_a)^{\frac{B_a}{2}}q_\sigma^{-\frac{\bar B_b}{12}}(z\bar \zeta_b)^{\frac{\bar B_b}{2}}}{\Theta(q_\tau^\frac{1-B_a}{2} z^{-1}\zeta_a;q_\sigma,q_\tau)_{B_a}\Theta(q_\tau^\frac{1-\bar B_b}{2} z\bar \zeta_b;q_\sigma,q_\tau)_{\bar B_b}}~,
%\end{split}
%\ee
%By taking $\eta<0$, in this case the JKRes operation just tells us to take the sum of the residues at the simple poles coming from the fundamental matter (mod $q_\sigma^\mathbb{Z}$). Such factors have poles only for $B_c=\ell-\bs r_c+1>0$, which are then at
\be
z=z_*=\zeta_c^{-1} q_\tau^{\frac{B_c-1-2k}{2}}=\zeta_c^{-1} q_\tau^{\frac{\ell+h_c-2k}{2}}~,\quad k=0,\ldots, \ell+h_c~, \quad c=1,\ldots, N_f~.
\ee
Therefore
\be\label{s2t2res}
Z_{\rm SQED}[S^2\times T^2]=e^{-\i\pi\mathcal{P}_{\rm SQED}}\sum_c\sum_{\ell\geq -h_c}\sum_{k=0}^{\ell+h_c}\Big\|\Upsilon^{\rm 4d}_{\rm SQED}\Big\|^2_A~,
\ee
%\be
%Z_{\rm SQED}[S^2\times T^2]=\sum_c\sum_{-\ell\geq \bs r_c}\sum_{k=0}^{-\ell-\bs r_c}e^{-2\pi\i\xi\ell}Z_{\rm 1-loop}(z_*,\zeta,\bar \zeta,B,\bar B)~,
%\ee
and we can replace
\be
\sum_{\ell\geq -h_c}\sum_{k=0}^{\ell+h_c}=\sum_{k_1,k_2\geq 0}~,\quad k_1=\ell+h_c-k~,\quad k_2=k~.
\ee
%Now it may be useful to rewrite
%\begin{multline}
%Z_{\rm 1-loop}(z_*,\zeta,\bar \zeta,B,\bar B)=\prod_{a,b=1}^{N_f}q_\sigma^{-\frac{B_a}{12}}(z_*^{-1}\zeta_a)^{\frac{B_a}{2}}q_\sigma^{-\frac{\bar B_b}{12}}(z_*\bar \zeta_b)^{\frac{\bar B_b}{2}}\times\\
%\times\frac{\Gamma(q_\tau^\frac{1-B_a}{2} z_*^{-1}\zeta_a;q_\sigma,q_\tau)\Gamma(q_\tau^\frac{1-\bar B_b}{2} z_*\bar \zeta_b;q_\sigma,q_\tau)}{\Gamma(q_\tau^\frac{1-B_a}{2} z_*^{-1}\zeta_a q_\tau^{B_a};q_\sigma,q_\tau)\Gamma(q_\tau^\frac{1-\bar B_b}{2} z_*\bar \zeta_b q_\tau^{\bar B_b};q_\sigma,q_\tau)}~,
%\end{multline}
Substituting $s_*=q^{\ell/2}z_*=q_\tau^{k_1}x_c$, $\tilde s_*=q^{-\ell/2}z_*=q_\tau^{-k_2}\tilde x_c$ into (\ref{s2t2res}), with the help of (\ref{thetafac}), (\ref{thetafacneg}), one can finally show that
%\begin{multline}
%Z_{\rm SQED}[S^2\times T^2]=e^{-\i\pi\mathcal{P}} \sum_c \zeta_c^{\frac{1}{2}(2N_f-\sum_{a,b}(\bs r_a+\bar{\bs r}_b))} \left((-1)^{N_f}e^{2\pi\i\xi}\prod_{a,b}\zeta_a^{-1/2}\bar\zeta_b^{-1/2}\right)^{ \bs r_c}\times\\
%\times\Big\|\prod_{a,b=1}^{N_f}\frac{\Gamma(x_c\bar x_b^{-1};q_\sigma, q_\tau)}{\Gamma(q_\tau x_c x_a^{-1};q_\sigma, q_\tau)}
%{}_{N_f}E_{N_f-1}\left(\begin{array}{c}x_c\bar x_b^{-1}\\q_\tau x_c x_a^{-1}\end{array};q_\tau, q_\sigma; u_{\rm 4d}\right)\Big\|^2_{A}~,
%\end{multline}
%where we defined 
%\be
%x_a=q_\tau^{\bs r_a/2}\zeta_a~,\quad \bar x_b=q_\tau^{-\bar{\bs r}_b/2}\bar \zeta_b^{-1}~,\quad u_{\rm 4d}=(-1)^{N_f}e^{2\pi\i\xi}q_\tau^{\frac{1}{4}(2N_f-\sum_{a,b}(\bs r_a+\bar{\bs r}_b))}\prod_{a,b}\zeta_a^{-1/2}\bar\zeta_b^{-1/2}~.
%\ee 
%Using
%\be
%\Big\|\Theta(-q_\tau^{\frac{1+B}{2}}  w  ;q_\tau)\Big\|^2_{A}=w^{-B}~,
%\ee
%we can also write
%\be
%\zeta_c^{\frac{1}{2}(2N_f-\sum_{a,b}(\bs r_a+\bar{\bs r}_b))} \left((-1)^{N_f}e^{2\pi\i\xi}\prod_{a,b}\zeta_a^{-1/2}\bar\zeta_b^{-1/2}\right)^{ \bs r_c}\propto\Big\|\frac{\Theta(x_c^{-1}u_{\rm 4d};q_\tau)}{\Theta(x_c^{-1};q_\tau)\Theta(u_{\rm 4d};q_\tau)}\Big\|^2_{A}~,
%\ee
%so that 
\be
Z_{\rm SQED}[S^2\times T^2]= e^{-\i\pi\mathcal{P}_{\rm SQED}}\sum_c  \Big\| \mathcal{B}^{\rm 4d}_c\Big\|_A^2~,
\ee
with the very same $\mathcal{B}^{\rm 4d}_c$ defined in (\ref{4dhbqed}). This is result agrees perfectly with the expected result following our analysis. 

The $SU(2)$ case is essentially the same, since the vector multiplet  does not bring new poles to the integrand. 
 We define
\be
\zeta_{a'}=(\zeta_a,\bar \zeta_b^{-1})=\bar\zeta_{b'}^{-1}~,\quad h_{a'}=(h_a,\bar h_b)=\bar h_{b'}~,
\ee
and $x_{a'}=(x_a,\bar x_{b}^{-1})=\bar x_{b'}^{-1}$ with the same parametrisation as in  (\ref{s2t2parap}).
The anomaly cancellation requires
\be
\prod_{a'}\zeta_{a'}=1~,\quad \sum_{a'}h_{a'}+2N_f-4=0~.
\ee
As expected also the SQCD can be expressed in terms of the blocks  $\mathcal{B}^{\rm 4d}_{c'}$ given in (\ref{sqcdblock})
\be
Z_{\rm SQCD}[S^2\times T^2]= e^{-\i\pi\mathcal{P}_{\rm SQCD}}\sum_{c'}  \Big\| \mathcal{B}^{\rm 4d}_{c'}\Big\|_A^2~.
\ee

\section{4d holomorphic blocks}\label{4dhb}

In this section we would like to develop a  formalism to compute  the holomorphic blocks from first principles by 
extending to 4d the 3d formalism introduced in \cite{hb}.
 We tentatively define  4d blocks via block integrals as
\be
\mathcal{B}^{\rm 4d}_c
=\oint_{\Gamma_c}\frac{\rd \bs s}{2\pi\i \bs s}\Upsilon^{\rm 4d}~,
\ee 
where $\Upsilon^{\rm 4d}$ is the ``square root" of the compact space integrand.
As we have seen in sections  \ref{sec_anomaly} and \ref{other},  when there are no obstructions from anomalies
it is always  possible to factorise the compact space integrand.
Alternatively one can assemble directly  $\Upsilon^{\rm 4d}$.
For each  chiral multiplet we insert a factor 
$\mathcal{B}^{\rm 4d}_{\rm D}$ or $\mathcal{B}^{\rm 4d}_{\rm N}$ and adding an appropriate ratio of
Theta functions associated to  $\mathcal{P}^{\rm 3d}_{\rm loc}$ to cancel the induced mixed CS units.
We then add $\mathcal{B}^{\rm 4d}_{\rm vec}$
for each  vector multiplet and in presence of $U(1)$ gauge factors we multiply by the FI contributions given in
(\ref{4dfi}).

Before discussing  the integration contour it is important to make the following observation.
In section \ref{sec_anomaly} we observed that as a result of invariance under large gauge transformations,  block integrals are semiclassically doubly periodic on the torus $T^2_\sigma$.
As we anticipated, at the quantum level there is a mild modification, that is
 under the shift  $s\to s q_\sigma$ the blocks are multiplied by 
 $q_\tau$-phases with unit $r,A$-square, representing the intrinsic ambiguity in their definition.

For example consider the SQCD block integrand
\be
\label{qup}
\Upsilon^{\rm 4d}_{\rm SQCD}(s)=
s\Theta(  s^{2};q_\sigma)
\prod_{a',b'}\frac{\Gamma( s\bar x_{b'}^{-1};q_\tau,q_\sigma)}
{\Gamma( q_\tau s x_{a'}^{-1};q_\tau,q_\sigma)}\,.
\ee
It is easy to check that the effect of the shift $s\to s q_\sigma$ is simply to multiply the integrand by the $q_\tau$-phase
\be
\frac{\Upsilon^{\rm 4d}_{\rm SQCD}(q_\sigma s)}{\Upsilon^{\rm 4d}_{\rm SQCD}(s)}=
%q_\sigma \frac{\Theta(q_\sigma^2 z^2;q_\sigma)}{\Theta(z^2;q_\sigma)}\prod_{a',b'}\frac{\Theta(z\bar x^{-1}_{b'};q_\tau)}{\Theta(q_\tau z x^{-1}_{a'};q_\tau)}=
s^{-4}\prod_{a',b'}\frac{\Theta(s\bar x^{-1}_{b'};q_\tau)}{\Theta(q_\tau s x^{-1}_{a'};q_\tau)}~,\quad
\Big\|  \frac{\Upsilon^{\rm 4d}_{\rm SQCD}(q_\sigma s)}{\Upsilon^{\rm 4d}_{\rm SQCD}(s)} \Big\|^2_{r,A}=1~.
\ee 
To see this we observe that thanks to the anomaly cancellation condition $\sum_{a'}(Q-2X_{a'})=4Q$ we have
\be
\prod_{a',b'}\Big\|\frac{\Theta( s\bar x^{-1}_{b'};q_\tau)}{\Theta(q_\tau  s x^{-1}_{a'};q_\tau)} \Big\|^2_r=\prod_{a'}e^{\frac{2\pi\i}{r\omega_1\omega_2}Z(Q-2X_{a'})}=e^{\frac{2\pi\i}{r\omega_1\omega_2}4ZQ}=\Big\| s^4\Big\|^2_r~,
\ee
 and similarly
\be
\prod_{a',b'}\Big\|\frac{\Theta( s\bar x^{-1}_{b'};q_\tau)}{\Theta(q_\tau  s x^{-1}_{a'};q_\tau)} \Big\|^2_A=\prod_{a'}z^{2+2h_{a'}}\zeta_{a'}^{2\ell}=\Big\| s^4\Big\|^2_A~,
\ee
for $\prod_{a'}\zeta_{a'}=1$, $\sum_{a'}h_{a'}+2N_f-4=0$.
As $q_\tau$-phases have trivial semiclassical limit, the  doubly periodicity is indeed restored in the semiclassical limit.

This observation will guide us in the definition of the integration contour.
For example the SQCD block integrand (\ref{qup}) has poles at $s= x_{c'}q_\tau^k  q_\sigma^{n+1}$ and $s=\bar x_{c'} q_\tau^{-k}q_\sigma^{-n}$, $k,n\in\mathbb{Z}_{\geq 0}$. However our  discussion indicates that we should   restrict to a $q_\sigma$ period. 
Indeed a shift by $q_\sigma^n$ (where $n$ may be negative)  would only multiply the integrand and the integrated result by a  $q_\tau$-phase.
We then suggest that the proper integration contour $\Gamma_c$
will encircle the poles located at  $s=x_{c'}q_\tau^k$ coming from the fundamental  chirals.
Indeed it is easy to check that 
\begin{multline}
\oint_{s=x_c q_\tau^k }\frac{\rd s}{2\pi\i s}\Upsilon^{\rm 4d}_{\rm SQCD}=x_{c'}\Theta(  x_{c'}^{2};q_\sigma)
\prod_{a',b'}\frac{\Gamma( x_{c'} \bar x_{b'}^{-1};q_\tau,q_\sigma)}
{\Gamma( q_\tau x_{c'} x_{a'}^{-1};q_\tau,q_\sigma)}\times\\
\frac{\Theta(x_{c'}^{2}q_\tau^{2k};q_\sigma)}{\Theta( x_{c'}^{2};q_\sigma)}
\prod_{a',b'}\frac{\Theta( x_{c'} \bar x_{b'}^{-1};q_\sigma,q_\tau)_k}
{\Theta( q_\tau x_{c'} x_{a'}^{-1};q_\sigma,q_\tau)_k}q_\tau^k~,
\end{multline}
and  integrating over $\Gamma_c$ we recover the SQCD blocks defined in (\ref{sqcdblock})
\be
\oint_{\Gamma_c} \frac{\rd s}{2\pi\i s}\Upsilon^{\rm 4d}_{\rm SQCD}=\mathcal{B}^{\rm 4d}_c\,.
\ee

In general determining convergent contours could be quite delicate. For example the analogy with the 3d case suggests that
by moving in the moduli  we could encounter Stokes walls where contours jump \cite{hb}. We leave the general discussion of integration contours
to future analysis.
However, we  can check that our prescription works also in the SQED case where blocks can be obtained by integrating 
the SQED integrand (\ref{4dusqed}) 
\be
\Upsilon^{\rm 4d}_{\rm SQED}(s)=
 \frac{\Theta(s^{-1}u_{\rm 4d} ;q_\tau)}{\Theta( u_{\rm 4d};q_\tau)\Theta( s^{-1};q_\tau)}
\prod_{a,b}\frac{\Gamma( s\bar x_b^{-1};q_\tau,q_\sigma)}
{\Gamma( q_\tau s x_a^{-1};q_\tau,q_\sigma)}~,
\ee
along the contour $\Gamma_c$  encircling  the poles located at  $z=x_{c}q_\tau^k$
\be
\oint_{\Gamma_c}\frac{\rd s}{2\pi\i s}\Upsilon^{\rm 4d}_{\rm SQED}=\mathcal{B}^{\rm 4d}_c~,
\ee
with $\mathcal{B}^{\rm 4d}_c$ defined in (\ref{4dhbqed}).
Notice that also  in this case we are using the prescription to restrict to a $q_\sigma$  period. However, in this case the FI term explicitly breaks
the periodicity already at the semiclassical level.
Nevertheless we find that also in this case  a $q_\sigma$-shift  has a trivial effect:
\be\label{upshift}
\frac{\Upsilon_{\rm SQED}(q_\sigma s)}{\Upsilon_{\rm SQED}(s)}=
 \frac{\Theta( q_\sigma^{-1}  s^{-1}u_{\rm 4d} ;q_\tau)  \Theta(s^{-1};q_\tau)}{\Theta(  s^{-1}u_{\rm 4d} ;q_\tau) \Theta(q_\sigma^{-1} s^{-1};q_\tau) }
\prod_{a,b} \frac{\Theta( s\bar x_b^{-1};q_\tau)}
{\Theta( q_\tau s x_a^{-1};q_\tau)}~.
  \ee
Indeed the second  factor is a $q_\tau$-phase
\be
\prod_{a,b}\Big\|\frac{\Theta( s\bar x_b^{-1};q_\tau)}
{\Theta( q_\tau s x_a^{-1};q_\tau)}\Big\|^2_{r,A}=1~,
\ee
once we impose all the anomaly cancellations.
The  first factor also has unit square
\be
\Big\| \frac{\Theta(  q_\sigma^{-n}  s^{-1}u_{\rm 4d} ;q_\tau)  \Theta( s^{-1};q_\tau)}{\Theta(s^{-1}u_{\rm 4d} ;q_\tau) \Theta( q_\sigma^{-n} s^{-1};q_\tau)}
\Big\|^2_r=
e^{-\frac{2\pi i}{r}  \xi^{\rm 4d} n   } =1~, \qquad \Big\| \frac{\Theta(  q_\sigma^{-n}  s^{-1}u_{\rm 4d} ;q_\tau)  \Theta( s^{-1};q_\tau)}{\Theta(s^{-1}u_{\rm 4d} ;q_\tau) \Theta( q_\sigma^{-n} s^{-1};q_\tau)}
\Big\|^2_A=1~,
\ee
since  $  \xi^{\rm 4d}/r$ is integer on the lens index.

Summarising we have argued  that for  $L_r\times S^1$ (which includes $S^3\times S^1$) and $S^2\times T^2$ we have the following  remarkable  Riemann bilinear-like relations
\begin{align}
 \sum_{\bs \ell}\oint_{T^{|G|}}\frac{\rd \bs z}{2\pi\i \bs z\prod_k |\mathcal{W}_k| }~\Big\| \Upsilon^{\rm 4d} \Big\|^2_r&=
\sum_{c} \Big\| \oint_{\Gamma_c} \frac{\rd \bs s}{2\pi\i \bs s}   \Upsilon^{\rm 4d} \Big\|^2_r~,\\
 \sum_{\bs \ell}\oint_{\rm J.K.}\frac{\rd \bs z}{2\pi\i \bs z |\mathcal{W}| }~\Big\| \Upsilon^{\rm 4d} \Big\|^2_A&=
\sum_{c} \Big\| \oint_{\Gamma_c} \frac{\rd \bs s}{2\pi\i \bs s}   \Upsilon^{\rm 4d} \Big\|^2_A~.
\end{align}
This  identities seem to be quite ubiquitous for these backgrounds  and it  would be important to have a 
deeper  understanding of their geometrical meaning.
Riemann-bilinear like identities    appear also in  the analytic continuation of Chern-Simons theory \cite{Witten:2010cx} and in the the study of
$tt^*$ geometries \cite{Cecotti:1991me}.

While 3d holomorphic blocks have been relatively well studied, here we have only initiated the study of 4d blocks
and there are various  directions to explore.
For example it would  be interesting to study the behaviour of 4d blocks under 4d dualities.
It should be  also fairly simple to re-derive our 4d block integrand prescription via localisation on $D^2\times T^2$,
however  the general  definition of integration contours seems quite challenging.
Another aspect to investigate  is the relation of 4d blocks to integrable systems and to CFT correlators.
 3d block integrals  have been identified with $q$-deformed  Virasoro free-field correlators in \cite{Aganagic:2013tta}, \cite{Aganagic:2014oia}.  
The possibility to interpret 4d  block integrals as free-field correlators in an elliptic deformation of the Virasoro algebra will be  investigated in \cite{fab}.

\section*{Acknowledgments}
SP would like to thank T~Dimofte for collaboration on related topics.
FB and SP would like to thank G~Bonelli and  F~Benini for discussions.
The work of FN is partially  supported  by the EPSRC - EP/K503186/1.

\appendix

\section{Special functions}\label{appSpecial}

\subsection{Bernoulli polynomials}\label{ber}
The quadratic Bernoulli polynomial $B_{22}$ is
\be
B_{22}(X|\omega_1,\omega_2)=\frac{1}{\omega_1\omega_2}\left(\left(X-\frac{Q}{2}\right)^2-\frac{\omega_1^2+\omega_2^2}{12}\right)~, \quad Q=\omega_1+\omega_2~.
\ee
Useful properties are
\ben 
B_{22}(\lambda X|\lambda \omega_1,\lambda \omega_2)&=&B_{22}(X| \omega_1, \omega_2)~,\quad \lambda\neq 0~,\\
B_{22}(X+\omega_2| \omega_1, \omega_2)&=&B_{22}(X| \omega_1, \omega_2)+\frac{2X-\omega_1}{\omega_1}~,\\
B_{22}(X| \omega_1, \omega_2)&=&B_{22}(Q-X| \omega_1, \omega_2)~.
\een

We define the combination
\begin{multline}\label{phi2}
\Phi_2(X)=B_{22}(X|Q,r\omega_1)+B_{22}(X+r\omega_2|Q,r\omega_2)=\\
=B_{22}(X+r\omega_1|Q,r\omega_1)+B_{22}(X|Q,r\omega_2)=\Phi_2(Q-X)~.
\end{multline}
We also have
\be
\Phi_2(X)=\frac{1}{r}B_{22}(X|\omega_1,\omega_2)+\frac{r^2-1}{6r}.
\ee

The cubic Bernoulli polynomial $B_{33}$ is
\be\label{B33}
B_{33}(X|\omega_1,\omega_2,\omega_3)=\frac{1}{\omega_1\omega_2\omega_3}\left(X-\frac{Q}{2}-\frac{\omega_3}{2}\right)\left(\left(X-\frac{Q}{2}\right)^2-\omega_3\left(X-\frac{Q}{2}\right)-\frac{\omega_1^2+\omega_2^2}{4}\right)~.
\ee
Useful properties are
\begin{align}
B_{33}(\lambda X|\lambda \omega_1,\lambda \omega_2,\lambda \omega_3)&=B_{33}(X| \omega_1, \omega_2,\omega_3)~,\quad \lambda\neq 0~,\\
B_{33}(X+\omega_3| \omega_1, \omega_2,\omega_3)&=B_{33}(X| \omega_1, \omega_2,\omega_3)+3B_{22}(X| \omega_1, \omega_2)~,\\
B_{33}(X|\omega_1,\omega_2,-\omega_3)&=-B_{33}(X+\omega_3|\omega_1,\omega_2,\omega_3)~.
%B_{33}(X| \omega_1, \omega_2,\omega_3)&=-B_{33}(\omega_3+Q-X| \omega_1, \omega_2,\omega_3)~.
\end{align}
We define the combination
\be
\label{phi3}
\Phi_3(X)=B_{33}(X|Q,r\omega_1,\omega_3)+B_{33}(X+r\omega_2|Q,r\omega_2,\omega_3)~.
\ee
We also have
\begin{align} \Phi_3(X)&=\frac{1}{r}B_{33}(X|\omega_1,\omega_2,\omega_3)+\frac{r^2-1}{4r\omega_3}(2X-Q)-\frac{r^2-1}{4r}~,\\
\Phi_3(X+\omega_3)&=3\Phi_2(X)+\Phi_3(X)=-\Phi_3(Q-X)~.
\end{align}

\subsection{Double Gamma and Sine functions}
The Barnes double Gamma function $\Gamma_2$ is defined as the $\zeta$-regularized product
\be
\Gamma_2(X|\omega_1,\omega_2)=\prod_{n_1,n_2\geq 0}\frac{1}{X+n_1\omega_2+n_2\omega_2}~.
\ee
It satisfies the functional relation
\be\label{G2ratio}
\frac{\Gamma_2(X+\omega_2|\omega_1,\omega_2)}{\Gamma_2(X|\omega_1,\omega_2)}=\frac{1}{\Gamma_1(X|\omega_1)}~,
\ee
where $\Gamma_1$ is simply related to the Euler $\Gamma$ function, $\Gamma_1(X|\omega_1)=\frac{\omega_1^{\frac{X}{\omega_1}-\frac{1}{2}}}{\sqrt{2\pi}}\Gamma\left(\frac{X}{\omega_1}\right)$.

The double Sine function $S_2$ is defined as the $\zeta$-regularized product
\be
S_2(X|\omega_1,\omega_2)=\prod_{n_1,n_2\geq 0}\frac{n_1\omega_1+n_2\omega_2+X}{n_1\omega_1+n_2\omega_2+Q-X}~,
\ee
where $Q=\omega_1+\omega_2$. The regularised expression is given by
\be\label{S2G2}
S_2(X|\omega_1,\omega_2)=\frac{\Gamma_2(Q-X|\omega_1,\omega_2)}{\Gamma_2(X|\omega_1,\omega_2)}~.
\ee
For irrational $\frac{\omega_1}{\omega_2}$, the $S_2$ has simple poles and zeros at 
\be
\begin{array}{ll}
\text{zeros }:& X=-n_1\omega_1-n_2\omega_2\\[5pt]
\text{poles }:& X=Q+n_1\omega_1+n_2\omega_2
\end{array}~,\quad n_1,n_2\in\mathbb{Z}_{\geq 0}~.
\ee
It enjoys the properties
\begin{align}
&S_2(X|\omega_1,\omega_2)S_2(Q-X|\omega_1,\omega_2)=1~,\\
&\frac{S_2(X+\omega_2|\omega_1,\omega_2)}{S_2(X|\omega_1,\omega_2)}=\frac{1}{S_1(X|\omega_1)}~,\label{S2ratio}\\
&S_2(\lambda X|\lambda \omega_1,\lambda \omega_2)=S_2(X| \omega_1, \omega_2)~,\quad \lambda\neq 0~,
\end{align}
where the $S_1$ function is simply related to the sine function, $S_1(X|\omega_1)=2\sin\left(\frac{\pi X}{\omega_1}\right)$.

For $n_1,n_2\in \mathbb{Z}_{\geq 0}$, formulas (\ref{G2ratio}), (\ref{S2ratio}) are generalized to
\begin{align}
\frac{\Gamma_2(X+n_1\omega_1+n_2\omega_2|\omega_1,\omega_2)}{\Gamma_2(X|\omega_1,\omega_2)}
&=\frac{\prod_{j=0}^{n_1-1}\prod_{k=0}^{n_2-1}(X+j\omega_1+k\omega_2)^{-1}}{\prod_{j=0}^{n_1-1}\Gamma_1(X+j\omega_1|\omega_2)\prod_{k=0}^{n_2-1}\Gamma_1(X+k\omega_2|\omega_1)}~,\\
\frac{\Gamma_2(X-n_1\omega_1-n_2\omega_2|\omega_1,\omega_2)}{\Gamma_2(X|\omega_1,\omega_2)}&=\frac{\prod_{j=1}^{n_1}\Gamma_1(X-j\omega_1|\omega_2)\prod_{k=1}^{n_2}\Gamma_1(X-k\omega_2|\omega_1)}{\prod_{j=1}^{n_1}\prod_{k=1}^{n_2}(X-j\omega_1-k\omega_2)}~,
\end{align}
and
\begin{align}
\frac{S_2(n_1\omega_1+n_2\omega_2+X|\omega_1,\omega_2)}{S_2(X|\omega_1,\omega_2)}&=\frac{(-1)^{n_1n_2}}{\prod_{j=0}^{n_1-1} S_1(j\omega_1+X|\omega_2)\prod_{k=0}^{n_2-1}S_1(k\omega_2+X|\omega_1)}\label{S2piu}~,\\
\frac{S_2(n_1\omega_1-n_2\omega_2+X|\omega_1,\omega_2)}{S_2(X|\omega_1,\omega_2)}&=(-1)^{n_1n_2}\frac{\prod_{k=0}^{n_2-1}S_1(k\omega_2+Q-X|\omega_1)}{\prod_{j=0}^{n_1-1} S_1(j\omega_1+X|\omega_2)}\label{S2meno}~.
\end{align}
%Useful factorised expressions for the $S_2$ can be obtained by introducing the $q$-Pochhammer symbols
%\be
%(x;q)_\infty=\prod_{k=0}^{\infty}(1-xq^k),\quad (x;q)_n=\prod_{k=0}^{n-1}(1-xq^k)~,
%\ee
%where $|q|<1$ is assumed for the convergence of $(x;q)_\infty$, and it is otherwise defined by the ``analytic continuation"
%\be\label{qanalytic}
%(x;q)_\infty\rightarrow \frac{1}{(q^{-1}x;q^{-1})_\infty}~,
%\ee
%which can be understood using the following $q$-series 
%\be\label{qPochhseries}
%(x;q)_\infty=\sum_{n\geq 0}(-1)^nq^{\frac{n(n-1)}{2}}\frac{x^n}{(q;q)_n}~,\quad \frac{1}{(x;q)_\infty}=\sum_{n\geq 0}\frac{x^n}{(q;q)_n}~.
%\ee
For ${\rm Im}\left(\frac{\omega_1}{\omega_2}\right)\neq 0$, using  the $q$-Pochhammer defined in eq. (\ref{zqinf})
we can express the double sine function in a factorised form:
\be\label{S2factor}
S_2(X|\omega_1,\omega_2)=e^{\frac{\i\pi}{2}B_{22}(X|\omega_1,\omega_2)}(e^{\frac{2\pi \i }{\omega_1}X};e^{2\pi \i\frac{\omega_2}{\omega_1}})_\infty(e^{\frac{2\pi \i }{\omega_2}X};e^{2\pi \i\frac{\omega_1}{\omega_2}})_\infty~.
\ee
In order to compute contour integrals, we will also be interested in the asymptotic behaviour of $S_2$ for $X\to\infty$
\be\label{asymptotics}
S_2(X|\omega_1,\omega_2)\sim \left\{\begin{array}{ll}
e^{\frac{\i\pi}{2}B_{22}(X)}&\quad{\rm if}\quad {\rm arg}(\omega_1)<{\rm arg}(X)<{\rm arg}(\omega_2)+\pi\\[5pt]
e^{-\frac{\i\pi}{2}B_{22}(X)}&\quad{\rm if}\quad {\rm arg}(\omega_1)-\pi<{\rm arg}(X)<{\rm arg}(\omega_2)
\end{array}\right.~.
\ee

Another useful function is the shifted double Sine function $s_b$
\be
s_b(X)=S_2(Q/2-\i X|\omega_1,\omega_2)~,
\ee
in which case it is usually assumed $\omega_2=\omega_1^{-1}=b$. 
%Given
%\be
%S_1(X|\omega_1)=ie^{-\frac{i\pi X}{\omega_1}}(1-e^{\frac{2\pi i X}{\omega_1}})~,
%\ee
%the identities (\ref{S2piu}), (\ref{S2meno}) can be written as
%\be
%\frac{S_2(n_1\omega_1+n_2\omega_2+X|\omega_1,\omega_2)}{S_2(X|\omega_1,\omega_2)}=
%\frac{(-1)^{n_1n_2+n_1+n_2}~i^{n_1+n_2}~e^{\frac{i\pi X}{\omega_2}n_1}e^{\frac{i\pi X}{\omega_1}n_2}~
%e^{\frac{i\pi \omega_1}{\omega_2}\frac{n_1(n_1-1)}{2}}e^{\frac{i\pi \omega_2}{\omega_1}\frac{n_2(n_2-1)}{2}}}{(e^{\frac{2\pi i X}{\omega_1}};e^{2\pi i\frac{\omega_2}{\omega_1}})_{n_2}(e^{\frac{2\pi i X}{\omega_2}};e^{2\pi i\frac{\omega_1}{\omega_2}})_{n_1}}
%\ee

\subsection{Generalised double Sine function}
The following $\zeta$-regularised product
\be\label{S2h}
S_{2,h}(X)=\prod_{\substack{n_1,n_2\geq 0\\ n_2-n_1=h\mod r} }\frac{n_1\omega_1+n_2\omega_2+X}{n_2\omega_1+n_1\omega_2+Q-X}~,
\ee
defines a generalisation of the $S_2$ function (which is recovered for $r=1$).\footnote{Another class of generalised multiple Sine functions has been extensively studied in \cite{tizwi}.}
The parameters $\omega_1$, $\omega_2$ and $r$ are not displayed amongst the arguments for compactness. For irrational $\frac{\omega_1}{\omega_2}$, it has simple zeros and poles at
\be
\begin{array}{ll}
\text{zeros }:& X=-n_1\omega_1-n_2\omega_2\\[5pt]
\text{poles }:& X=Q+n_1\omega_2+n_2\omega_1
\end{array}~,\quad n_2-n_1=h\mod r~,\quad n_1,n_2\in\mathbb{Z}_{\geq 0}~. 
\ee
We can rewrite $S_{2,h}$ in terms of the ordinary $S_2$ as follows. First of all, we can resolve the constraint $n_2-n_1=h\mod r$ as
\be
n_2=n_1+[h]+kr\geq 0,\quad k\in \mathbb{Z}~,
\ee
where $[h]$ denotes the smallest non negative number $\text{mod}~r$.\footnote{For positive $h$ we have $h=[h]+r\lfloor h/r\rfloor$, while for negative $h$ we have $h=[h]+r(\lceil h/r\rceil-1)$. Also, for non-zero $h$ we have $[-h]=r-[h]$. In any case, we have $h=[h]+rn_h$, $[h]\geq 0$ for a suitable $n_h\in \mathbb{Z}$.} Then we can write (\ref{S2h}) as
\ben\label{S2hsequence}
S_{2,h}(X)&=&\prod_{n_1\geq 0}\prod_{k\geq -\lfloor\frac{n_1+[h]}{r}\rfloor}\frac{n_1\omega_1+(n_1+[h]+kr)\omega_2+X}{(n_1+[h]+kr)\omega_1+n_1\omega_2+Q-X}=\nn\\
&=&\frac{\prod_{s\geq 0}\prod_{k\geq-\lfloor \frac{s}{r}\rfloor}\frac{(s-[h])\omega_1+(s+kr)\omega_2+X}{(s+kr)\omega_1+(s-[h])\omega_2+Q-X}}
{\prod_{s=0}^{[h]-1}\prod_{k\geq-\lfloor \frac{s}{r}\rfloor}\frac{(s-[h])\omega_1+(s+kr)\omega_2+X}{(s+kr)\omega_1+(s-[h])\omega_2+Q-X}}~,
\een
where we set $s=n_1+[h]$. Moreover, for a generic sequence of functions $f_{s,k}$ we have
\be
\frac{\prod_{s\geq 0}\prod_{k\geq -\lfloor s/r\rfloor}f_{s,k}}{\prod_{s=0}^{[h]-1}\prod_{k\geq -\lfloor s/r\rfloor}f_{s,k}}=
\frac{\prod_{s,k\geq 0}f_{s,k+1}f_{s+kr,-k}}{\prod_{s=0}^{[h]-1}\prod_{k\geq 0}f_{s,k}}~,
\ee
where in the last step we used that in the denominator $s\in [0,r-1]<r$ so that $\lfloor s/r\rfloor=0$. Substituting the actual expression (\ref{S2hsequence}) for $f_{s,k}$, we finally get
\be\label{S2hS2}
S_{2,h}(X)=S_2(\omega_1(r-[h])+X|Q,r\omega_1)S_2(\omega_2[h]+X|Q,r\omega_2)~,
\ee
where we used the definition (\ref{S2G2}) of $S_2$ and repeatedly used the relation (\ref{G2ratio}). 
It is easy to check  the following reflection property
\be\label{reflection}
S_{2,h}(X)S_{2,-h}(Q-X)=1\,.
\ee
From (\ref{S2hS2}) we see that zeros and poles are located at
\be
\label{rpoles}
\begin{array}{llll}
\text{zeros }:& X=-\omega_1(p-[h])-kQ-nr\omega_1~,\quad & X=-\omega_2[h]-Qk-np\omega_2~,\\[5pt]
\text{poles }:& X=Q+\omega_1[h]+kQ+nr\omega_1~,\quad & X=Q+\omega_2(r-[h])+kQ+nr\omega_2~,
\end{array}
\ee
for $k,n\in\mathbb{Z}_{\geq 0}$, which are all simple and distinct as long as $\frac{\omega_1}{\omega_2}$ is irrational.
 Using (\ref{S2factor}) we can  obtain the factorised form
\be\label{S2hfact}
S_{2,h}(X)=e^{-\frac{\i\pi}{2r}[h](r-[h])}e^{\frac{\i\pi}{2}\Phi_2(X)}(e^{\frac{2\pi \i}{r\omega_1}(X-[h]\omega_1)};e^{2\pi \i\frac{Q}{r\omega_1}})_\infty (e^{\frac{2\pi \i}{r\omega_2}(X+[h]\omega_2)};e^{2\pi \i\frac{Q}{r\omega_2}})_\infty~.
\ee
This leads us to define the $r$-pairing
\be\label{ppairing}
\Big\|f(\omega_1,\omega_2,[h])\Big\|^2_r=\Big\|f(\omega_1,\omega_2,[h])\Big\|^2_{\substack{\omega_1\leftrightarrow\omega_2\\ h\leftrightarrow r-h}}=f(\omega_1,\omega_2,[h])f(\omega_2,\omega_1,r-[h])~,
\ee
exchanging $\omega_1$, $\omega_2$ and reflecting the holonomy variable, so that $S_{2,h}$ can be compactly represented as
\be\label{S2hfactorised}
S_{2,h}(X)=e^{-\frac{\i\pi}{2r}[h](r-[h])}e^{\frac{\i\pi}{2}\Phi_2(X)}\Big\|(e^{\frac{2\pi \i}{r\omega_1}(X-[h]\omega_1)};e^{2\pi \i\frac{Q}{r\omega_1}})_\infty\Big\|^2_r\,.
\ee
Notice we may remove the $[\cdot]$ inside the $q$-Pochhammer symbols because of the periodicity. Moreover, the asymptotic behaviour of $S_{2,h}$ for $X\to \infty$ can be deduced from (\ref{asymptotics})
\be\label{asymptotics2}
S_{2,h}(X)\sim \left\{\begin{array}{ll}
e^{-\frac{i\pi}{2r}[h](r-[h])} e^{\frac{\i\pi}{2}\Phi_{2}(X)}&\quad{\rm if}\quad {\rm arg}(\omega_1)<{\rm arg}(X)<{\rm arg}(\omega_2)+\pi\\[5pt]
e^{\frac{i\pi}{2r}[h](r-[h])}e^{-\frac{\i\pi}{2}\Phi_2(X)}&\quad{\rm if}\quad {\rm arg}(\omega_1)-\pi<{\rm arg}(X)<{\rm arg}(\omega_2)
\end{array}\right.~\,.
\ee
In the main text we need also to introduce an improved $S_{2,h}$, defined by
\be
\hat S_{2,h}(X)=\sigma(h)S_{2,h}(X),\quad \sigma(h)=e^{\frac{\i\pi}{2r}([h](r-[h])-(r-1)h^2)}~,
\ee
where $\sigma(h)$ is a sign factor, namely $\sigma(h)=\pm 1$ depending on the value of $h$. Also, it is convenient to introduce the improved $s_b$ function 
\ben
\label{imds}
\hat s_{b,-h}(X)=\hat S_{2,h}(Q/2- \i X|\omega_1,\omega_2)~,
\een
satisfying the reflection property
\be\label{sbreflection}
\hat s_{b,h}(X)\hat s_{b,-h}(-X)=1~.
\ee

In the particular case $r=1$ (and hence $h=0$), we obtain an interesting  identity for the ordinary $S_2$. In fact, for $r=1$ the product in (\ref{S2h}) is not actually restricted, and we obtain the relation
\be
S_{2,0}(X)|_{r=1}=S_2(X|\omega_1,\omega_2)=S_2(\omega_1+X|Q,\omega_1)S_2(X|Q,\omega_2)~,
\ee
or, in terms of the modular parameter $\tau=\frac{\omega_2}{\omega_1}$ 
\be\label{zagierformula}
S_2(\chi|1,\tau)=S_2\left(1+\chi|1,1+\tau\right)S_2\left(\frac{\chi}{1+\tau}|1,\frac{\tau}{1+\tau}\right)~,
\ee
where we rescaled $\chi=X/\omega_1$. This identity  
appears in  eq. (3.38) of \cite{Dimofte:2009yn}, where
\be
e^{-\frac{i\pi}{2}B_{22}(z|1,\tau)}S_2(z|1,\tau)=\Phi\left(z-\frac{1+\tau}{2};\tau\right)~
\ee
in their notation.
\subsection{Elliptic functions}
The short Jacobi Theta function is defined by
\be\label{Theta}
\Theta(x;q)=(x;q)_\infty(qx^{-1};q)_\infty~.
\ee
Useful properties are 
\be\label{thetashift}
\frac{\Theta(q^mx;q)}{\Theta(x;q)}=(-xq^{(m-1)/2})^{-m},\quad 
\frac{\Theta(q^{-m}x;q)}{\Theta(x;q)}=(-x^{-1}q^{(m+1)/2})^{-m}~,
\ee
where $m\in\mathbb{Z}_{\geq 0}$. 
We will be using the generalised modular transformation property of the theta function
\be\label{thetamod}
\Theta(e^{\frac{2\pi i}{r\omega_1}X}e^{\frac{2\pi\i}{r}h};e^{2\pi\i\frac{ Q}{r\omega_1}})\Theta(e^{\frac{2\pi i}{r\omega_2}X}e^{-\frac{2\pi\i}{r}h};e^{2\pi\i\frac{ Q}{r\omega_2}})=e^{-\i\pi\Phi_2(X)}e^{\frac{\i\pi}{r}h(r-h)}~,
\ee
For $r=1$ this formula reduce to the standar modular transformation of the theta function (see for example \cite{naru}).

The elliptic Gamma function is defined by
\be
\Gamma(x;p,q)=\frac{(pq x^{-1};p,q)_\infty}{(x;p,q)_\infty}~,
\ee
where the double $q$-Pochhammer symbol is defined by
\be
(x;p,q)_\infty=\prod_{j,k=0}^\infty(1-xp^jq^k)~.
\ee
It is assumed $|p|,|q|<1$ for convergence, and it can be extended to $|q|>1$ by means of
\be\label{pqcont}
(x;p,q)_\infty\to \frac{1}{(q^{-1}x;p,q^{-1})_\infty}~.
\ee
The elliptic Gamma function $\Gamma(x;p,q)$ has zeros and poles outside and inside the unit circle at
\be
\begin{array}{ll}
\text{zeros }:& x=p^{m+1}q^{n+1}~,\\[5pt]
\text{poles }:& x=p^{-m}q^{-n}~,
\end{array}\quad m,n\in \mathbb{Z}_{\geq 0}~. 
\ee
For $m,n \in\mathbb{Z}_{\geq 0}$, useful properties of the elliptic Gamma function are
\begin{align}
&\Gamma(x;p,q)\Gamma(pqx^{-1};p,q)=1~,\\
&\frac{\Gamma(p^mq^nx)}{\Gamma(x)}=(-xp^{(m-1)/2}q^{(n-1)/2})^{-mn}\Theta(x;p,q)_n\Theta(x;q,p)_m~,\label{gammapiu}\\
&\frac{\Gamma(p^mq^{-n}  x)}{\Gamma(x)}=(-xp^{(m-1)/2}q^{-(n+1)/2})^{mn}\frac{\Theta(x;q,p)_m}{\Theta(pqx^{-1};p,q)_n}~,\label{gammashiftpiu}\\
&\text{Res}_{x=t_ip^mq^n}\frac{\Gamma(t_ix^{-1})}{x}=\text{Res}_{x=1}\Gamma(x)~\frac{(-pq~q^{(n-1)/2}p^{(m-1)/2})^{mn}}{\Theta(pq;p,q)_n\Theta(pq;q,p)_m}~,
\end{align}
where we introduced the $\Theta$-factorial
\be\label{thetafac}
\Theta(x;p,q)_n=\frac{\Gamma(q^nx;p,q)}{\Gamma(x;p,q)}=\left\{\begin{array}{ll}\prod_{k=0}^{n-1}\Theta(xq^k;p)&\quad \textrm{ if } n\geq 0\\[5pt]
\prod_{k=0}^{|n|-1}\Theta(q^{-1}xq^{-k};p)^{-1}&\quad \textrm{ if } n<0\end{array}\right.~.
\ee
A useful propety which can be derived from the definition is
\be\label{thetafacneg}
\Theta(x;p,q)_{-n}=\Theta(q^{-n}x;p,q)_{n}^{-1}=\Theta(q^{-1}x;p,q^{-1})_n^{-1}~.
\ee
The elliptic Gamma function has a very non-trivial behaviour under  modular transformations  \cite{felderv, naru}
\be\label{gammamod}
\Gamma(e^{\frac{2\pi\i }{\omega_1}X};e^{2\pi\i\frac{\omega_2}{\omega_1}},e^{2\pi\i\frac{\omega_3}{\omega_1}})
\Gamma(e^{\frac{2\pi\i }{\omega_2}X};e^{2\pi\i\frac{\omega_1}{\omega_2}},e^{2\pi\i\frac{\omega_3}{\omega_2}})
\Gamma(e^{\frac{2\pi\i }{\omega_3}X};e^{2\pi\i\frac{\omega_1}{\omega_3}},e^{2\pi\i\frac{\omega_2}{\omega_3}})
=e^{-\frac{\i\pi}{3}B_{33}(X|\omega_1,\omega_2,\omega_3)}~,
\ee
Expression (\ref{gammamod}) is valid for ${\rm Im}\left(\frac{\omega_i}{\omega_{j\neq i}}\right)\neq 0$.
In particular, by assuming ${\rm Im}\left(\frac{\omega_1}{\omega_3},\frac{\omega_2}{\omega_3}\right)>0$ we get
\be\label{gammamod2}
\Gamma(e^{\frac{2\pi\i }{\omega_3}X};e^{2\pi\i\frac{\omega_1}{\omega_3}},e^{2\pi\i\frac{\omega_2}{\omega_3}})=e^{\frac{\i\pi}{3}B_{33}(X|\omega_1,\omega_2,-\omega_3)}\Gamma(e^{\frac{2\pi\i }{\omega_1}X};e^{2\pi\i\frac{\omega_2}{\omega_1}},e^{-2\pi\i\frac{\omega_3}{\omega_1}})\Gamma(e^{\frac{2\pi\i }{\omega_2}X};e^{2\pi\i\frac{\omega_1}{\omega_2}},e^{-2\pi\i\frac{\omega_3}{\omega_2}})~.
\ee

\subsection*{Basic hypergeometric identities}
The $q$-hypergeometric function 
\be
{}_2\Phi_1\left(\begin{array}{cc}a&b\\c&q\end{array};u\right)=\sum_{k\geq 0}\frac{(a;q)_k(b;q)_k}{(c;q)_k(q;q)_k}u^k~,
\ee
for $|q|<1$ satisfies the following identities
\begin{align}
\label{id1}
{}_2\Phi_1\left(\begin{array}{cc}a&b\\c&q\end{array};u\right)&=\frac{(b;q)_\infty (au;q)_\infty }{( u;q)_\infty (c;q)_\infty } {}_2\Phi_1\left(\begin{array}{cc}cb^{-1}&u\\au&q\end{array};b\right)~,\\
\label{id2}
{}_2\Phi_1\left(\begin{array}{cc}a&b\\c&q\end{array};u\right)&=\frac{(b;q)_\infty (ca^{-1};q)_\infty }{( c;q)_\infty (ba^{-1};q)_\infty } \frac{(au;q)_\infty (qa^{-1}u^{-1};q)_\infty }{( u;q)_\infty (qu^{-1};q)_\infty } {}_2\Phi_1\left(\begin{array}{cc}a&qac^{-1}\\q ab^{-1}&q\end{array}; \frac{qc}{abu}\right   )+\nn \\ 
&+
\frac{(a;q)_\infty (cb^{-1};q)_\infty }{( c;q)_\infty (ab^{-1};q)_\infty } \frac{(bu;q)_\infty (qb^{-1}u^{-1};q)_\infty }{( u;q)_\infty (qu^{-1};q)_\infty } {}_2\Phi_1\left(\begin{array}{cc}b&qbc^{-1}\\q ba^{-1}&q\end{array}; \frac{qc}{abu}   \right)~.
\end{align}
Now consider ${}_2\Phi_1{\tiny \left(\begin{array}{lr}a&b\\c&\tilde q\end{array};u\right)}$ with $|\tilde q|>1$. 
In this case we have 
\begin{multline}
\label{id4}
 {}_2\Phi_1\left(\begin{array}{cc}a&b\\c&\tilde q\end{array};u\right)%{}_2\Phi_1\left(\begin{array}{cc}a^{-1}&b^{-1}\\c^{-1}&\tilde q^{-1}\end{array};\frac{a b u}{\tilde q c}\right)=\\
 = \frac{(\tilde q  c^{-1} ;\tilde q)_\infty  (\tilde q ab^{-1} ;\tilde q)_\infty }{(\tilde q b^{-1} ;\tilde q)_\infty  (\tilde q ac^{-1} ;\tilde q)_\infty }
\frac{(a b c^{-1}u ;\tilde q)_\infty (\tilde q ca^{-1} b^{-1} u^{-1};\tilde q)_\infty }{( b c^{-1}u;\tilde q)_\infty (\tilde  qcb^{-1}u^{-1};\tilde q)_\infty }
 {}_2\Phi_1\left(\begin{array}{cc}a&\tilde q ac^{-1}\\ \tilde q ab^{-1}&\tilde q\end{array}; \frac{ \tilde q c}{a b u}   \right)+\\
+ \frac{(\tilde q  c^{-1} ;\tilde q)_\infty  (\tilde q ba^{-1} ;\tilde q)_\infty }{(\tilde q a^{-1} ;\tilde q)_\infty  (\tilde q bc^{-1} ;\tilde q)_\infty }
\frac{( a b c^{-1}u ;\tilde q)_\infty (\tilde qca^{-1} b^{-1}u^{-1});\tilde q)_\infty }{(a c^{-1}u;\tilde q)_\infty (\tilde qca^{-1}u^{-1});\tilde q)_\infty }
 {}_2\Phi_1\left(\begin{array}{cc}b&\tilde q bc^{-1}\\ \tilde q ba^{-1}&\tilde q\end{array};\frac{\tilde q   c}{abu} \right)~.
 \end{multline}
Also, for $|q|>1$ we have the following identity
\be
\label{id3}
{}_2\Phi_1\left(\begin{array}{cc}a&b\\c&q\end{array};u\right)=\frac{(abc^{-1} u  ;q)_\infty (q c^{-1};q)_\infty }{( q b^{-1};q)_\infty (bc^{-1} u;q)_\infty } 
{}_2\Phi_1\left(\begin{array}{cc}cb^{-1}&qc a^{-1} b^{-1}u^{-1}\\ qcb^{-1}u^{-1}&q\end{array};\frac{q a}{c}\right)\,.
\ee

\subsection{Elliptic series}
Let us consider the elliptic hypergeometric series \cite{spirtheta}
\be\label{ellseries1}
{}_{N}E_{N-1}\left(\begin{array}{c}\vec{x}\\ \vec{y}\end{array};q_\tau,q_\sigma;u\right)=\sum_{n\geq 0}\prod_{i,j=1}^N\frac{\Theta(x_i;q_\sigma,q_\tau)_n}{\Theta(y_j;q_\sigma,q_\tau)_n}u^n~,\quad y_N=q_\tau~.
\ee
This series is usually considered to be balanced, namely
\be\label{ellbal1}
\prod_{i,j}x_i y_j^{-1}=1~.
\ee
We now introduce the parametrisation 
\be
q_\tau=e^{2\pi\i\tau}~,\quad q_\sigma=e^{2\pi\i\sigma}~,\quad x_i=e^{2\pi\i X_i}~,\quad y_j=e^{2\pi\i Y_j}~,
\ee 
and  study  the modular properties of the series under
\be
\sigma\to -\frac{1}{\sigma}~,\quad \tau\to-\frac{\tau}{\sigma}~,\quad X_i\to -\frac{X_i}{\sigma}~,\quad Y_j\to-\frac{Y_j}{\sigma}~.
\ee
Using the modular transformation property 
\be
\Theta(e^{-\frac{2\pi\i}{\sigma}X};e^{-\frac{2\pi\i}{\sigma}})=e^{\i\pi B_{22}(X|1,\sigma)}\Theta(e^{2\pi\i X};e^{2\pi\i\sigma})~,
\ee
we get
\begin{multline}
\prod_{i,j=1}^N\frac{\Theta(e^{-2\pi\i \frac{X_i}{\sigma}};e^{-\frac{2\pi\i}{\sigma}},e^{-\frac{2\pi\i\tau}{\sigma}})_n}{\Theta(e^{-2\pi\i \frac{Y_j}{\sigma}};e^{-\frac{2\pi\i}{\sigma}},e^{-\frac{2\pi\i\tau}{\sigma}})_n}=\prod_{i,j=1}^N\frac{\Theta(e^{2\pi\i X_i};e^{2\pi\i\sigma},e^{2\pi\i\tau})_n}{\Theta(e^{2\pi\i Y_j};e^{2\pi\i\sigma},e^{2\pi\i\tau})_n}\times\\
\times \prod_{i,j=1}^N e^{\frac{\i\pi n}{\sigma}\left((X_i^2-Y^2_j)+(\tau(n-1)-\sigma-1)(X_i-Y_i)\right)}~.
\end{multline}
Once the balancing condition (\ref{ellbal1}) $\sum_{i,j}(X_i-Y_j)=0$ is imposed, the series can be made modular invariant either by imposing 
\be\label{ellmod1}
\sum_{i,j}(X_i^2-Y^2_j)=0~,
\ee 
or by a suitable transformation of the expansion parameter $u$ .

%
%It is interesting to study also the periodicity with respect to the shifts $x_i\to q_\sigma x_i$, $i=1,\ldots, N$, $y_j\to q_\sigma y_j$, $j=1,\dots, N-1$. The series is not invariant. However, in our applications we actually have $x_i\to x_{c}\bar x_{b}^{-1}$, $y_j\to q_\tau x_{c}x_{a}^{-1}$, and we have the obvious periodicity with respect to $x_a\to q_\sigma x_a$, $\bar x_b\to q_\sigma \bar x_{b}$.

Next, let us consider the very-well-poised elliptic hypergeometric series \cite{spirtheta}
\be\label{ellseries2}
{}_{N+1}E_N(t_0;\vec{t};q_\tau,q_\sigma;u)=\sum_{n=0}^{\infty}\frac{\Theta(t_0^2q_\tau^{2n};q_\sigma)}{\Theta(t_0^2;q_\sigma)}\prod_{i=0}^{N-4}\frac{\Theta(t_0t_i;q_\sigma,q_\tau)_n}{\Theta(q_\tau t_0t_i^{-1};q_\sigma,q_\tau)_n}(q_\tau u)^n~,
\ee
subjected to the balancing condition
\be\label{ellbal2}
\prod_{i=0}^{N-4}t_i=q^{\frac{N-7}{2}} ~.
\ee
In this case, proceeding as above, it is easy to see that the series is automatically modular invariant.

\section{Computations}

\subsection{Fundamental Abelian relation}\label{far}

The free chiral theory with $-1/2$ Chern-Simons units has a mirror given by the $U(1)$ theory with 1 chiral
and $1/2$ Chern-Simons units (also for the holonomies).

At the level of  lens space partition functions the duality reads (up to a trivial proportionality constant)
\be\label{Zdeltafund}
\sum_{\ell=0}^{r-1}\int_\mathbb{R}\frac{\rd Z}{2\pi \i}\,e^{-\frac{\i\pi}{r}(Z^2+2Z(\xi-\i Q/2))}e^{-(r-1)\frac{\i\pi}{r}(\ell^2+2\ell\theta)}\,Z_\Delta(Z,\ell)=Z_\Delta(\xi,\theta)~,
\ee
where we have also turned on the FI and $\theta$ terms. To prove this identity we evaluate the l.h.s. integral by closing the contour in the lower half plane (assuming $\xi>0$) and taking the sum of the residues at the poles of $Z_\Delta$. By using (\ref{rpoles})  we can see that there are two sets of  poles  located at
\be
\begin{array}{l}Z=Z_{(1)}=-\i\omega_1\ell-\i jQ-\i k r\omega_1~,\\[5pt]
Z=Z_{(2)}=-\i\omega_2(r-\ell)-\i jQ-\i kr\omega_2~,\end{array}\quad\quad j,k\in \mathbb{Z}_{\geq 0}~.
\ee
The integral is then given by 
\be
\label{go1}\sum_{\ell=0}^{r-1}\oint\frac{\rd Z}{2\pi \i}\,e^{-\frac{\i\pi}{r}(Z^2+2\xi Z-\i QZ)}e^{-(r-1)\frac{\i\pi}{r}(\ell^2+2\ell\theta)}\,Z_\Delta(Z,\ell)= I_1+I_2\,,
\ee
with
\be
I_1=\Big\|(q;q)_\infty\Big\|^2_r \sum_{\ell=0}^{r-1} \sum_{j,k\geq 0} \frac{q^{j(j-1)/2}}{(q;q)_j}\frac{\tilde q^{(\ell+kr+j)(\ell+kr+j-1)/2}}{(\tilde q;\tilde q)_{\ell+kr+j}}\left(-q e^{-\frac{2\pi\xi}{r\omega_1}}e^{-\frac{2\pi\i }{r}\theta}\right)^j
\left(  - \tilde q e^{-\frac{2\pi\xi}{r\omega_2}}e^{\frac{2\pi\i }{r}\theta}\right)^{\ell+kr+j}~.
\ee
The sum of residues at the second set of poles is simply obtained by $\omega_1\leftrightarrow \omega_2$ and $\ell\leftrightarrow r-\ell$, $\theta\leftrightarrow r-\theta$.
Combining the two sums we see that the original integral (\ref{go1}) has the schematic form 
\be
I_1+I_2=\sum_{\ell=0}^{r-1} \sum_{j,k\geq 0}  f_{j,j+\ell+kr}+\sum_{\ell=0}^{r-1} \sum_{j,k\geq 0}  f_{r-\ell+kr+j,j}.
\ee
Since $\ell+kr$ runs from $0$ to $\infty$ while $r-\ell+kr$ runs from $1$ to $\infty$, we can replace $r-\ell\rightarrow \ell+1$, set $j''=j+\ell+kr$, and write
\begin{multline}\label{sums}
I_1+I_2=\sum_{j,j''\geq j}f_{j,j''}+\sum_{j,j''\geq j}f_{j''+1,j}=\sum_{j,j''\geq j}f_{j,j''}+\sum_{j,j''\geq j+1}f_{j'',j}=\\
=\sum_{j,j''\geq j}f_{j,j''}+\sum_{j'',j\geq j''+1}f_{j,j''}=\sum_{j,j''\geq 0}f_{j,j''}~,
\end{multline}
so that we find as expected
\be
I_1+I_2\propto\Big\|(q  e^{-\frac{2\pi\xi}{r\omega_1}}e^{-\frac{2\pi\i }{r}\theta};q)_\infty\Big\|^2_r=Z_\Delta(\xi,\theta)\,.
\ee

\subsection{SQED lens space partition function}\label{qedcomp}

Here we compute the residues  at the poles given in eq. (\ref{SQEDpoles}) of the partition function
 \be
 Z_{\rm SQED}=e^{-\i\pi \mathcal{P}}\sum_{i=1,2}\sum_{\ell=0}^{r-1}\sum_{\{Z_c^{(i)}\}}\textrm{Res}_{Z=Z_c^{(i)}}e^{\frac{2\pi \i}{r}Z\xi_{\rm eff}}e^{\frac{2\pi \i}{r}\ell\theta_{\rm eff}}\;\prod_{a,b=1}^{N_f}~\Big\|\frac{(e^{\frac{2\pi }{r\omega_1}(\i Q+Z-X_a-\i\omega_1[\ell+H_a])};q)_\infty}{(e^{\frac{2\pi }{r\omega_1}(Z-\bar X_b-\i\omega_1[\ell+\bar H_b])};q)_\infty}\Big\|^2_r~,
\ee
where 
\be
q=e^{2\pi\i\frac{Q}{r\omega_1}}=q_1~,\quad \tilde q=e^{2\pi\i\frac{Q}{r\omega_2}}=q_2~,
\ee
and
\be
\xi_{\rm eff}=\xi -\frac{1}{2}\sum_{a,b}(\bar X_b- X_a)   -\i \frac{N_f}{2}Q~,\quad
 \theta_{\rm eff}=\theta+\frac{r-1}{2}\sum_{a,b}(\bar H_b-H_a)~.
\ee
The latter must be integer (we can add contact terms to ensure that it is). The exponential prefactor is
\be\label{3dprefactor}
e^{-\i\pi\mathcal{P}}=e^{\frac{\i\pi}{2r}\sum_{a,b}(\bar X_b^2-X_a^2)}e^{-\frac{Q\pi}{2r}\sum_{a,b}(X_a+\bar X_b)}e^{\frac{\i\pi}{2r}(r-1)\sum_{a,b}(\bar H_b^2-H^2_a)}~,
\ee
representing background CS terms. We rewrite the classical part evaluated at the first set of poles  $Z_{(1)}$ as follows\footnote{We use $\omega_1 \omega_2=1$,  $[\ell+H_c]-[H_c]=[\ell]\mod r$, and 
$\theta_{\rm eff} \ell=[\theta_{\rm eff}][\ell]\mod r $, this is why we need $\theta_{\rm eff}$ to be integer.}
\begin{multline}
e^{\frac{2\pi \i}{r}Z_{(1)}\xi_{\rm eff}}e^{\frac{2\pi \i}{r}\ell\theta_{\rm eff}}= e^{\frac{2\pi \i}{r}\xi_{\rm eff} X_c } e^{-\frac{2\pi \i}{r}[\theta_{\rm eff}][H_c]  }e^{-\frac{2\pi }{r}\xi_{\rm eff} ( \frac{[\ell+H_c]+k r+j}{\omega_2 } +\frac{j}{\omega_1} ) } 
e^{\frac{2\pi \i}{r}[\theta_{\rm eff}]([\ell+H_c]+kr+j-j)  }=\\
=e^{\frac{2\pi \i}{r}\xi_{\rm eff} X_c }  e^{-\frac{2\pi \i}{r}[\theta_{\rm eff}][H_c]  } u_1^{j}\;u_2^{[\ell+H_c]+kr+j}~,
\end{multline}
where
\be
u=e^{-\frac{2\pi}{r\omega_1}\xi_{\rm eff}}e^{-\frac{2\pi\i}{r}\theta_{\rm eff}}=u_1~,\quad 
\tilde u=e^{-\frac{2\pi}{r\omega_2}\xi_{\rm eff}}e^{\frac{2\pi\i}{r}\theta_{\rm eff}}=u_2~,
\ee
and similarly for the second set of poles $Z_{(2)}$.
Summing over (\ref{SQEDpoles}) yields \footnote{
It is understood that we are taking the residue of the $a=c$ term.}
 \begin{multline}
 Z_{\rm SQED}=e^{-\i\pi\mathcal{P}}\sum_{c=1}^{N_f}e^{\frac{2\pi \i}{r}(X_c\xi_{\rm eff}-H_c\theta_{\rm eff})}
  \sum_{\ell=0}^{r-1}
 \sum_{j,k\geq 0}\Big\{u_1^{j} u_2^{[\ell+H_c]+kr+j} \times\\ 
 \times
 \prod_{a,b=1}^{N_f}
 \frac{(e^{\frac{2\pi }{r\omega_1}(\i Q+X_{ca}+\i \omega_1[H_{ca}]}q_1^{j};q_1)_\infty(e^{\frac{2\pi }{r\omega_2}(\i Q+X_{ca}-\i\omega_2[H_{ca}]}q_2^{(j+[\ell+H_c]+kr)};q_2)_\infty}{(e^{\frac{2\pi }{r\omega_1}X_{c\bar b}+\i\omega_1[H_{c\bar b}]}q_1^{j};q_1)_\infty(e^{\frac{2\pi }{r\omega_2}(X_{c\bar b}-\i\omega_2[H_{c\bar b}]} q_2^{(j+[\ell+H_c]+kr)};q_2)_\infty}+\\
+u_1^{r-[\ell+H_c]+kr+j}u_2^{j}\times\\
\times \prod_{a,b=1}^{N_f}\frac{(e^{\frac{2\pi }{r\omega_1}(\i Q+X_{ca}+\i\omega_1[H_{ca}])}
 q_1^{r-[\ell+H_c]+kr +j};q_1)_\infty(e^{\frac{2\pi }{r\omega_2}(\i Q+X_{ca}-\i\omega_2[H_{ca}])} q_2^{j};q_2)_\infty}{(e^{\frac{2\pi }{r\omega_1}(X_{c\bar b}+\i\omega_1[H_{c\bar b}])}  q_1^{r-[\ell+H_c]+kr +j}
 ;q_1)_\infty(e^{\frac{2\pi }{r\omega_2}( X_{c\bar b}-\i\omega_2[H_{c\bar b}])} q_2^{j};q_2)_\infty}
\Big\}~,
\end{multline}
where we defined 
\be
X_{ca}=X_c-X_a\,, \quad X_{c\bar b}=X_c-\bar X_b\,, \quad H_{ca}=H_c-H_a\,, \quad H_{c\bar b}=H_c-\bar H_b\,.
\ee

Using $ (q^nx;q)_\infty=\frac{(x;q)_\infty}{(x;q)_n}$, we get
 \begin{multline}
 Z_{\rm SQED}=e^{-\i\pi\mathcal{P}}\sum_{c=1}^{N_f}e^{\frac{2\pi \i}{r}(X_c\xi_{\rm eff}-H_c\theta_{\rm eff})}\times\\
\times \prod_{a,b=1}^{N_f}
 \frac{(e^{\frac{2\pi }{r\omega_1}(\i Q+X_{ca}+\i\omega_1[H_{ca}]};q_1)_\infty
 (e^{\frac{2\pi }{r\omega_2}(\i Q+X_{ca}-\i\omega_2[H_{ca}]}
 ;q_2)_\infty}{(e^{\frac{2\pi }{r\omega_1}(X_{c\bar b}+\i\omega_1[H_{c\bar b}]}q_1)_\infty(e^{\frac{2\pi }{r\omega_2}(X_{c\bar b}-\i\omega_2[ H_{c\bar b}]};q_2)_\infty}\times\\ 
 \times\Big\{  \sum_{\ell=0}^{r-1}
 \sum_{j,k\geq 0}
 u_1^{j}\;u_2^{[\ell+H_c]+kr+j}\times\\
 \times \prod_{a,b=1}^{N_f}
 \frac{(e^{\frac{2\pi }{r\omega_1}(X_{c\bar b}+\i\omega_1[H_{c\bar b}]}q_1)_{j}(e^{\frac{2\pi }{r\omega_2}(X_{c\bar b}-\i\omega_2[H_{c\bar b}]};q_2)_{[\ell+H_c]+kr +j } }{(e^{\frac{2\pi }{r\omega_1}(\i Q+X_{ca}+\i\omega_1[H_{ca}]};q_1)_{j}
 (e^{\frac{2\pi }{r\omega_2}(\i Q+X_{ca}-\i\omega_2[H_{ca}]}
 ;q_2)_{[\ell+H_c]+kr+j } }
  +\\
+  \sum_{\ell=0}^{r-1}
 \sum_{j,k\geq 0}u_1^{r-[\ell+H_c]+kr+j}u_2^{j}\times\\
\times \prod_{a,b=1}^{N_f}\frac{(e^{\frac{2\pi }{r\omega_1}(X_{c\bar b}+\i\omega_1[H_{c\bar b}])}  ;q_1)_{r-[\ell+H_c] +kr+j}(e^{\frac{2\pi }{r\omega_2}(X_{c\bar b}-\i\omega_2[H_{c\bar b}])} ;q_2)_{j}}
{(e^{\frac{2\pi }{r\omega_1}(\i Q+X_{ca}+\i\omega_1[H_{ca}])}
;q_1)_{r-[\ell+H_c] +kr+j}(e^{\frac{2\pi }{r\omega_2}(\i Q+X_{ca}-\i\omega_2[H_{ca}])};q_2)_{j}}\Big\}~.
\end{multline}
We see that the first term in brakets is a sequence $f_{j,j+[\ell+H_c]+kr}$, whereas the second one is $f_{j+r-[\ell+H_c]+kr,j}$.
Since $[\ell]+kr$ runs from $0$ to $+\infty$ while $r-[\ell+H_c]+kr$ runs from $1$ to $+\infty$, we can replace $r-[\ell+H_c]\rightarrow [\ell+H_c]+1$, set $j''=[\ell+H_c]+kr$, and write
\begin{multline}
\{\ldots\}=\sum_{j,j''\geq j}f_{j,j''}+\sum_{j,j''\geq j}f_{j''+1,j}=\sum_{j,j''\geq j}f_{j,j''}+\sum_{j,j''\geq j+1}f_{j'',j}=\\
=\sum_{j,j''\geq j}f_{j,j''}+\sum_{j'',j\geq j''+1}f_{j,j''}=\sum_{j,j''\geq 0}f_{j,j''}~.
\end{multline}
Therefore we find $Z_{\rm SQED}$ can be expressed in terms of the $r$-square of the $q$-hypergeometric series 
\be
{}_N\Phi_{N-1}\left(\begin{array}{l}\vec {x}\\ \vec {y}\end{array};u\right)=\sum_{k\geq 0}\prod_{i,j=1}^N\frac{(x_i;q)_k}{(y_j;q)_k}u^k~,\quad y_N=q~,
\ee
namely
\begin{multline}
 Z_{\rm SQED}=e^{-\i\pi\mathcal{P}}\sum_{c=1}^{N_f}e^{\frac{2\pi \i}{r}(X_c\xi_{\rm eff}-H_c\theta_{\rm eff})}\times\\
 \times\Big\|\prod_{a,b=1}^{N_f}\frac{(qe^{\frac{2\pi }{r\omega_1}X_{ca}}e^{\frac{2\pi\i}{r}H_{ca}};q)_\infty}{(e^{\frac{2\pi }{r\omega_1}X_{c\bar b}}e^{\frac{2\pi\i}{r} H_{c\bar b}};q)_\infty}{}_{N_f}\Phi_{N_f-1}\left(\begin{array}{c}e^{\frac{2\pi }{r\omega_1}X_{c\bar b}}e^{\frac{2\pi\i}{r} H_{c\bar b}}\\
qe^{\frac{2\pi }{r\omega_1}X_{ca}}e^{\frac{2\pi\i}{r}H_{ca}}\end{array};u\right)\Big\|^2_{\substack{\omega_1\leftrightarrow \omega_2\\ H\leftrightarrow r-H}}~.
\end{multline}

\subsection{Twisted superpotential}\label{twistreg}
In this appendix we briefly review how the double sum defining the twisted superpotential (\ref{dpw}) can be regularized in two steps, first regularizing the sum over $m$, and then over $n$.\footnote{We verified the 1-step regularization by means of double Gamma functions yields the same result.} In order to regularise the sum over $m$, let us consider the exponential derivative
\be
e^{\frac{\rd \widetilde{\mathcal{W}}}{\rd a}}=\exp\left(\frac{\rd }{\rd a}\sum_{m\in \mathbb{Z}}\left(a+ \frac{\i}{R_1}m\right) \left(\ln(a+\frac{\i}{R_1}m)-1\right)\right)=\prod_{m\in \mathbb{Z}}\left(a+\frac{\i}{R_1}m\right)~.
\ee
By using the definition
\be 
\prod_{m\in \mathbb{Z}}\left(a+\frac{\i}{R_1}m\right)=2\sinh\left(\pi R_1 a\right)\,,
\ee
by  integrating we find
\be\label{Wreg}
\sum_{m\in \mathbb{Z}}\left(a+ \frac{\i}{R_1}m \right) \left(\ln(a+\frac{\i}{R_1}m)-1\right)=\frac{1}{2\pi R_1}{\rm Li}_2(e^{-2\pi R_1a})+\frac{\pi R_1}{2}a^2~,
\ee
up to linear terms. Next, we shift $a\to a+\frac{\i}{R_1}n\sigma $ and compute
\begin{multline}
\frac{1}{2\pi R_1}\sum_{n\in \mathbb{Z}}{\rm Li}_2(e^{2\pi\i(n\sigma +\i R_1a)})+\sum_{n\in\mathbb{Z}}\frac{\pi R_1}{2}\left(a+\frac{\i}{R_1}n\sigma \right)^2=
%&=\frac{1}{2\pi R}\sum_{n\geq 1}\left({\rm Li}_2(e^{2\pi\i((n-1)\sigma +\i R a)})+{\rm Li}_2(e^{-2\pi\i(n\sigma-\i R a)})\right)+\sum_{n\in\mathbb{Z}}\frac{\pi R}{2}\left(a+\frac{\i}{R}n\sigma \right)^2\nn\\
\frac{1}{2\pi R_1}\sum_{k\neq 0}\frac{e^{-2\pi R_1 a k}}{k^2(1-q_\sigma^k)}+\\
+\frac{1}{2\pi R_1}\sum_{n\geq 1}\left(\frac{\pi^2}{3}+2\pi^2(n\sigma -\i R_1 a)+2\pi^2(n\sigma -\i R_1 a)^2\right)
+\sum_{n\in\mathbb{Z}}\frac{\pi R_1}{2}\left(a+\frac{\i}{R_1}n\sigma \right)^2~,
\end{multline}
where we used
\be
{\rm Li}_2(e^{-X})=-{\rm Li}_2(e^X)+\frac{\pi^2}{3}-\i\pi X-\frac{X^2}{2}~.
\ee
We regularize the other infinite sums by means of Hurwitz $\zeta$-function\footnote{$\zeta(s,X)=\sum_{n\geq 0}(X+n)^{-s}$, $\zeta(-1,X)=-\frac{X^2}{2}+\frac{X}{2}-\frac{1}{12}~,\quad \zeta(-2,X)=-\frac{X^3}{3}+\frac{X^2}{2}-\frac{X}{6}$.}
and we get
\be
\frac{1}{2\pi R_1}\sum_{n\geq 1}\left(\frac{\pi^2}{3}+2\pi^2(n\sigma -\i R_1 a)+2\pi^2(n\sigma -\i R_1a)^2\right)+\sum_{n\in\mathbb{Z}}\frac{\pi R_1}{2}\left(a+\frac{\i}{R_1}n\sigma \right)^2
=\mathcal{P}_3(\i R_1 a)~.
\ee

\subsection{SQED lens index}\label{SQEDindex}
In this appendix we provide the explicit derivation of (\ref{ZSQED4dbody}), which amounts to the evaluation of the residues of the integrand (\ref{modtra}) on the poles (\ref{poles4qsqed}) given in the main text. 
%%\SP{We perform the modular transformation  (\ref{4dchiral}) using   eq. (\ref{4dchiral1})  for the anti-chirals and  (\ref{4dchiral2}) for the the chirals CHECK}
% and get\begin{multline}
%\prod_{a,b}{\mathcal{I}}^{(R)}_\chi(z^{-1} \zeta_a,\ell+H_a)\hat{\mathcal{I}}^{(R)}_\chi(z\bar \zeta_b^{-1},-\ell-\bar H_b)=\\
%=e^{-\i\pi\mathcal{P}_{\rm gl}}\prod_{a,b}\frac{e^{\frac{\i\pi}{2r}(\ell+\bar H_b)^2(r-1)}e^{-\frac{\i\pi}{2}\Phi_2(Z-\bar X_b)}}{e^{\frac{\i\pi}{2r}(\ell+ H_a)^2(r-1)}e^{-\frac{\i\pi}{2}\Phi_2(Q+Z-X_a)}}\frac{\mathcal{G}(Z-\bar X_b,-\ell-\bar H_b)}{\mathcal{G}(Q+Z-X_a,-\ell-H_a)}~,
%\end{multline}
%where we used the vanishing of the gauge anomaly polynomials discussed in subsection \ref{sec_anomaly}. 
First of all, expanding the polynomials $\Phi_2$ we get the exponential factor
\begin{multline}
\prod_{a,b}\frac{e^{\frac{\i\pi}{2r}(\ell+\bar H_b)^2(r-1)}e^{-\frac{\i\pi}{2}\Phi_2(Z-\bar X_b)}}{e^{\frac{\i\pi}{2r}(\ell+ H_a)^2(r-1)}e^{-\frac{\i\pi}{2}\Phi_2(Q+Z-X_a)}}=\\
=e^{\frac{\i\pi}{2r}(r-1)\sum_{a,b}(\bar H_b^2-H_a^2)}e^{\frac{\i\pi}{2r\omega_1\omega_2}\sum_{a,b}(M_a^2-\bar M_b^2)}e^{\frac{\i\pi}{2r\omega_1\omega_2}\sum_{a,b}(M_a+\bar M_b)Q(R-1)}\times\\
\times e^{-\frac{2\pi\i}{r}\ell\frac{(r-1)}{2}\sum_{a,b}(H_a-\bar H_b)}e^{-2\pi\i Z\frac{1}{2r\omega_1\omega_2}(Q(R-1)N_f+\sum_{a,b}(M_a-\bar M_b))}~.\label{B22U1}
\end{multline}
The first line represent the global prefactor $e^{-\i\pi\mathcal{P}^{\rm 3d}_{\rm gl}}$.
In the second line the dynamical term $e^{-\frac{2\pi i }{r}\ell\frac{(r-1)}{2}\sum_{a,b}(H_a-\bar H_b) }$ can be absorbed into a renomalisation of $\theta$
\be
\theta_{\rm eff}=\theta-\frac{(r-1)}{2}\sum_{a,b}(H_a-\bar H_b)~,
\ee
provided $\theta_{\rm eff}$ is integer, while $e^{-2\pi\i Z\frac{1}{2r\omega_1\omega_2}(Q(R-1)N_f+\sum_{a,b}(M_a-\bar M_b))}$ goes into a renormalisation of $\xi^{\rm 4d}$
\be
\frac{\xi^{\rm 4d}_{\rm eff}}{r\omega_3}=\frac{\xi^{\rm 4d}}{r\omega_3}+\frac{1}{2r\omega_1\omega_2}(N_fQ(R-1)+\sum_{a,b}(M_a-\bar M_b))~.
\ee
%\SP{FIX THIS
%Actually, we may forget about these renormalisations because we are also implicitly assuming that the $U(1)$ flavour charge is opposite between fundamentals and antifundamentals (see (\ref{Z2anomaly})), namely $\sum_{a,b}(M_a-\bar M_b)=0$, $\sum_{a,b}(H_a-\bar H_b)=0\mod r$. }
Then, the residues series reads as\footnote{
It is understood that we are taking the residue of the $a=c$ term.}
\be
I_{\rm SQED}=e^{-\i\pi(\mathcal{P}_{\rm gl}+\mathcal{P}_{\rm gl}^{\rm 3d})}\sum_{c}\sum_{s=1,2}\sum_{\ell=0}^{r-1}\sum_{j,k\geq 0} e^{-\frac{2\pi\i}{r\omega_3}\xi_{\rm eff}Z_{(s)}}e^{\frac{2\pi\i}{r}\ell\theta_{\rm eff}} ~
\prod_{a,b}\frac{\mathcal{G}(Z_{(s)}-\bar X_b,-\ell-\bar H_b)}{\mathcal{G}(Q+Z_{(s)}-X_a,-\ell-H_a)}~.
\ee
Using the definition (\ref{modga}) of $\mathcal{G}$ and the properties in appendix \ref{appSpecial}, on the first family of poles the ratio of $\mathcal{G}$ functions yields 
\begin{multline}
\prod_{a,b}\frac{\mathcal{G}(X_{c\bar b},H_{c \bar b})}{\mathcal{G}(Q+X_{ca},H_{ca})}\times \frac{\Theta(e^{\frac{2\pi\i}{r\omega_1}(X_{c\bar b}+\omega_1H_{c\bar b})};e^{-2\pi\i\frac{\omega_3}{r\omega_1}},e^{2\pi\i\frac{Q}{r\omega_1}})_j}{\Theta(e^{\frac{2\pi\i}{r\omega_1}(Q+X_{ca}+\omega_1H_{ca})};e^{-2\pi\i\frac{\omega_3}{\omega_1}},e^{2\pi\i\frac{Q}{r\omega_1}})_{j}}\times\\
\times\frac{\Theta(e^{\frac{2\pi\i}{r\omega_2}(X_{c\bar b}-\omega_2H_{c\bar b})};e^{-2\pi\i\frac{\omega_3}{r\omega_2}},e^{2\pi\i\frac{Q}{r\omega_2}})_{j+kr+[\ell+H_c]}}
{\Theta(e^{\frac{2\pi\i}{r\omega_2}(Q+X_{ca}-\omega_2H_{ca})};e^{-2\pi\i\frac{\omega_3}{r\omega_2}},e^{2\pi\i\frac{Q}{r\omega_2}})_{j+kr+[\ell+H_c]}}~,
\end{multline}
while on the second family of poles we simply have $j\rightarrow j+kr+r-[\ell+H_c]$ and $j+kr+[\ell+H_c]\rightarrow j$ in the subindex of the $\Theta$-factorials. The FI terms on the first family read as
\be\label{xieff}
e^{-\frac{2\pi\i}{r\omega_3}\xi^{\rm 4d}_{\rm eff}Z_{(1)}}e^{\frac{2\pi\i}{r}\ell\theta_{\rm eff}}=e^{-\frac{2\pi\i}{r\omega_3}\xi^{\rm 4d}_{\rm eff} X_c}e^{-\frac{2\pi\i}{r}\theta_{\rm eff} H_c}\left(e^{-2\pi\i\frac{\omega_2}{r\omega_3}\xi^{\rm 4d}_{\rm eff}}e^{-\frac{2\pi\i}{r}\theta_{\rm eff}}\right)^j\left(e^{-2\pi\i\frac{\omega_1}{r\omega_3}\xi^{\rm 4d}_{\rm eff}}e^{\frac{2\pi\i}{r}\theta_{\rm eff}}\right)^{j+kr+[\ell+H_c]}~,
\ee
and similarly on the second family.
%\be
%e^{\frac{2\pi\i}{r\omega_3}\xi_{\rm eff}Z_{(2)}}e^{\frac{2\pi\i}{r}\ell\theta_{\rm eff}}=e^{\frac{2\pi\i}{r\omega_3}\xi_{\rm eff} X_c}e^{-\frac{2\pi\i}{r}\theta_{\rm eff} H_c}\left(e^{2\pi\i\frac{\omega_2}{r\omega_3}\xi^{\rm 4d}_{\rm eff}}e^{-\frac{2\pi\i}{r}\theta_{\rm eff}}\right)^{r-[\ell+H_c]+j+kr}\left(e^{2\pi\i\frac{\omega_1}{r\omega_3}\xi^{\rm 4d}_{\rm eff}}e^{\frac{2\pi\i}{r}\theta_{\rm eff}}\right)^{j}~.
%\ee
We can now resolve the sum by using (\ref{sums}) as in 3d, and we find $I_{\rm SQED}$ can be written in terms of the $r$-square of the elliptic hypergeometric series ${}_{N}E_{N-1}$ defined in (\ref{ellseries1})
\begin{multline}\label{ZSQED4d}
I_{\rm SQED}=e^{-\i\pi(\mathcal{P}_{\rm gl}+\mathcal{P}_{\rm gl}^{\rm 3d})}\sum_c~e^{-\frac{2\pi\i}{\omega_3}\frac{\xi^{\rm 4d}_{\rm eff}}{r} X_c}e^{-\frac{2\pi\i}{r}\theta_{\rm eff} H_c}
\prod_{a,b}\frac{\mathcal{G}(X_{c\bar b},H_c-\bar H_b)}{\mathcal{G}(Q+X_{ca},H_c-H_a)}
\times\\
\times~\Big\|{}_{N_f}E_{N_f-1}\left(\begin{array}{c}
e^{\frac{2\pi\i}{r\omega_1}X_{c\bar b}}e^{\frac{2\pi\i}{r}H_{c\bar b}}\\
e^{\frac{2\pi\i}{r\omega_1}(Q+X_{ca})}e^{\frac{2\pi\i}{r}H_{ca}}
\end{array};e^{2\pi\i\frac{Q}{r\omega_1}},e^{-2\pi\i\frac{\omega_3}{r\omega_1}};e^{-\frac{2\pi\i}{r\omega_1}\frac{\omega_1\omega_2}{\omega_3}\xi^{\rm 4d}_{\rm eff}}e^{-\frac{2\pi\i}{r}\theta_{\rm eff}}\right)\Big\|^2_{\substack{\omega_1\leftrightarrow\omega_2\\
 H\leftrightarrow r-H}}~.
\end{multline}

\subsection{SQCD lens index}\label{SQCDindex}
Here we present the derivation of (\ref{ZSQCD4dbody}). For the chiral multiplets  the discussion parallels the SQED case, so we focus on the vector multiplet. From (\ref{4dvector}) we find
\ben
\label{ragogo}
\hat{\mathcal{I}}_V(z^{\mp 2},\pm 2\ell)=e^{-\i\pi\sum_{\alpha}\mathcal{P}_\alpha}\times e^{-\frac{2\pi\i Z Q}{r\omega_1\omega_2}}\times \frac{\mathcal{G}(Q+2Z,-2\ell)}{\mathcal{G}(2Z,-2\ell)}~,
\een
where we used the reflection property (\ref{refl}). The first factor can be neglected as it contributes to the vanishing of the total gauge anomaly. The factor $e^{-\frac{2\pi\i ZQ}{r\omega_1\omega_2}}$ combines with an analogue contribution from the chiral multipltes 
(\ref{B22U1})
\be
e^{-\frac{\i\pi Z}{r\omega_1\omega_2}(Q(R-1)2N_f+\sum_{a',b'}(M_{a'}-\bar M_{b'}))}~,
%=e^{\frac{2\pi\i ZQ}{r\omega_1\omega_2}\frac{T_2(ad)}{T_2(f)+T_2(\bar f)}}~,
\ee
to given a total contribution
\be
e^{\frac{2\pi\i Z}{r\omega_1}}e^{\frac{2\pi\i Z}{r\omega_2}}~,
\ee
when anomaly cancellation conditions $R=\frac{N_f-2}{N_f}$, and  $\sum_{a'}M_{a'}=-\sum_{b'}\bar M_{b'}=0.$
When evaluated on  the first family of poles, these exponential factors give the expansion parameters
\be
e^{\frac{2\pi\i X_{c'}}{r\omega_1}}e^{\frac{2\pi\i X_{c'}}{r\omega_2}}\left(e^{2\pi \i\frac{Q}{r\omega_1}}\right)^j\left(e^{2\pi\i \frac{Q}{r\omega_2}}\right)^{j+kr+[\ell+H_{c'}]}~,
\ee
while the ratio of the $\mathcal{G}$ functions in (\ref{ragogo}) yields
%\begin{multline}
%\frac{\mathcal{G}(Q+2Z_{c'},2H_{c'})}{\mathcal{G}(2Z_{c'},2H_{c'})}
%\times
%\frac{\Theta(e^{\frac{2\pi\i}{r\omega_1}(Q+2Z_{c'}+2\omega_1H_{c'})};e^{-2\pi\i\frac{\omega_3}{r\omega_1}},e^{2\pi\i\frac{Q}{r\omega_1}})_{2j}}{\Theta(e^{\frac{2\pi\i}{r\omega_1}(2Z_{c'}+2\omega_1H_{c'})};e^{-2\pi\i\frac{\omega_3}{r\omega_1}},e^{2\pi\i\frac{Q}{r\omega_1}})_{2j}}\times\\
%\times\frac{\Theta(e^{\frac{2\pi\i}{r\omega_2}(Q+2Z_{c'}-2\omega_2H_{c'})};e^{-2\pi\i\frac{\omega_3}{r\omega_2}},e^{2\pi\i\frac{Q}{r\omega_2}})_{2(j+kr+[\ell+H_{c'}])}}{\Theta(e^{\frac{2\pi\i}{r\omega_2}(2Z_{c'}-2\omega_2H_{c'})};e^{-2\pi\i\frac{\omega_3}{r\omega_2}},e^{2\pi\i\frac{Q}{r\omega_2}})_{2(j+kr+[\ell+H_{c'}])}}~,
%\end{multline}
%where we used the definition (\ref{modga}) of $\mathcal{G}$ and the properties in appendix \ref{appSpecial}.
%Using
%\be
%\frac{\Theta(q_\tau x;q_\sigma,q_\tau)_{n}}{\Theta(x;q_\sigma,q_\tau)_{n}}=\frac{\Theta(q_\tau ^n x;q_\sigma)}{\Theta(x;q_\sigma)}~,
%\ee
%the previous expression reduces to
\begin{multline}
%\frac{\mathcal{G}(2Z_{(1)}+Q,-2\ell)}{\mathcal{G}(2Z_{(1)},-2\ell)}=
\frac{\mathcal{G}(2X_{c'}+Q,2H_{c'})}{\mathcal{G}(2X_{c'},2H_{c'})}\times\\
\times\frac{\Theta(e^{2\pi\i\frac{Q}{r\omega_1}\cdot 2j}e^{\frac{2\pi\i}{r\omega_1}2X_{c'}}e^{\frac{2\pi\i}{r}2H_{c'}};e^{-2\pi\i\frac{\omega_3}{r\omega_1}})
\Theta(e^{2\pi\i\frac{Q}{r\omega_1}\cdot 2(j+kr+[\ell+H_{c'}])}e^{\frac{2\pi\i}{r\omega_2}2X_{c'}}e^{-\frac{2\pi\i}{r}2H_{c'}};e^{-2\pi\i\frac{\omega_3}{r\omega_2}})}{\Theta(e^{\frac{2\pi\i}{r\omega_1}2X_{c'}}e^{\frac{2\pi\i}{r}2H_{c'}};e^{-2\pi\i\frac{\omega_3}{r\omega_1}})\Theta(e^{\frac{2\pi\i}{r\omega_2}2X_{c'}}e^{-\frac{2\pi\i}{r}2H_{c'}};e^{-2\pi\i\frac{\omega_3}{r\omega_2}})}~.
\end{multline}
Similar results hold also for the other family of poles, we have just to consider the substitutions $j\rightarrow j+kr+r-[\ell+H_{c'}]$ and $j+kr+[\ell+H_{c'}]\rightarrow j$. By the usual argument for resolving the sums we find $I_{\rm SQCD}$ can be written in terms of the $r$-square of a very-well-poised elliptic hypergeometric series ${}_{N+1}E_{N}$ defined in (\ref{ellseries2})
\begin{multline}
I_{\rm SQCD}=e^{-\i\pi(\mathcal{P}_{\rm gl}+\mathcal{P}_{\rm gl}^{\rm 3d})}\sum_{c'} e^{\frac{2\pi\i X_{c'}}{r\omega_1}}e^{\frac{2\pi\i X_{c'}}{r\omega_2}}\times\frac{\mathcal{G}(Q+2X_{c'},2H_{c'})}{\mathcal{G}(2X_{c'},2H_{c'})}\prod_{a',b'}\frac{\mathcal{G}(X_{c'\bar b'},H_{c'\bar b'})}
{\mathcal{G}(Q+X_{c'a'},H_{c'a'})}\times\\
\times\Big\|{}_{2N_f+4}E_{2N_f+3}\left(e^{\frac{2\pi\i}{r\omega_1}X_{c'}}e^{\frac{2\pi\i}{r}H_{c'}};e^{\frac{2\pi\i}{r\omega_1}X_{a'}}e^{\frac{2\pi\i}{r}H_{a'}};e^{2\pi\i\frac{Q}{r\omega_1}},e^{-2\pi\i\frac{\omega_3}{r\omega_1}};1\right)\Big\|^2_{\substack{\omega_1\leftrightarrow \omega_2\\ H\leftrightarrow r-H}}~.
\end{multline}

\bibliography{nlensref}

\end{document}